\documentclass[aps,prd,reprint,twocolumn,preprintnumbers,floatfix,nofootinbib]{revtex4-2}
\usepackage[utf8]{inputenc}
\usepackage[dvips]{graphicx}
\usepackage[dvipsnames]{xcolor}
\usepackage{color}
\usepackage{relsize}
\usepackage{graphics}
\usepackage{epstopdf}
\usepackage{hyperref}
\usepackage{mathrsfs}
\usepackage{amssymb}
\usepackage{physics}
\usepackage{booktabs}
\usepackage{float}

\usepackage[normalem]{ ulem }
\usepackage{amsthm}
\usepackage{amsmath}
\usepackage{cancel}
\usepackage{fontawesome} 
\usepackage[caption=false]{subfig}
\usepackage{tabularx,ragged2e} 
\newcolumntype{L}{>{\RaggedRight\arraybackslash}X}

\usepackage{hyperref}

\newcommand{\TBH}{T}
\newcommand{\MBH}{M}
\newcommand{\MBHi}{M_{\rm in}}

\newcommand{\as}{a_\star}
\newcommand{\asi}{a_\star^{\rm in}}

\definecolor{palatd}{RGB}{104, 36, 109}
\definecolor{palatb}{RGB}{0, 56, 168}
\definecolor{palatr}{rgb}{0.745,0.118,0.176}
\newcommand\myshade{80}
\colorlet{mylinkcolor}{palatr}
\colorlet{mycitecolor}{palatb}
\colorlet{myurlcolor}{palatd}

\hypersetup{
	linkcolor  = mylinkcolor!\myshade!black,
	citecolor  = mycitecolor!\myshade!black,
	urlcolor   = myurlcolor!\myshade!black,
	colorlinks = true
}

\usepackage{pgfplots}

\pgfplotsset{compat=1.17}

\begin{document}
	\sloppy  
	
	\preprint{IPPP/23/36}
	
	\title{Identifying Spin Properties of Evaporating Black Holes through\\ Asymmetric Neutrino and Photon Emission}
	
	\author{Yuber F. Perez-Gonzalez}
	\email{yuber.f.perez-gonzalez@durham.ac.uk}
	
	\affiliation{Institute for Particle Physics Phenomenology, Durham University, South Road DH1 3LE, Durham, United Kingdom}
	
	\begin{abstract}
		Kerr black holes radiate neutrinos in an asymmetric pattern, preferentially in the lower hemisphere relative to the black hole's rotation axis, while antineutrinos are predominantly produced in the upper hemisphere. 
		Leveraging this asymmetric emission, we explore the potential of high-energy, $E_\nu \gtrsim 1$ TeV, neutrino and antineutrino detection to reveal crucial characteristics of an evaporating primordial black hole at the time of its burst when observed near Earth. 
		We improve upon previous calculations by carefully accounting for the non-isotropic particle emission, as Earth occupies a privileged angle relative to the black hole's rotation axis. 
		Additionally, we investigate the angular dependence of primary and secondary photon spectra and assess the evaporating black hole's time evolution during the final explosive stages of its lifetime. 
		Since photon events outnumber neutrinos by about three orders of magnitude, we find that a neutrino measurement can aid in identifying the initial angular momentum and the black hole hemisphere facing Earth only for evaporating black holes within our solar system, at distances $\lesssim 10^{-4}$ pc, and observed during the final 100 s of their lifetime.
		Codes used in this work will be publicly available in \href{https://github.com/yfperezg/evapbh}{\faGithub}.
	\end{abstract}
	
	\maketitle
	
	\section{Introduction}

	The Early Universe's high density provides an ideal environment for the formation of black holes (BH) with masses significantly smaller than those observed in Gra\-vi\-ta\-tio\-nal Waves or direct astrophysical measurements~\cite{LIGOScientific:2016aoc,LIGOScientific:2017vwq,EventHorizonTelescope:2019dse}. 
	These \emph{primordial} black holes (PBHs) could profoundly impact the evolution of our Universe~\cite{Zeldovich:1967lct,Hawking:1971ei,Carr:1974nx}.
	Given their potentially tiny masses, as small as the Planck mass, it becomes imperative to take into account quantum effects in the evolution of PBHs.
	Using a semi-classical approximation, Hawking demonstrated that BHs evaporate by emitting a flux of particles with a thermal spectrum~\cite{Hawking:1974rv,Hawking:1974sw}. 
	As a consequence of energy conservation arguments, the BH mass decreases at a rate $\dot{M} \sim M^{-2}$, initially leading to a slow evaporation. 
	However, particle emission accelerates, triggering a runaway effect that culminates in an explosive stage.
	Such evaporation process, assuming the Standard Model's (SM) degrees-of-freedom (dofs), enables to estimate the initial mass of a PBH whose lifetime matches the age of the Universe, yielding a value of $\sim 10^{15}$ g. 
	As a result, PBHs with masses smaller than this value would have already evaporated, establishing constraints to be derived on the initial PBH abundance if they evaporated during the Big-Bang Nucleosynthesis~\cite{Carr:2009jm,Carr:2020gox,Keith:2020jww}, reionization~\cite{He:2002vz,Mack:2008nv}  or the formation of the CMB~\cite{Carr:2009jm,Carr:2020gox}.
	Furthermore, due to the u\-ni\-ver\-sa\-li\-ty of Hawking evaporation, PBHs could have produced dofs that do not interact with the SM sector, potentially contributing to the observed Dark Matter (DM)~\cite{Cheek:2021odj, Cheek:2021cfe, Hooper:2019gtx, Masina:2020xhk,Morrison:2018xla, Auffinger:2020afu, Khlopov:2004tn, Allahverdi:2017sks, Lennon:2017tqq, Gondolo:2020uqv, Baldes:2020nuv, Bernal:2020bjf, Bernal:2020ili, Masina:2021zpu, Kitabayashi:2021hox, Bernal:2021bbv,Bernal:2020kse,Bernal:2022oha,Cheek:2022mmy}, to the relativistic dofs, parameterized by $\Delta N_{\rm eff}$~\cite{Cheek:2022dbx, Hooper:2019gtx, Hooper:2020evu, Masina:2020xhk, Arbey:2021ysg, Masina:2021zpu, Cheek:2022dbx}, or modify the generation of the matter-antimatter asymmetry~\cite{Perez-Gonzalez:2020vnz, JyotiDas:2021shi,Datta:2020bht, Morrison:2018xla, Fujita:2014hha, Granelli:2020pim, Hook:2014mla, Hamada:2016jnq,Hooper:2020otu,Chaudhuri:2020wjo,Bernal:2022pue,Barman:2022pdo,Gehrman:2022imk,Agashe:2022phd,Gehrman:2023esa,Friedlander:2023qmc}. 
	Conversely, if their initial mass exceeds $10^{15}$ g, these PBHs could potentially contribute to a portion of the observed DM~\cite{Carr:2016drx,Carr:2020gox,Green:2020jor,Villanueva-Domingo:2021spv,Friedlander:2022ttk}.
	For a comprehensive review, see Ref.~\cite{Auffinger:2022khh}.

	There remains the possibility that some of the PBHs would have an initial mass such that they would be evaporating today.
	These evaporating primordial black holes (EPBHs) would emit a large flux of particles during the final explosive stages of their lifetimes, presenting a potential opportunity for observation in different facilities.
	Thus, experiments designed to search for gamma-ray bursts, such as H.E.S.S~\cite{Glicenstein:2013vha,Tavernier:2019exh}, Milagro~\cite{Abdo:2014apa}, VERITAS~\cite{Archambault:2017asc}, HAWC~\cite{HAWC:2013kzm,HAWC:2019wla}, and Fermi-LAT~\cite{Fermi-LAT:2018pfs}, have already placed constraints on the existence of PBHs in our local region.
	However, it is worth noting that other particles, such as neutrinos~\cite{Halzen:1995hu,Dave:2019epr,Capanema:2021hnm}, could also be measurable if an EPBH happens to occur close enough to Earth. 
	Undoubtedly, observing PBH evaporation in its final stages would not only be a triumph for theoretical physics but also hold the potential to unlock a wealth of new physics. 
	For instance, this unique phenomenon could provide invaluable insights into the complete spectrum of particles, including those beyond-the-Standard-Model (BSM), that exist nature~\cite{Ukwatta:2015iba, Baker:2021btk, Baker:2022rkn,Boluna:2023jlo}.

	Another avenue to explore BSM physics in EPBHs involves determining whether these objects had a substantial angular momentum at the onset of their final bursts. 
	In a scenario that includes only the SM set of particles, even a close-to-maximally rotating PBH should have lost most of its angular momentum in the early stages of the Hawking evaporation~\cite{Page:1976df,Page:1976ki}. 
	However, the presence of an abundant number of scalar dofs, as expected in string axiverse models~\cite{Arvanitaki:2009fg}, could significantly modify this behaviour. 
	In such cases, the BH's angular momentum might not deplete as rapidly as in the standard scenario since scalar particles only reduce the BH mass but not its angular momentum~\cite{Chambers:1997ai, Taylor:1998dk}.
	This leads to PBHs approaching their final stages with substantial angular momenta~\cite{Calza:2021czr,Calza:2022ljw}. 
	Thus, investigating the angular momentum properties of EPBHs offers an intriguing window into potential BSM physics.
	
	\begin{figure}[!t]
		\centering
		\includegraphics[width=0.75\linewidth]{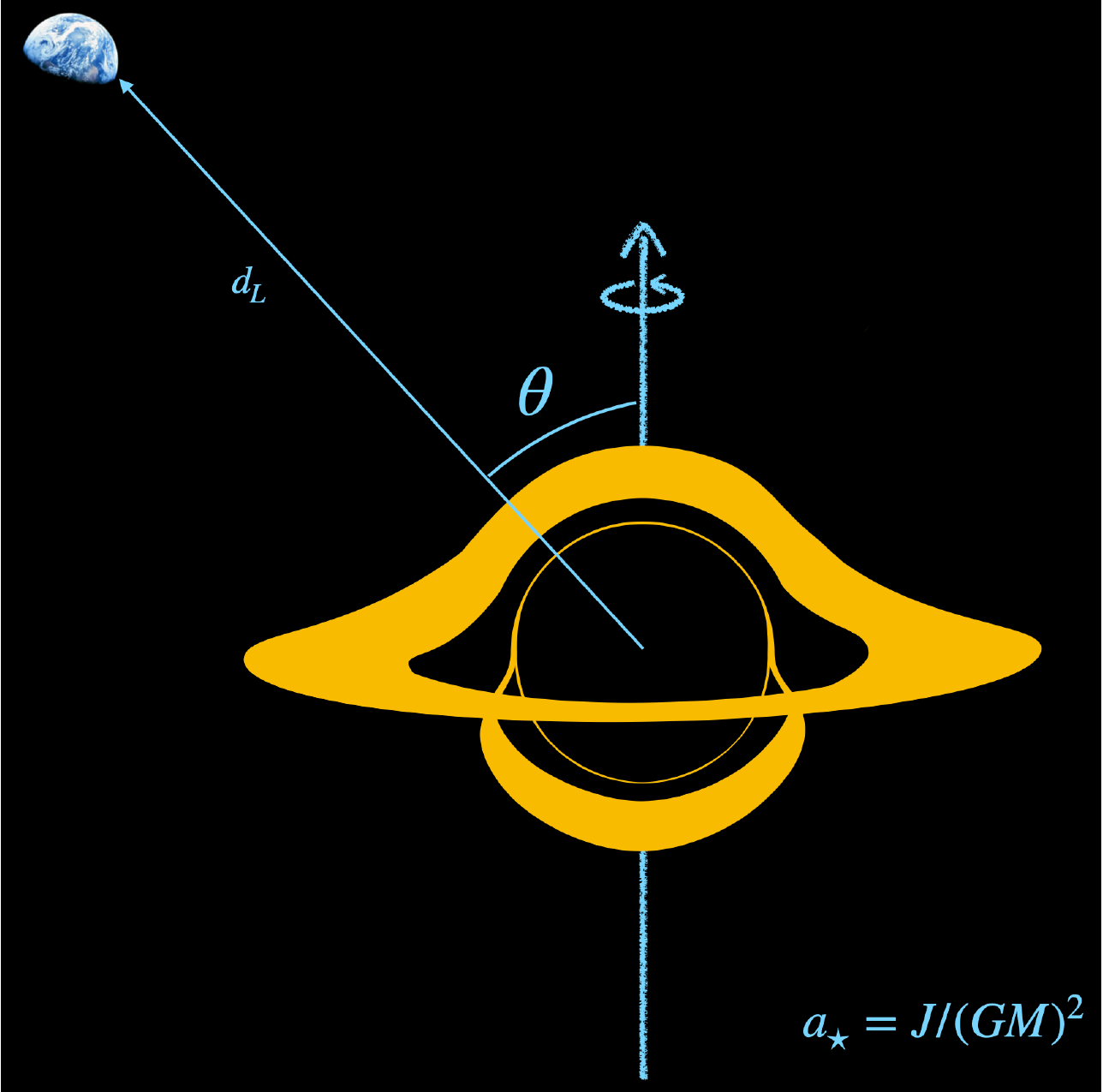}
		\caption{\label{fig:Ills} An illustration of the Earth-Evaporating Kerr Black Hole system considered in this work. Earth is placed at an angle $\theta$ with respect to the rotation axis and at a distance $d_L$. Earthrise picture reproduced from~\cite{nasa_earthrise}.}
	\end{figure}
	Previous studies have examined the impact of non-zero angular momentum on the photon and neutrino emissions from an EPBH~\cite{Capanema:2021hnm,Calza:2022ljw}. 
	These works used the {\tt BlackHawk} code~\cite{Arbey:2019mbc, Arbey:2021mbl}, which calculates both the primary particle spectrum directly emitted during evaporation and the secondary spectrum originating from the decay of unstable particles produced by the black hole. 
	However, it is important to note that {\tt BlackHawk} computes the angle-integrated spectrum, which is not suitable for a Kerr EPBH since rotating EPBHs exhibit a non-isotropic particle emission due to their axisymmetric nature. 
	Consequently, if an EPBH were to burst close to Earth, observatories would measure the particle flux at a unique angle with respect to the rotation axis, leading to distinct observational characteristics, see Fig.~\ref{fig:Ills} for an artistic depiction of the Earth--EPBH system considered in this work.
	Additionally, in the specific case of neutrinos, it has been observed that due to the parity violation in weak interactions, their emission from an EPBH follows an asymmetric pattern: neutrinos are preferentially emitted in the EPBH's ``southern'' hemisphere, while antineutrinos are predominantly produced in the ``northern'' hemisphere~\cite{Unruh:1973bda, Vilenkin:1978is,  Leahy:1979xi, Vilenkin:1979ui, Vilenkin:1980ft}. 
	This well-established behavior opens up new avenues for exploring the angular momentum of an EPBH since neutrino emission is intrinsically linked to the black hole's spin and varies with the polar angle relative to its rotation axis.

	In this work, we conduct a comprehensive analysis of the angular distribution of neutrino and photon emissions, both primary and secondary, from Kerr EPBHs. 
	Additionally, we investigate the potential of gamma-ray and neutrino observatories to discern crucial EPBH properties, such as its angular momentum and the hemisphere facing Earth.
	Despite the photon events outnumbering neutrino events by approximately three orders of magnitude, the measurement of neutrinos is instrumental in determining the EPBH hemisphere facing Earth, particularly in the scenario where such an object is in close proximity to our planet.

	The structure of this paper is as follows. In Sec.~\ref{sec:AngDisK}, we provide a detailed description of the neutrino asymmetric emission and the photon angular distribution.  Furthermore, within the same section, we first outline a numerical procedure to compute the angular dependence of the secondary particle emission. Moving forward, in Sec.~\ref{sec:TimEv}, we delve into the time evolution of a Kerr EPBH during its final burst, critically assessing its impact on the neutrino emission asymmetry. Sec.~\ref{sec:Detect} is exclusively dedicated to exploring the detection prospects, with specific emphasis on the IceCube~\cite{IceCube:2018pgc,IceCube:2019cia} and HAWC~\cite{HAWC:2019wla} observatories. Moreover, we thoroughly examine the feasibility of determining the initial characteristics of EPBHs through the combination of neutrino and photon measurements.  In Sec.~\ref{sec:conc}, we draw our conclusions. We have included two  appendices: App.~\ref{ap:FerKBs} provides a short review on the Dirac equation in the Kerr spacetime, and App.~\ref{ap:TeuKBs} contains the effective potentials for scalar and vectors used to obtain absorption probabilities. We consider natural units where $\hbar = c = k_{\rm B} = 1$, and define the Planck mass to be $M_p=1/\sqrt{G}$, with $G$ the gravitational constant, throughout this manuscript.

	\section{Angular distribution of Particle Emission from Kerr Black Holes}\label{sec:AngDisK}
	
	Let us consider the Hawking radiation emitted from a Kerr BH with an instantaneous mass $\MBH$ and a dimensionless spin parameter $\as \equiv {\rm J} /(G\MBH)^2\in[0, 1)$, where ${\rm J}$ represents the BH angular momentum. 
	To analyse the angular distribution of particles, we adopt the Boyer-Lindquist coordinate system $(t,r,\theta,\phi)$. 
	In these coordinates, the line element for the Kerr spacetime is given by~\cite{Boyer:1966qh}
	\begin{widetext}
		\begin{align}
			\dd s^2=\frac{\Delta}{\Sigma}(\dd t - a \sin^2\theta \dd\phi)^2 - \frac{\sin^2\theta}{\Sigma}(-a \dd t +(r^2+a^2)\dd\phi)^2-\frac{\Sigma}{\Delta}\dd r^2 - \Sigma \dd\theta^2,
		\end{align}
	\end{widetext}
	where $\Delta \equiv r^2 - 2 G\MBH r + a^2$, $\Sigma = r^2 + a^2\cos\theta$, with $a = \as GM$.
	It has been demonstrated that the equations of motion for bosons and fermions are separable, leading to the well-known Teukolsky master equations~\cite{Teukolsky:1973ha,Press:1973zz,Teukolsky:1974yv}.
	Such a separability will allow us to analyse the angular distribution of particles, and thus the possible determination of the EPBH initial spin at the onset of a burst.
	Next, we will consider the primary neutrino emission and its dependence on the polar angle $\theta$, as well as the photon emission. 
	Subsequently, we will examine the secondary spectra resulting from the decay of unstable particles produced by the BH evaporation.
	
	\subsection{Primary Neutrino Emission and Asymmetry} 
	
	Early studies focused on neutrinos as a prototype for understanding the properties of massless fermions in the Kerr spacetime~\cite{Unruh:1973bda}. 
	This choice was influenced by the prevailing belief at that time that neutrinos were massless. 
	However, we now know that neutrinos are indeed massive particles that exhibit mixing. 
	Taking this into account, we have recently investigated the implications of neutrino mass on the process of Hawking evaporation for Schwarzschild BHs~\cite{Lunardini:2019zob}. 
	Here, we extend the discussion of massive neutrinos to Kerr BHs, and examine their asymmetric emission as a function of the polar angle.

	The Dirac equation and its separability for massive fermions in the Kerr background has been extensively studied in Refs.~\cite{Chandrasekhar:1976ap,Page:1976jj,Dolan:2009kj,Dolan:2015eua}.
	For sake of completeness, we briefly describe the properties of this equation in App.~\ref{ap:FerKBs}.
	After proposing an ansatz, we obtain equations for radial $R_{1,2}(r)$ and angular functions $S_{1,2}(\theta)$ appearing in the spinor components, depending on the energy $\omega$, and $l,m$ the total and axial angular momentum quantum numbers~\cite{Dolan:2015eua} (for details see App.~\ref{ap:FerKBs})
	\begin{widetext}
		\begin{subequations}\label{eq:DREqs}
			\begin{align}
				\sqrt{\Delta}(\partial_r - i K/\Delta) R_1 &= (\!\,_{\frac{1}{2}}A_{lm} + i \mu r) R_2, \\
				\sqrt{\Delta}(\partial_r + i K/\Delta) R_2 &= (\!\,_{\frac{1}{2}}A_{lm} - i \mu r) R_1,
			\end{align}
		\end{subequations}
		and
		\begin{subequations}\label{eq:DSEqs}
			\begin{align}
				\left(\partial_\theta - \frac{1}{2}\cot\theta - m \csc\theta + a \omega \sin\theta \right)S_{1}(\theta) &= (+\!\,_{\frac{1}{2}}A_{lm} + a\mu\cos\theta)S_{2}(\theta),\\
				\left(\partial_\theta - \frac{1}{2}\cot\theta + m \csc\theta - a \omega \sin\theta \right)S_2(\theta) &= (-\!\,_{\frac{1}{2}}A_{lm} + a\mu\cos\theta)S_1(\theta),
			\end{align}
		\end{subequations}
	\end{widetext}
	where  $K \equiv (r^2 + a^2)\omega - a m$ and $\mu$ is the fermion mass. 
	The quantities $\!\,_{\frac{1}{2}}A_{lm}$ represent the angular separation constants, which are simply $\!\,_{\frac{1}{2}}A_{lm} = {\cal P}(l+1/2)$, ${\cal P}=\pm 1$, for a Schwarzschild BH. 
	
	The solutions to the radial equations, Eqs.~\eqref{eq:DREqs}, allow us to determine the absorption probabilities $\!\,_{\frac{1}{2}}\Gamma_{lm}$, which are crucial for calculating the Hawking emission rate as they describe the effects of centrifugal and gravitational potentials on particle production ~\cite{Hawking:1974rv,Hawking:1974sw,Page:1976df,Page:1976ki,Page:1977um}.
	Numerical methods are required to find these solutions. 
	We employ the general procedure for Kerr-Newman BHs established in Ref.~\cite{Page:1977um}, where a transformation of variables is applied to simplify the numerical treatment. 
	Once the solutions have been obtained, we calculate the absorption probabilities $\!\,_\frac{1}{2}\Gamma_{lm}$ using the expressions provided in the appendix of Ref.~\cite{Page:1977um}.
	
	To investigate the dependence of neutrino production on the polar angle $\theta$ from an EPBH, it is also necessary to solve the angular equations, Eqs.~\eqref{eq:DSEqs}. Several numerical methods have been proposed in the literature to solve these equations, see e.g.~\cite{Batic:2004sz,Batic:2005va,Dolan:2009kj,Neznamov:2016qej}. 
	In this work, we closely follow the approach described in Ref.~\cite{Dolan:2009kj}.
	The method involves employing a series expansion for the eigenfunctions  $\!\,_{\pm \frac{1}{2}}S_{lm}\equiv S_{1,2}$ using associated Legendre polynomials, leading to a continued fraction equation. 
	By numerically solving this equation, we obtain the angular eigenvalues $\!\,_{\frac{1}{2}}A_{lm}$. Subsequently, the coefficients of the series expansion for the eigenfunctions are computed using these angular eigenvalues. 
	Through this numerical procedure, we find the angular eigenfunctions up to a normalization factor.
	To fix such a factor, we impose the normalization condition,
	\begin{align}
		\int |\!\,_{\pm \frac{1}{2}}S_{lm}(\theta)|^2\, \dd\Omega = 1.
	\end{align}
	
	The rates of neutrino emission as a function of time, energy, and solid angle $\Omega$, are computed after considering the quantization of fermions in a Kerr background. 
	The quantum field theory for massless fermions has been extensively studied in the literature, and interested readers can find detailed discussions in Refs.~\cite{Leahy:1979xi,Unruh:1973bda, Casals:2006xp,Casals:2012es,Winstanley:2013hua}. 
	However, to the best of our knowledge, currently there is no available treatment for the quantization of massive fermions in the Kerr geometry; for potential issues related to this particular case, see Ref.~\cite{Casals:2012es}. 
	Nevertheless, it is worth noting that the neutrino mass $m_\nu \lesssim 1~\text{eV}$ is negligible compared to the energies observed in neutrino telescopes, $E_\nu \gtrsim \text{100 GeV}$. 
	Consequently, we can reasonably \emph{assume} that the quantization procedure used for massless fermions is applicable with a high level of accuracy to neutrinos originating from an EPBH, a supposition that we will adopt from now on.
	
	The main physical quantity of interest is the expectation value of the current operator\footnote{Note that the commutator in the current definition should be taken only with respect to the quantum operators and not the spinor structure. }
	\begin{align}
		J^\mu = \frac{1}{2}[\overline{\nu},\gamma^\mu P_L \nu]
	\end{align}
	where $\nu$ represents the neutrino field.
	Here, $P_L=\frac{1}{2}(1-\gamma^5)$ is the left chiral projector defined in terms of the chirality matrix $\gamma^5$, see App.~\ref{ap:FerKBs}. 
	As discussed in Ref.~\cite{Lunardini:2019zob}, neutrinos and antineutrinos are expected to be emitted as mass eigenstates, $\nu_{1,2,3}$.
	The neutrino field $\nu$ will represent generally any of the three fields associated to the mass eigenstates henceforth.
	The radial component of the current $J^r$ describes the flow of neutrinos minus antineutrinos~\cite{Leahy:1979xi,Casals:2012es}.
	The net neutrino minus antineutrino flux, ${\cal A}\equiv N_{\nu} - N_{\overline{\nu}}$, per unit time and solid angle in the radial direction \emph{away} from the BH corresponds to the expectation value~\cite{Leahy:1979xi}
	\begin{align}
		\frac{\dd^2{\cal A}}{r^2\dd\Omega \dd t} = \langle U^- | J^r | U^- \rangle,
	\end{align}
	where $|U^- \rangle$ represents the past Unruh vacuum~\cite{Unruh:1976db}. 
	This vacuum state, defined as the state that does not contain any particles incoming from the null past infinity ${\cal J}^-$, is the relevant for our purposes since we are considering the evaporation of a BH originated from a gravitational collapse event. 
	
	In terms of the angular functions and absorption probabilities defined previously, the net neutrino emission flux can be expressed as follows~\cite{Unruh:1973bda, Vilenkin:1978is,Leahy:1979xi,Vilenkin:1979ui,Casals:2012es}
	\begin{widetext}
		\begin{align}\label{eq:d2A}
			\frac{{\dd^2 {\cal A}}}{\dd\Omega \dd t } = \frac{1}{4\pi}\sum_{l=1/2}\sum_{m=-l}^l \int_0^\infty \dd\omega \frac{\!\,_\frac{1}{2}\Gamma_{lm}}{\exp(\varpi/\TBH)+1} \{|\!\,_{-\frac{1}{2}}S_{lm}(\theta)|^2 - |\!\,_{+\frac{1}{2}}S_{lm}(\theta)|^2\},
		\end{align}
		where $\varpi = \omega - m \vartheta$ with $\vartheta = a_\star/(2G\MBH(1+\sqrt{1-a_\star^2}))$ representing the horizon's angular velocity. 
		From the form of the emission asymmetry, we can interpret the term proportional to $\!\,_{-\frac{1}{2}}S_{lm}(\theta)$ as the contribution coming from neutrinos, while $\!\,_{-\frac{1}{2}}S_{lm}(\theta)$ corresponds to the antineutrino term, with the caveat that such interpretation is valid only far from the EPBH~\cite{Casals:2006xp}.
		In Eq.~\eqref{eq:d2A}, it is necessary to write explicitly the sum over the angular momentum quantum numbers as the PBH spin breaks the spherical symmetry, given the explicit dependence of the Hawking rate on $m$. 
		The Hawking temperature $\TBH$ is given by
		\begin{align}
			\TBH = \frac{1}{4\pi G \MBH}\, \frac{\sqrt{1-a_\star^2}}{1+\sqrt{1-a_\star^2}}.
		\end{align}
		Thus, we can define the net particle emission rate within time $\dd t$, solid angle $\dd\Omega$ and energy $[\omega, \omega+\dd\omega]$ as
		\begin{align}\label{eq:d3Anu}
			\frac{{\dd^3 {\cal A}}}{\dd\omega \dd t \dd\Omega} = \frac{1}{4\pi}\sum_{l=1/2}\sum_{m=-l}^l \frac{\!\,_\frac{1}{2}\Gamma_{lm}}{\exp(\varpi/\TBH)+1} \{|_{-\frac{1}{2}}S_{lm}(\theta)|^2 - |_{+\frac{1}{2}}S_{lm}(\theta)|^2\}.
		\end{align}
		Notice that the integration of the rate over the solid angle produces a vanishing net particle emission, as expected.
		Analogously, let us define the total emission rate, i.e., neutrino plus antineutrino rate as
		\begin{align}\label{eq:d3Nnu}
			\frac{\dd^3 N_{\nu+\overline{\nu}}}{\dd\omega \dd t \dd\Omega} = \frac{1}{4\pi}\sum_{l=1/2}\sum_{m=-l}^l \frac{\!\,_\frac{1}{2}\Gamma_{lm}}{\exp(\varpi/\TBH)+1} \{|_{-\frac{1}{2}}S_{lm}(\theta)|^2 + |_{+\frac{1}{2}}S_{lm}(\theta)|^2\}.
		\end{align}
	\end{widetext}
	This total flux of neutrinos comes directly from the evaporation of the BH, thus constituting what is referred to as the primary spectrum in the literature. 
	
	\begin{figure*}
		\centering
		\includegraphics[width=\linewidth]{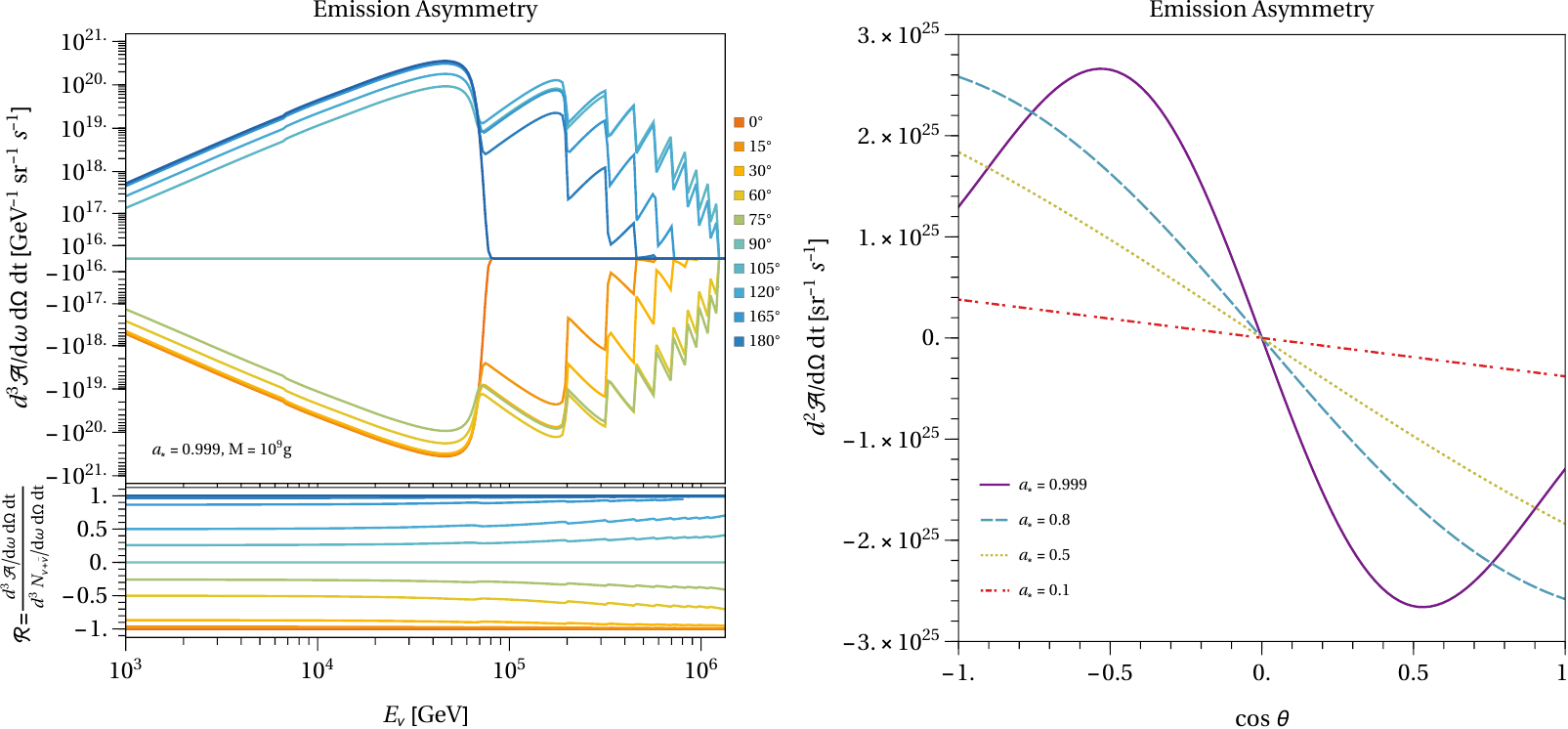}
		\caption{\label{fig:d3A} (Top left) Instantaneous neutrino minus antineutrino net particle flux,  ${\cal A}= N_{\nu} - N_{\overline{\nu}}$,  as function of energy for a BH having a mass of $\MBH = 10^{9}~{\rm g}$ and spin parameter $\as = 0.999$ for different values of the polar angle $\theta\in[0,\pi]$. (Bottom left) Instantaneous asymmetry ratio ${\cal R}$ between the net neutrino minus antineutrino flux to the total emission rate $\dd^3 N_{\nu+\overline{\nu}}/\dd\omega \dd t \dd\Omega$, cf.~Eq.~\eqref{eq:d3Nnu}. (Right) Energy-integrated instantaneous neutrino emission asymmetry as function of the polar angle $\theta$ for different values of the spin parameter, $\as = 0.999$ (purple full), $\as=0.8$ (dashed blue), $\as=0.5$ (dotted dark yellow), and $\as=0.1$ (dot-dashed red).}
	\end{figure*}
	On the top left of Fig.~\ref{fig:d3A}, we present the instantaneous neutrino minus antineutrino net flux $\dd^3{\cal A}/\dd\omega \dd\Omega \dd t$ as function of the neutrino energy for a BH having a mass of $\MBH = 10^{9}~{\rm g}$ and spin parameter $\as = 0.999$.
	The polar angle at which the observer is placed varies from $0^\circ$ to $180^\circ$.
	When $\theta=0$, we observe that the net rate is negative and presents a single peak.
	This distinctive feature arises because only the angular eigenfunctions with $l=m=\frac{1}{2}$ contribute at the poles, while the remaining modes vanish there.
	Moreover, the negative sign indicates that more antineutrinos than neutrinos are emitted.
	As we increase the polar angle, the contribution from higher $l$ modes becomes significant, resulting in additional peaks in the net flux. 
	The maximum emission asymmetry occurs at approximately $\theta \approx 60^\circ$, which aligns with previous findings~\cite{Leahy:1979xi}.
	For polar angles within the range $60^\circ\lesssim \theta < 180^\circ$, the net flux is reduced, and at $\theta=\pi$, the net flux completely vanishes due to a cancellation between the angular eigenfunctions $\!\,_{\pm\frac{1}{2}}S_{lm}(\theta)$ at the equator. 
	In the southern hemisphere, $90^\circ \leq \theta \leq 180^\circ$, the net flux becomes positive, indicating a higher number of emitted neutrinos compared to antineutrinos. 
	Despite this sign change, the behavior in the southern hemisphere is analogous to that in the northern hemisphere, $0^\circ \leq \theta \leq 90^\circ$,. 
	The net flux reaches its maximum value around $\theta \approx 120^\circ$, while at $\theta = 180^\circ$, only the $l=m=\frac{1}{2}$ mode contributes.

	To quantify the EPBH emission asymmetry with respect to the total instantaneous neutrino flux, we introduce the asymmetry ratio ${\cal R}$ which compares the net neutrino minus antineutrino flux to the total emission rate,
	\begin{align}
		{\cal R} \equiv \frac{\frac{{\dd^3 {\cal A}}}{\dd\omega \dd t \dd\Omega}}{\frac{\dd^3 N_{\nu+\overline{\nu}}}{\dd\omega \dd t \dd\Omega}}.
	\end{align}
	We present on the bottom left of Fig.~\ref{fig:d3A} the asymmetry ratio ${\cal R}$ for the same values for the polar angles $\theta$ as in the top pannel.
	The results indicate that at the poles ($\theta = 0^\circ, 180^\circ$), the asymmetry ratio approaches -1 and 1, respectively. 
	This implies that at these angles, the flux is predominantly composed of antineutrinos and neutrinos.
	For angles different from the poles, the absolute value of the ratio $|{\cal R}|$ remains below 1, and it is constant for neutrino energies $E_\nu\lesssim 5\times 10^4$ GeV, where the dominating mode is $l=m=\frac{1}{2}$. 
	However, at higher energies, the contribution of higher-$l$ modes becomes evident, resulting in ripples in the ratio.
	For the specific case of $\theta = 60^\circ$, the asymmetry ratio ${\cal R} \approx -0.5$ for energies $E_\nu\lesssim 5\times 10^4$ GeV. This confirms that at this angle, there are approximately $50\%$ more antineutrinos than neutrinos being emitted.
	Conversely, at $\theta = 120^\circ$, we observe the opposite behavior, as expected from the previous discussion.
	Finally, at the equator, $\theta = 90^\circ$, the ratio ${\cal R}$ is precisely 0, indicating that the number of emitted neutrinos and antineutrinos is equal at this angle.
	
	To comprehend the dependence of the neutrino-antineutrino net flux on the BH spin parameter, we present in the right panel of Fig.~\ref{fig:d3A} the energy-integrated emission asymmetry, $\dd^2{\cal A}/\dd\Omega \dd t$, cf.~Eq.\eqref{eq:d2A}, for varying spin parameter values, $\as = 0.999$ (purple full), $\as=0.8$ (dashed blue), $\as=0.5$ (dotted dark yellow), and $\as=0.1$ (dot-dashed red).
	For a close-to-maximally rotating BH with $\as=0.999$, we observe that the net emission is maximal at a value $|\cos\theta|\approx 0.5$, whereas for other spin parameter values, the emission asymmetry peaks at the poles. This behavior is a consequence of higher-$l$ modes increasingly dominating for $\as\gtrsim 0.8$, which in turn contribute away from the poles. 
	On the other hand, for $\as\lesssim 0.8$, the $l$ mode that predominantly dominates is the $l=m=\frac{1}{2}$, resulting in a maximum emission at the poles. 
	As a result, the net emission asymmetry peaks at the poles for these values of the spin parameter.
	For all spin parameter values presented, we observe that the net flux is negative for $\cos\theta > 0$, indicating a preference for antineutrino emission over neutrinos in the northern hemisphere. 
	Conversely, for negative values of $\cos\theta$, the behavior is opposite, with a preference for neutrino emission. 
	At the equator, the net emission asymmetry vanishes, ensuring that the numbers of emitted neutrinos and antineutrinos coincide.
	This overall pattern demonstrates that antineutrinos are preferentially emitted in the northern hemisphere, whereas neutrinos are predominantly produced in the southern hemisphere. 
	
	Neutrino physics holds an unresolved question regarding the fundamental fermionic nature of these particles.
	Neutrinos can be either their own antiparticles, known as Majorana fermions, or as distinct entities from their antiparticles, referred to as Dirac neutrinos. 
	In Ref.~\cite{Lunardini:2019zob}, we investigated the impact of each scenario on Hawking radiation from Schwarzschild black holes. Now, the question arises as to the main effect for Kerr black holes.
	In the case of Dirac neutrinos, additional dofs come into play, right-handed neutrinos and left-handed antineutrinos. 
	Consequently, the black hole would emit these states asymmetrically, with right-handed neutrinos predominantly emitted in the northern hemisphere and left-handed antineutrinos in the southern hemisphere. 
	This raises the possibility that the previously mentioned overall asymmetry might be removed. 
	However, the detection of these additional states via weak interactions is hindered by a helicity factor of $m_\nu/E_\nu \sim 10^{-11}$, for the energy range of interest. 
	As a result, the presence of the additional states becomes negligible, and the asymmetry remains intact.
	
	For Majorana neutrinos, only two dofs exist per active mass eigenstate, corresponding to positive and negative helicities.
	While there is technically no distinct state known as an antineutrino, weak interactions differentiate between the helicity states due to parity violation.
	Consequently, a positive-helicity state exhibits a different interaction compared to a negative-helicity one in the ultra-relativistic limit, which is of interest to us here.
	Thus, in such a limit, it is customary to denote the negative-helicity states as ``neutrinos'' and the positive-helicity ones as ``antineutrinos''~\cite{Kayser:1982br,Kayser:1981nw}.
	As a result, the asymmetric emission is also present for Majorana neutrinos.
	Consequently, detecting a discrepancy between the numbers of neutrinos and antineutrinos in a future  detection of an EPBH could indicate the presence of a nonzero angular momentum.
	
	\subsection{Primary Photon Emission}
	
	The emission of particles with higher spin, such as photons, is enhanced for rotating BHs~\cite{Page:1976df}. 
	Consequently, we anticipate that photon emission will be strongly influenced by both the EPBH spin and the polar angle $\theta$.
	To investigate this dependence, we consider the equations for the radial $\psi_s$ and the angular functions $S(\theta)$ for a massless field with spin $s$,
	\begin{widetext}
		\begin{subequations}\label{eq:TeuEqs}
			\begin{align}
				\frac{\dd^2 \psi_s}{\dd r_\star^2} + (\omega^2 - V_s(r)\psi_s &= 0,\label{eq:TeuREqs}\\
				\frac{1}{\sin\theta}\frac{\dd}{\dd\theta}\left(\sin\theta \frac{\dd S}{\dd\theta}\right) + [(c\cos\theta - s)^2 - (m\csc\theta + s\cot\theta)^2 - s(s-1) + \!\,_sA_{lm}]\, S &=0 ,\label{eq:TeuSEqs}
			\end{align}
		\end{subequations}
	\end{widetext}
	where $\lambda_{s} \equiv \!\,_{s}A_{lm} + c^2 - 2 m c$, being $c = a \omega$, and $\!\,_{s}A_{lm}$ are the angular eigenvalues, and $r_\star$ is the  Eddington–Finkelstein radial coordinate
	\begin{align}
		\frac{\dd r_\star}{\dd r} = \frac{r^2+a^2+am/\omega}{\Delta}.
	\end{align}
	The Schr\"odinger-like equation Eq.~\eqref{eq:TeuREqs} is obtained through the Chandrashekar-Detweiler method as presented in Refs.~\cite{Chandrasekhar:1975zz,Chandrasekhar:1976zz,Chandrasekhar:1977kf}.
	The potentials $V_s$ depend on the spin of the particle, which, for completeness, are provided in App.~\ref{ap:TeuKBs}. 
	To compute the absorption probabilities $\!\,_s\Gamma_{lm}$ for massless bosons, we solve the wave equation, Eq.~\eqref{eq:TeuREqs}, and compute the transmission coefficient for a purely ingoing wave\footnote{It is worth noting that we have verified the consistency of our results for the absorption probabilities $\!\,_s\Gamma_{lm}$  with those obtained from the {\tt BlackHawk} code~\cite{Arbey:2019mbc,Arbey:2021mbl}.}.
	On the other hand, for the angular eigenvalues and eigenfunctions, we adopt the approach outlined in Refs.~\cite{Leaver:1985ax,Berti:2005gp}, where the angular separation constants are obtained by numerically solving a continued fraction equation, similar to the case of massive fermions.
	
\begin{figure}
		\centering
		\includegraphics[width=0.95\linewidth]{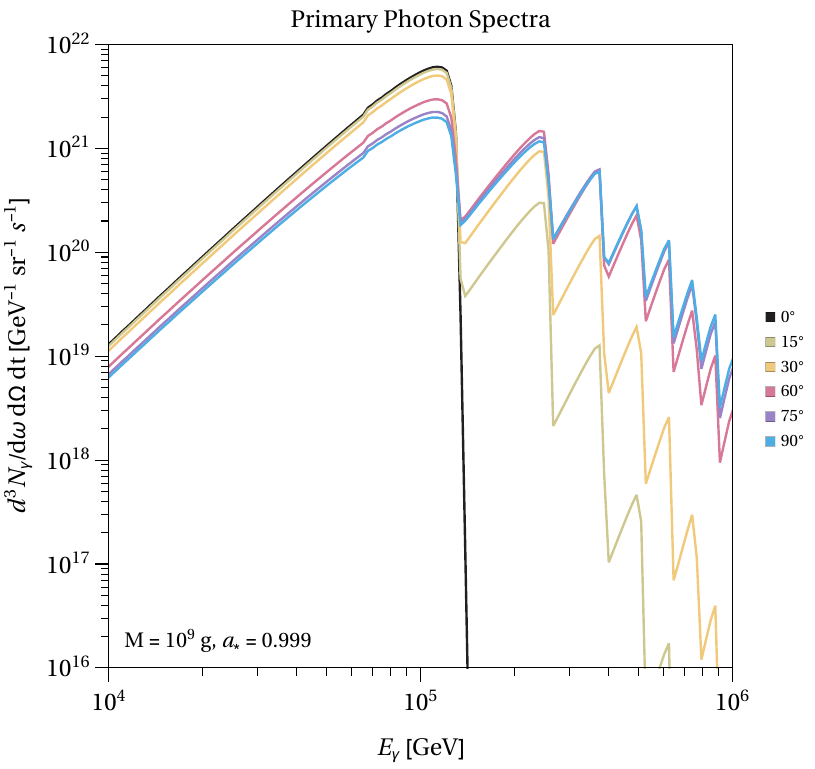}
		\caption{\label{fig:d3Ng} Primary instantaneous photon emission spectrum from a BH having a mass $M=10^{9}$ g and spin parameter $\as = 0.999$ for different values of the polar angle, $\theta = 0.^\circ$ (black), $15^\circ$ (beige), $30^\circ$ (light orange), $60^\circ$ (fucsia), $75^\circ$ (purple), and $90^\circ$ (blue).}
	\end{figure}

	Similarly to the neutrino case, we define the total photon emission rate as function of energy, time and solid angle as~\cite{Duffy:2005ns,Casals:2005sa}
	\begin{align}\label{eq:d3Ng}
		\frac{\dd^3 N_{\gamma}}{\dd\omega \dd t \dd \Omega} = \frac{1}{4\pi}\sum_{l=s}\sum_{m=-l}^l&\frac{\!\,_{1}\Gamma_{lm}}{\exp(\varpi/\TBH) - 1}\times\notag\\
		&\{|\!\,_{-1}S_{lm}(\theta)|^2 + |\!\,_{+1}S_{lm}(\theta)|^2\}\,.
	\end{align}
	Fig.~\ref{fig:d3Ng} illustrates the instantaneous photon emission rate for various polar angles of an EPBH with a mass of $\MBH=10^{9}$ g and a spin of $\as=0.999$. 
	The overall emission rate behavior is similar to that of neutrino-antineutrino emission asymmetry.
	At a polar angle of $\theta = 0$ (black curve), only the $l=1$ mode contributes to the emission. 
	As the polar angle increases, the contribution from other angular modes becomes noticeable, while the contribution from the $l=1$ mode diminishes.
	An important observation is that the emission rate exhibits a symmetry under the transformation $\theta \to \pi-\theta$, as the angular modes satisfy~\cite{Berti:2005gp,Casals:2005sa},
	\begin{align*}
		_{-s}S_{lm}(\theta) = (-1)^{-l-m}\!\,_{s}S_{lm}(\pi - \theta).
	\end{align*}
	Consequently, this symmetry implies that the measurement of photons emitted by an EPBH cannot determine, in principle, the hemisphere pointing towards Earth.
	Upon careful examination of the total emission rate, Eq.~\eqref{eq:d3Ng}, we observe that the two different angular eigenfunctions possess distinct polarisations. 
	Therefore, in principle, measuring the polarisation of the gamma rays could determine the value of $\theta$.
	Polarisation is commonly measured through the Compton scattering angle of photons for energies below electron-positron pair production, $E_\gamma \lesssim 1$ MeV~\cite{Eingorn:2015oga}. 
	For energies up to $E_\gamma \lesssim 10$ MeV, the event distribution of electron-positron pairs can be analysed~\cite{DEPAOLA1999175,Morselli:2013ntc,Bernard:2013jea}. 
	However, at higher energies, these techniques face limitations due to factors like multiple Coulomb scatterings. 
	Some works have proposed ideas for measuring polarisation up to energies of $E_\gamma \sim$ 30 GeV~\cite{DEPAOLA1999175,Morselli:2013ntc,Bernard:2013jea,Eingorn:2015oga,GROS201730}. 
	Nonetheless, it remains uncertain if these techniques are applicable to energies of interest here, $E_\gamma \gtrsim 1$ TeV.
	Furthermore, Earth's magnetic field may influence the polarisation of incoming photons from the EPBH.
	Thus, in this study, we adopt a conservative approach and assume that future experiments will not measure gamma-ray polarisation.
	To resolve the ambiguity in measuring $\theta$, we propose a multimessenger approach that incorporates the detection of neutrinos.
	
	\subsection{Secondary spectra}
	
	Since the EPBH also emits other SM dofs, most of which are unstable, there is an additional production of neutrinos/photons arising from their subsequent decay. 
	This additional contribution is known as the neutrino/photon secondary spectrum.
	To obtain this secondary spectrum, we convolve the primary spectrum of each particle species, denoted by $i$, with the number of daughter neutrinos/photons resulting from their decay as a function of energy and solid angle. 
	The expression for the secondary spectrum is given by
	\begin{align}
		\frac{\dd^3 N_{\nu (\gamma)}^{\rm sec}}{\dd\omega \dd t \dd\Omega}  = \int_0^\infty \dd\omega^\prime \int \dd\Omega^\prime \sum_i  
		\frac{\dd^3 N_i}{\dd\omega^\prime \dd t \dd\Omega^\prime} \frac{\dd^2 n_{i \to \nu (\gamma)}}{\dd\omega \dd\Omega}(\omega,\omega^\prime).
	\end{align}
	where $\dd^2 n_{i \to \nu(\gamma)}/\dd\omega \dd\Omega$ denotes the energy and angular distribution of neutrinos (photons) resulting from the decay of the $i$-th particle, and $d^3 N_i/d\omega^\prime dt d\Omega^\prime$ represents the primary emission rate of the $i$-th particle species defined as
	\begin{align}
		\frac{\dd^3 N_{i}}{\dd\omega \dd t \dd\Omega} = \frac{g_i}{4\pi}\sum_{l=s_i}\sum_{m=-l}^l&\frac{\!\,_{s_i}\Gamma_{lm}}{\exp(\varpi/\TBH) - (-1)^{2s_i}}\times\notag\\
		&\{|\!\,_{s_i}S_{lm}(\theta)|^2 + |\!\,_{-s_i}S_{lm}(\theta)|^2\},
	\end{align}
	being $s_i$ the spin and $g_i$ the internal dofs of the particle species $i$.
	To compute the secondary spectrum, we thus need to determine the angular and energy distributions of neutrinos resulting from the decay of each SM particle. 
	In our analysis, we have employed a similar approach to that used by {\tt BlackHawk}~\cite{Arbey:2019mbc,Arbey:2021ysg} which uses the {\tt PYTHIA} event generator~\cite{Bierlich:2022pfr} to calculate these distributions. 
	We use {\tt PYTHIA} since it has the full information of the four momenta of the produced particles, which is crucial for our purposes.
	
	The numerical strategy employed to determine $\dd^2 n_{i \to \nu(\gamma)}/\dd\omega \dd\Omega$ can be summarized as follows.
	First, we set the center-of-mass energy of the collision in {\tt PYTHIA} to be twice the energy of the primary particle $\omega^\prime$. 
	We generate a large number of events, typically around $10^5$, for a specific channel, $e^+ + e^- \rightarrow i + \overline{i} \rightarrow \cdots$, where $i$ represents the primary particle of interest. 
	For each event, we record the energy of the primary particle and calculate the relative angle between each daughter particle and the primary particle.  
	To facilitate the analysis, we perform a three-dimensional rotation that aligns the z-axis with the direction of the primary particle. 
	This rotation is then applied to all the final state particles, ensuring consistent orientation. 
	Next, we construct a two-dimensional histogram to capture the distribution of energy and relative angle for each value of the center-of-mass energy. 
	Each entry in the histogram represents the frequency of occurrence for a particular combination of energy and angle, normalised by the total number of events. 
	We repeat this process by systematically varying the center-of-mass energy in the range of interest, corresponding to the range of $1~{\rm GeV} - 10^6~{\rm GeV}$. 
	Finally, to convert the dependence on relative angle to a dependence on the solid angle of the particle distributions, we perform a rebinning of the histograms. 
	This rebinning takes into account the different possible orientations of the polar and azimuthal angles of the decay products for a given angle $\theta$ of the primary particle.
	
	\begin{figure}
		\centering
		\includegraphics[width=\linewidth]{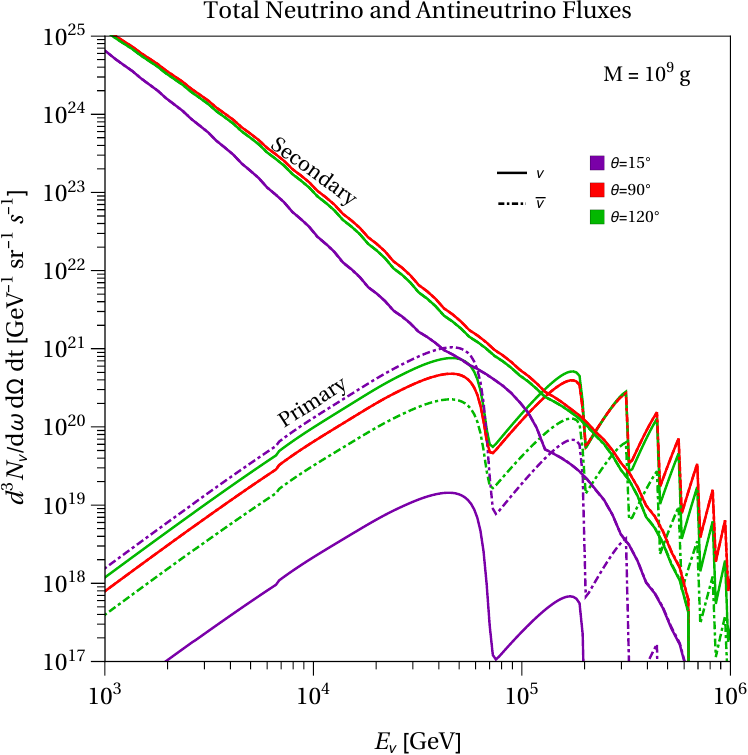}
		\caption{\label{fig:seconds} Total neutrino (full) and antineutrino (dot-dashed) instantaneous emission rates for three different values of the polar angle, $\theta = 15^\circ$ (purple), $\theta = 90^\circ$ (red), and $\theta = 130^\circ$ (green) and for a BH having a mass of $\MBH = 10^{9}~{\rm g}$ and spin parameter $\as = 0.999$. Note that neutrino and antineutrino lines for the secondary spectra and for the primary spectra for $\theta = 90^\circ$ lie on top of each other.}
	\end{figure}
	In Fig.~\ref{fig:seconds}, we present the total---summed over all neutrino states---primary and secondary neutrino (full) and antineutrino (dot-dashed) emission rates $d^3 N_{\nu(\overline{\nu})}/d\omega d\Omega dt$, obtained by adding and subtracting Eqs.~\eqref{eq:d3Anu} and \eqref{eq:d3Nnu}, for $\theta = 15^\circ$ (purple), $\theta = 90^\circ$ (red), and $\theta = 120^\circ$ (green).
	The primary spectra presents the peaks related to the $l$ modes as expected from the neutrino-antineutrino emission rate in Fig.~\ref{fig:d3A}.
	We observe that for the polar angle $\theta = 15^\circ$, belonging to the upper hemisphere, the primary emission rate of antineutrinos is $\sim 2$ orders of magnitude larger than the one from neutrinos.
	Meanwhile, the rates at the equator, $\theta=90^\circ$ are equal.
	For $\theta = 120^\circ$, the behaviour is the opposite, the neutrino primary emission rate is larger, although only a factor of $\sim 3.5$. 
	Regarding the secondary spectra, there are no significant differences between the neutrino and antineutrino fluxes. 
	This is expected since these neutrinos originate from the decay of other particles that do not exhibit the same emission asymmetry in their production that primary neutrinos present~\cite{Vilenkin:1980ft}.
	The main distinction arises from the dependence on the polar angle.
	Since the secondary spectrum arises from integrating the primary spectrum over energy and angle, with weights determined by the decay distributions, higher $l$ modes play a more significant role, especially at elevated energies.
	Consequently, it becomes evident that the most substantial secondary spectrum is produced from the equatorial region; meanwhile, for $\theta=15$, there is a observable reduction in flux compared to other values.

	\section{Time evolution}\label{sec:TimEv}
	
	Once a PBH reaches its final stages of life, particle emission intensifies, culminating in a potential observable burst for future observatories. 
	These observatories will track the temporal evolution of a PBH, with an initial mass linked to the observed duration of the burst. 
	Consequently, it becomes essential to understand the underlying time evolution. 
	To derive the evolution equations for the PBH mass and spin, we multiply the integrated emission rate in solid angle,
	\begin{align}
		\frac{\dd^2N_{i}}{\dd \omega \dd t} &= \int \dd\Omega\, \frac{\dd^3 N_{i}}{\dd\omega \dd t \dd\Omega}\notag\\
		& = \frac{g_i}{2\pi}\sum_{l=s_i}\sum_{m=-l}^l\frac{\!\,_{s_i}\Gamma_{lm}}{\exp(\varpi/\TBH) - (-1)^{2s_i}},
	\end{align}
	by the total energy of a given particle $\omega$ or by the $m$ quantum number, and then we integrate over the phase space . 
	Defining the evaporation functions for mass and angular momentum, $\epsilon_i(\MBH, a_\star)$ and $\gamma_i(\MBH, a_\star)$ per particle $i$, respectively, as
	\begin{subequations}    
		\begin{align}
			\varepsilon_i(\MBH, a_\star) &= \frac{g_i}{2\pi}\int_{0}^\infty \sum_{l=s_i}\sum_{m=-l}^l\frac{\omega \,_{s_i}\Gamma_{lm}}{\exp(\varpi/\TBH) - (-1)^{2s_i}}\,d\omega\,,\\
			\gamma_i(\MBH, a_\star) &= \frac{g_i}{2\pi}\int_{0}^\infty \sum_{l=s_i}\sum_{m=-l}^l \frac{m\,_{s_i}\Gamma_{lm}}{\exp(\varpi/\TBH) - (-1)^{2s_i}}\, d\omega\,,
		\end{align}
	\end{subequations}
	and summing over \emph{all} existing species, we obtain the following system of coupled equations for the time evolution~\cite{Page:1976df,Page:1976ki,Cheek:2021odj, Cheek:2022dbx}
	\begin{subequations}    
		\begin{align}\label{eq:dynamicsevaporation}
			\frac{\dd \MBH}{\dd t} &= - \epsilon(\MBH, a_\star)\frac{M_p^4}{\MBH^2}\,,\\
			\label{eq:dynamicsspin}
			\frac{\dd a_\star}{\dd t} &= - a_\star[\gamma(\MBH, a_\star) - 2\epsilon(\MBH, a_\star)]\frac{M_p^4}{\MBH^3}\,.
		\end{align}
	\end{subequations}    
	Assuming only the presence of SM dofs, the numerical solution of these equations reveals that a nearly maximal spinning BH loses its angular momentum at a rate significantly faster than its mass.
	However, if there exists a large sector of scalar particles, such as those in the string axiverse~\cite{Arvanitaki:2009fg}, the BH spin does not completely evaporate but tends towards an asymptotic value~\cite{Chambers:1997ai, Taylor:1998dk, Calza:2021czr}. 
	This emphasizes the importance of determining the angular momentum of the BH prior to its evaporation as a means to constrain these models.
	In what follows, and in order to be model-independent, we will assume that the EPBH follows the time evolution given the SM dofs at the onset of the final burst.
	The dependence with BSM scenarios will be considered elsewhere.
	
	Hence, it becomes crucial to investigate whether the neutrino-antineutrino emission asymmetry could be observed in the final PBH burst after taking into account the BH time evolution. 
	We integrate the neutrino (antineutrino) particle flux over time taking into account the time evolution of both mass and angular momentum of the BH,
	\begin{align}
		\frac{\dd^2 N_{\nu(\overline{\nu})}}{\dd\omega \dd\Omega} =\int_{0}^\tau \, \dd t\, \frac{\dd^3 N_{\nu(\overline{\nu})}}{\dd\omega \dd t \dd\Omega} (\MBH(t), \as(t)),
	\end{align}
	where $\tau$ represents the remaining lifetime of the evaporating black hole, assumed to be equal to the observed burst duration. 
	The parameter $\tau$ also determines the PBH mass at the onset of the burst, given a specific initial $\asi$. 
	For instance, if the burst duration is $\tau=100$ s and we consider a nearly maximal spin case, $\asi=0.999$, the initial mass is $\MBHi\sim 8.3\times 10^{9}$ g. 
	Conversely, for a non-rotating Schwarzschild BH, $\asi = 0$, the initial mass would be smaller, approximately $\MBHi\sim 6.3\times 10^{9}$ g, due to the increased lifetime of a non-rotating BH.
	\begin{figure}
		\centering
		\includegraphics[width=\linewidth]{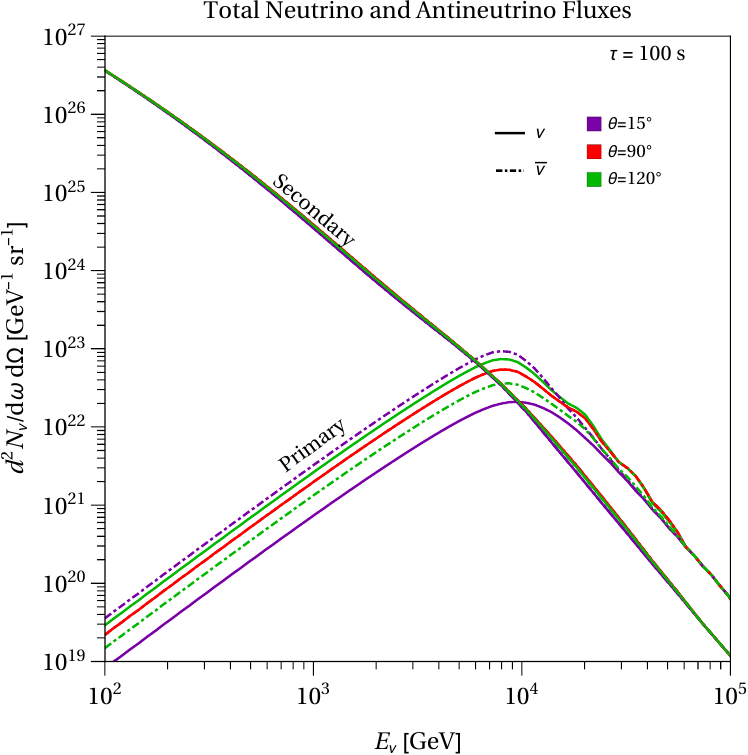}
		\caption{\label{fig:d2Nnu} Total time integrated neutrino (full) and antineutrino (dot-dashed) emission rates for two different values of the polar angle, $\theta = 15^\circ$ (purple), $\theta = 90^\circ$ (red), and $\theta = 130^\circ$ (green), and for a burst duration of $\tau = 100$ s and initial spin of $\asi = 0.999$.}
	\end{figure}
	
	The total time integrated neutrino and antineutrino fluxes for a burst duration of $\tau = 100$ s and assuming an initial $\asi=0.999$ are presented in Fig.~\ref{fig:d2Nnu}.
	Consistent with previous figures, we present the fluxes for polar angles $\theta = 15^\circ$ (purple), $\theta = 90^\circ$ (red), and $\theta = 120^\circ$ (green), distinguishing between the contributions of the primary and secondary spectra.
	Despite the spin is depleted faster than the mass, we observe an asymmetry in the emission for neutrino and antineutrino fluences depending on the polar angle. For instance, at $\theta = 15^\circ$, the antineutrino fluence is greater than the neutrino fluence by a factor of $\sim 5$ at an energy of $E_\nu \sim 8$ TeV. 
	However, we observe that the peak structure tends to vanish for this value of the polar angle at higher energies, which in turn will modify the number of events in neutrino telescopes.
	For other values of the polar angle the behaviour is similar to the one encountered for the instantaneous spectra in Fig.~\ref{fig:seconds}; however, the peak structure is less pronounced.
	Moreover, the secondary spectra does not present any asymmetry between neutrinos and antineutrinos.
	These fluence profiles suggest potential variations in the number of neutrino vs antineutrino events in neutrino telescopes, which could enable the determination of the EPBH's spin during its final stages. 
	This topic is further explored in the subsequent section.

	\section{Detection Prospects}\label{sec:Detect}
	
	PBHs could have been formed in the Early Universe due to different mechanisms, with initial masses $M_{\rm in}$ spanning from $\sim {\cal O}(1)$ g to values several orders of magnitude larger than the Solar mass~\cite{Carr:2020gox,Carr:2020xqk}. 
	The present-day distribution of these PBHs depends not only on their formation mechanisms but also on the subsequent evolution of the Universe, as PBHs may cluster around galaxies, among other possibilities.
	Moreover, the PBH number density is subject to stringent constraints contingent upon their initial masses, given that they could be undergoing evaporation, producing observable particle fluxes, such as $\gamma$-rays, today~\cite{Carr:2020gox,Carr:2020xqk}.
	Let us estimate the number density of PBHs that could exist in the Solar neighborhood, neglecting the effect of clustering, i.e., assuming a uniform PBH distribution.
	Denoting $\beta^\prime$ as the ratio of PBH energy density at formation $\rho_{\rm PBH}$ to the total energy density $\rho_{\rm tot}$, 
	\begin{align*}
		\beta^\prime = \alpha^\frac{1}{2}\left(\frac{g_{*, in}}{106.75}\right)^{-\frac{1}{4}}\left(\frac{\rho_{\rm PBH}}{\rho_{\rm tot}}\right),
	\end{align*}
	with $\alpha$ the gravitational collapse factor and $g_{*, in}$ the relativistic dofs, we obtain the PBH number density~\cite{Carr:2020gox}
	\begin{align}
		n_{\rm PBH} \approx 0.35 \left(\frac{\beta^\prime}{10^{-29}}\right)\left(\frac{10^{15}~{\rm g}}{M_{\rm in}}\right)^{\frac{3}{2}}~{\rm pc}^{-3},
	\end{align}
	under the assumption of standard cosmological evolution. 
	Consequently, on average, $\sim 1.5$ PBHs with initial mass of $10^{15}~{\rm g}$ are anticipated to exist within a spherical volume with radius of 1 parsec, for $\beta^\prime=10^{-29}$, consistent with current constraints.
	Notably, this number could be higher if clustering indeed influenced the distribution of PBHs.
	This implies that the average distance between PBHs is
	\begin{align}
		d_{\rm sep} \sim 0.88 \left(\frac{10^{-29}}{\beta^\prime}\right)^\frac{1}{3}\left(\frac{M_{\rm in}}{10^{15}~{\rm g}}\right)^{\frac{1}{2}}~{\rm pc},
	\end{align}
	If a PBH were to exist in the Solar neighborhood at a distance of approximately $\sim 10^{-3}$ pc from Earth, the nearest neighboring PBH would be situated at a distance of roughly $\sim 1$ pc. 
	Consequently, the particle flux originating from this additional PBH would be reduced by a factor of $\sim 10^{-6}$, rendering its contribution to the observed flux effectively negligible.
	Therefore, let us assume that an EPBH resides in close proximity to the Solar System, and focus on the possibility of determining the angular momentum and polar angle orientation of such an EPBH in both neutrino and photon detectors.
	
	\subsection{Neutrino Telescopes}
	
	Current and future neutrino telescopes hold the potential to observe the final stages of PBH evaporation by detecting the flux of high-energy neutrinos emitted during this process. 
	Let us consider the observation of $\mu$-tracks in the IceCube observatory resulting from an EPBH burst occurring in close proximity to Earth. 
	The exceptional angular resolution of IceCube for high-energy tracks, achieving an impressive precision of $\lesssim 1^\circ$ for energies in the TeV range~\cite{IceCube:2018pgc,IceCube:2019cia}, will play a crucial role in determining the origin of these events and distinguishing them from background signals.
	Furthermore, IceCube has the capability to differentiate between neutrinos and antineutrinos based on their characteristic inelasticity distributions, particularly for neutrino energies below $\lesssim 10$ TeV~\cite{IceCube:2018pgc}. This energy range is of particular interest to us.
	
	To accurately estimate the number of $\mu$-track events detected on Earth, we must account for neutrino flavor oscillations that occur during their journey from the source to the detector. 
	As mentioned before, the primary neutrinos and antineutrinos are assumed to be produced as mass eigenstates. 
	Consequently, the fluence of muon neutrinos (antineutrinos) originating from the primary spectra at Earth can be expressed as 
	\begin{align}
		F_{ \nu_\mu(\overline{\nu}_\mu)}^{\rm primary}(\asi, \theta, d_L) =\frac{1}{d_L^2}\sum_{i=1,2,3}|U_{\mu i}|^2\frac{d^2 N_{\nu_i(\overline{\nu}_i)}}{d\omega d\Omega},
	\end{align}
	where $U$ denotes the Pontecorvo-Maki-Nakagawa-Sakata mixing matrix, $d^2 N_{\nu_i(\overline{\nu}_i)}/d\omega d\Omega$ represents the time-integrated fluence of primary neutrinos (antineutrinos) for mass eigenstate $i$, and $d_L$ corresponds to the distance between the PBH and Earth. 
	Notably, the absence of the usual $4\pi$ factor in the denominator is due to the lack of spherical symmetry of particle emission from a Kerr EPBH, as emphasized previously. 
	Since the energies at which these primary neutrinos are emitted are significantly higher than the neutrino masses, the primary spectra for all three mass eigenstates are identical.
	
	In contrast, secondary neutrinos are generated through weak interactions, leading them to be produced in flavor eigenstates. 
	Neutrino oscillations over the PBH-Earth distances present decoherence since these distances are expected to greatly exceed the standard oscillation lengths. 
	Consequently, the fluence for the secondary component can be expressed as
	\begin{align}
		F_{\nu_\mu(\overline{\nu}_\mu)}^{\rm secondary}(\asi, \theta, d_L) =\frac{1}{d_L^2}\sum_{i=1}^{3}\sum_{\alpha=e}^{\tau}|U_{\mu i}|^2|U_{\alpha i}|^2\frac{d^2 N_{\nu_\alpha(\overline{\nu}_\alpha)}^{\rm sec}}{d\omega d\Omega}.
	\end{align}
	The number of muon neutrino(antineutrino) events in IceCube for a given zenith angle $\zeta$, corresponding to the location of the EPBH, involves the following expression
	\begin{align}\label{eq:nu_evs}
		N_{\nu_\mu(\overline{\nu}_\mu)}(\asi, \theta, d_L) = \int_{\omega_{\rm min}}^{\omega_{\rm max}} F_{\nu_\mu(\overline{\nu}_\mu)} \mathscr{A}_{\rm eff} (\omega, \zeta) d\omega,
	\end{align}
	where $F_{\nu_\mu(\overline{\nu}_\mu)} = F_{\nu_\mu(\overline{\nu}_\mu)}^{\rm primary}+F_{\nu_\mu(\overline{\nu}_\mu)}^{\rm secondary}$ 
	represents the total neutrino fluence at the detector, and $\mathscr{A}_{\rm eff} (\omega, \zeta)$ denotes the effective area of IceCube~\cite{IceCube:2019cia}.
	In principle, the effective areas should be different for neutrinos and antineutrinos.
	However, we use the publicly available effective area~\cite{IceCube:2019cia}, which corresponds to the averaged area for neutrinos and antineutrinos.
	The energy integration is performed over the range between IceCube's threshold energy of $\omega_{\rm min} = 100~{\rm GeV}$ and the maximum energy of the neutrino fluence, which, in turn, depends on the duration of the burst~\cite{Capanema:2021hnm}.
	The main background in IceCube that could affect the measurement of the neutrinos from an EPBH corresponds to high-energy atmospheric neutrinos creating observable tracks.
	Nevertheless, the events from these neutrinos is of order $10^{-4}$ for a time interval of $100$ s~\cite{Capanema:2021hnm}, making such a background negligible.
	
\begin{figure}
		\centering
		\includegraphics[width=\linewidth]{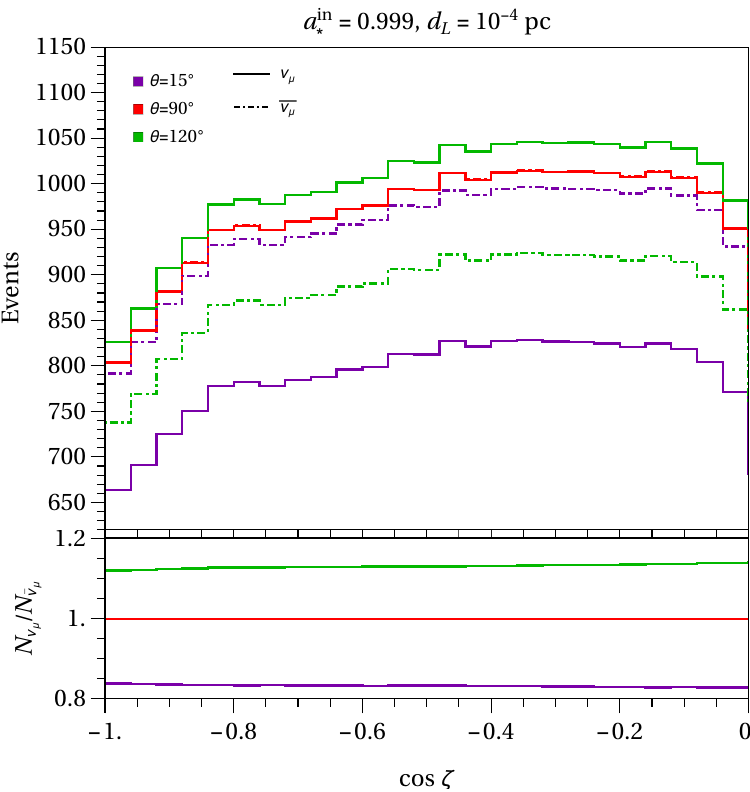}
		\caption{\label{fig:Evs_IC}  Muon neutrino (full) and antineutrino (dot-dashed) events in IceCube (top), and the neutrino-to-antineutrino ratio (bottom) as function of the zenith angle for an EPBH at a distance of $d_L=10^{-4}$ pc and a burst duration of $\tau = 100$ s for different values of the polar angle: $\theta = 15^\circ$ (purple), $\theta = 90^\circ$ (red), and $\theta = 120^\circ$ (green), representing the orientation of the Earth relative to the axis of rotation of the PBH.}
	\end{figure}

	Figure \ref{fig:Evs_IC} illustrates the variation of neutrino and antineutrino events in IceCube as a function of the zenith angle (top) and the ratio of neutrinos to antineutrinos (bottom) for an EPBH having an initial $\asi = 0.999$, a burst duration of $\tau = 100$ s at distance of $d_L=10^{-4}$ pc.
	Similar to previous figures, we consider the polar angles $\theta = 15^\circ$ (purple), $\theta = 90^\circ$ (red), and $\theta = 120^\circ$ (green), representing the orientation of the Earth relative to the axis of rotation of the PBH.
	For the case of $\theta = 15^\circ$, it is observed that muon antineutrino events dominate, while neutrino events are only $\sim 80\%$ of those for antineutrinos across various zenith angles. 
	Conversely, at $\theta = 120^\circ$, where the Earth is positioned in the southern hemisphere of the EPBH, neutrino events exceed antineutrino events by $\sim 12\%$.
	In line with expectations, when the Earth aligns with the EPBH's equatorial plane, neutrino and antineutrino events coincide.
	
	\subsection{Gamma Ray Experiments}
	
	Gamma rays, emitted abundantly during the final stages of PBH lifetime, can be effectively detected by experiments specifically designed for high-energy photon searches, such as the High Altitude Water Cherenkov observatory (HAWC)~\cite{HAWC:2013kzm,HAWC:2019wla}. 
	HAWC is a very-high-energy (VHE) air shower array located on the slopes of the Sierra Negra volcano at an altitude of 4100 m above sea level.
	HAWC's main array comprises 300 cylindrical water tanks, each equipped with photomultiplier tubes that detect the Cherenkov light produced by secondary particles resulting from the interaction of VHE gamma rays. With a wide field of view of approximately 2 sr, HAWC is capable of detecting photons with energies ranging from $10^2$ to $10^5$ GeV~\cite{HAWC:2019wla}. 
	Additionally, HAWC boasts excellent angular resolution, ranging from approximately $\sim 0.2^\circ$ to $1^\circ$, rendering it well-suited for capturing transient events such as EPBH bursts~\cite{HAWC:2019wla}.
	
	To estimate the number of gamma-ray events from the burst of a EPBH, we follow a similar approach as with neutrinos,
	\begin{align}\label{eq:gamma_evs}
		N_{\gamma}(\asi, \theta, d_L) = \int_{\omega_{\rm min}}^{\omega_{\rm max}} F_{\gamma} (\asi, \theta, d_L)\mathscr{A}_{\rm eff}^{\rm HAWC} (\omega, \zeta)\, d\omega,
	\end{align}
	where $F_{\gamma} (\theta)$ encompasses both primary and secondary photon emissions, and $\mathscr{A}_{\rm eff}^{\rm HAWC} (\omega, \zeta)$ represents the effective area of HAWC~\cite{HAWC:2013kzm}. 
	Given the larger cross-section of interaction for photons, we anticipate that the number of photon events will surpass that of neutrinos.
	However, as previously mentioned, the symmetric emission of photons from both PBH hemispheres introduces an ambiguity in determining the orientation of the axis of rotation relative to Earth.
	Consequently, a simultaneous measurement of both neutrinos and photons would provide valuable insights, not only regarding the initial angular momentum of the EPBH at the onset of the burst but also enabling determination of the axis of rotation's orientation relative to Earth.
	
	\subsection{Sensitivity}
	
	To demonstrate the feasibility of a multimessenger approach, we conduct a sensitivity analysis of combined neutrino/antineutrino and photon measurements. This analysis aims to understand the potential of such an approach in determining the initial parameters of the EPBH at the beginning of the burst. 
	We employ a rate analysis utilizing the following test statistics
	\begin{widetext}
		\begin{align}
			\chi^2 = {\rm min}_{\alpha}\left\{\frac{(N_{\nu_\mu}(\asi, \theta, d_L, \alpha) - N_{\nu_\mu}^{\rm b})^2}{N_{\nu_\mu}^{\rm b}} + \frac{(N_{\overline{\nu}_\mu}(\asi, \theta, d_L, \alpha) - N_{\overline{\nu}_\mu}^{\rm b})^2}{N_{\overline{\nu}_\mu}^{\rm b}} + \frac{(N_{\gamma}(\asi, \theta, d_L, \alpha) - N_{\gamma}^{\rm b})^2}{N_{\gamma}^{\rm b}} + \frac{\alpha^2}{\sigma_\alpha^2}\right\},
		\end{align}
	\end{widetext}
	where $N_{\nu_\mu}(\asi, \theta, d_L, \alpha), N_{\overline{\nu}_\mu}(\asi, \theta, d_L, \alpha) , N_{\gamma}(\asi, \theta, d_L, \alpha) $ represent the predicted events of muon neutrinos, muon antineutrinos, and photons, respectively, for a given set of initial parameters $\asi,\theta$ for a EPBH at a distance $d_L$, measured via some independent parallax technique.
	We include a pull parameter $\alpha$ related to the measurement of the EPBH-Earth distance by performing the substitution 
	\begin{align*}
		d_L \to (1+\alpha)d_L,
	\end{align*}
	in the determination of the number of events, Eqs.~\eqref{eq:nu_evs} and \eqref{eq:gamma_evs}.
	We assume an uncertainty of $\sigma_\alpha = 10\%$ on the measurement of the distance $d_L$.
	$N_{\nu_\mu}^{\rm b}$, $N_{\overline{\nu}\mu}^{\rm b}$, and $N{\gamma}^{\rm b}$ are the corresponding benchmark event values.
	For our analysis, we assume benchmark EPBH parameter values of $\asi = 0.5$, $\theta = 45^\circ$, and a distance of $d_L= 10^{-4}$ pc, 
	yielding values of $N_{\nu_\mu}^{\rm b} = 800.65$, $N_{\overline{\nu}\mu}^{\rm b} = 873.09$, and $N_{\gamma}^{\rm b} = 2.81\times 10^6$ for $\tau = 100$ s.
	Additionally, we assume an ideal scenario with no backgrounds and perfect discrimination between neutrinos and antineutrinos. 
	While these assumptions are quite optimistic, they provide a best-case scenario for assessing the feasibility of determining the properties of an EPBH using neutrinos.
	\begin{figure*}
		\centering
		\includegraphics[width=\linewidth]{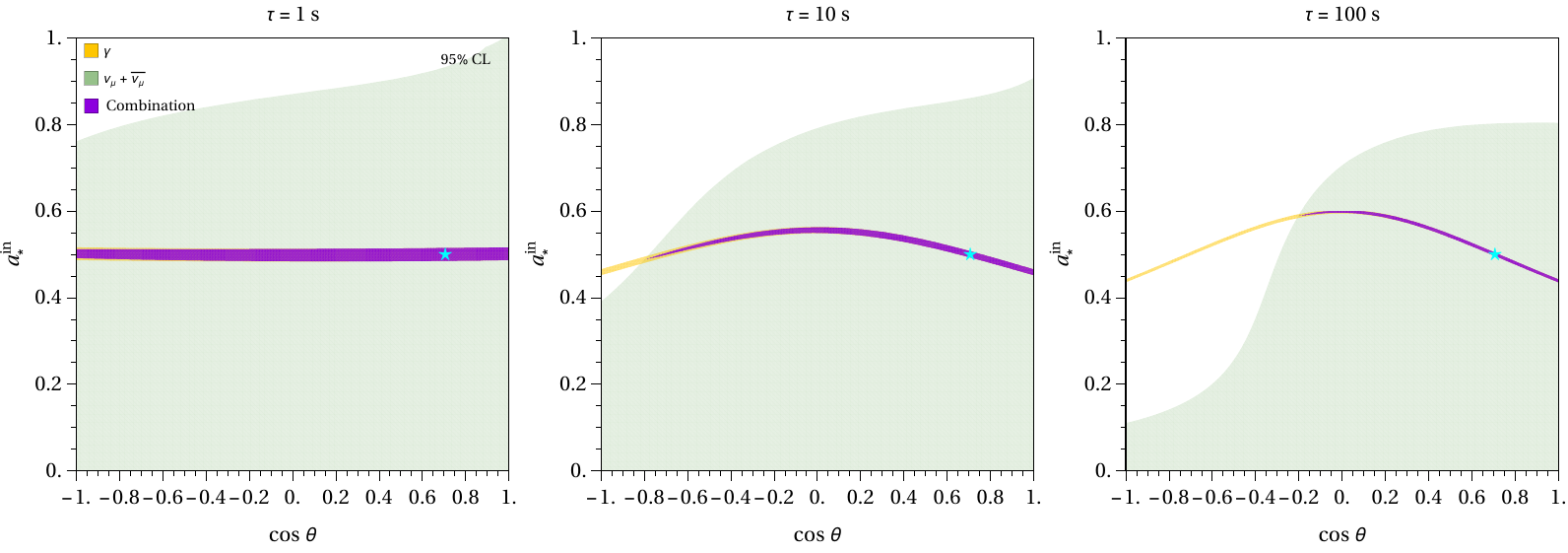}
		\caption{\label{fig:chisq} Sensitivity to the EPBH's initial parameters at the onset of a burst having three durations of $\tau = 1$ s (left), $\tau = 10$ s (center), and $\tau = 100$ s (right). The allowed regions are at the 95$\%$ CL from photons (yellow), neutrinos (green), and their combination (purple). We assumed a distance of $d_L=10^{-4}$ pc with a $10\%$ uncertainty, an initial $\asi=0.5$ and the Earth placed at a polar angle of $\theta = 45^\circ$ with respect to the EPBH axis of rotation.}
	\end{figure*}
	
	We perform a two-dimensional $\chi^2$ analysis varying the initial characteristics of the EPBH at the onset of the burst, specifically its angular momentum and orientation with respect to Earth, $\asi$ and $\cos\theta$. 
	Moreover, we marginalize over the additional pull parameter $\alpha$ to obtain the allowed regions.
	This analysis is performed for three distinct durations, $\tau = 1$ s (left panel), $\tau = 10$ s (center panel), and $\tau = 100$ s (right panel). 
	The outcomes of this analysis are visually represented in Fig.~\ref{fig:chisq}.
	The contributions to the $\Delta \chi^2$ from individual measurements of photons (yellow)
	and neutrinos (green), as well as their combination (purple), are presented, with all regions corresponding to a $95\%$ CL, while the input benchmark value is represented by the cyan star.
	Analyzing the photon measurement alone, we observe that the large number of photon events facilitates a precise determination of the initial EPBH spin at the onset of the observed burst.
	However, the allowed regions exhibit a degeneracy between the hemispheres, resulting in a symmetric region under the transformation $\cos\theta \to -\cos\theta$, and exhibits a convex-like structure for the larger burst durations, $\tau = 10~{\rm s}, 100~{\rm s}$. 
	This structure arises due to the enhanced emission of photons at the EPBH poles, resulting in similar gamma-ray event rates for EPBHs with lower spin parameters in the polar region as compared to higher-spin BHs at different polar angles pointing towards Earth.
	For the smaller burst time of $\tau=1$ s, such a structure is absent since the number of events in the poles differ from those at the equator at less than $\sim 1\%$, due to the short duration of the burst.
	
	Given the precision of the photon-only measurement in determining $\asi$, one might wonder about the impact of uncertainties in other parameters, such as the EPBH-Earth distance, $d_L$.
	We tested a range of distance measurement uncertainties, ranging from $\sigma_\alpha = 10\%$ to $100\%$, and observed that even at the upper limit, the allowed regions were affected by only approximately $10\%$.
	
	In contrast, the neutrino measurement yields a broader region for all burst durations primarily due to the lower number of neutrino events, which is approximately $10^{-3}$ smaller than that of photons.
	However, we observe that for the shortest burst duration ($\tau=1$ s), neutrinos alone can exclude an initially close-to-maximally rotating EPBH ($\asi=0.999$) for values of $\cos\theta\lesssim 1$ at more than 95$\%$ CL. 
	For the same burst duration, neutrinos could exclude $\asi\gtrsim 0.75$ for an EPBH with its northern pole pointing towards Earth. 
	This is because, in the considered benchmark scenario, there are more antineutrino events than neutrino events. 
	Conversely, for $\cos\theta = -1$, the behavior is the opposite, with neutrino events being approximately 14$\%$ larger than those for the benchmark, while antineutrinos are reduced by about 5$\%$.
	
	Longer observation times significantly contribute to better determination of the initial EPBH parameters, as both the number of observed neutrinos and photons increase with the burst duration.
	For a burst duration of $\tau = 10$ s, it becomes apparent that the asymmetry in neutrino-antineutrino emission begins to impact the combined sensitivity, leading to a reduction in the degeneracy associated with determining the EPBH hemisphere that is oriented towards Earth. 
	Notably, the previously degenerate solution at $\theta = 135^\circ$ is now disfavored with a $\Delta\chi^2=5.78$.
	Moreover, the neutrino measurement alone excludes values of $\asi \gtrsim 0.6$ for $\cos\theta \lesssim -0.5$ at a 95 $\%$ confidence level, while for $\cos\theta\gtrsim 0.5$, the exclusion occurs exclusively for highly rotating EPBHs, with $\asi\gtrsim 0.8$ at the same confidence level. 
	Such a sensitivity is attained through the enhanced emission of fermions for higher rotating BHs, consequently leading to an increased net particle flux.
	
	The exclusion of the solution in the second quadrant ($\theta = 135^\circ$) at level exceeding 95$\%$ is ultimately achieved for the longer burst duration of $\tau = 100$ s. 
	Furthermore, the neutrino measurement independently excludes a significant portion of the parameter space, particularly for values $\cos\theta\lesssim -0.1$ and $\asi \gtrsim 0.1$.
	For $\asi \lesssim 0.1$, we observe that the sensitivity becomes nearly independent of $\cos\theta$. 
	This is because these EPBHs start to resemble a Schwarzschild BH, which is spherically symmetric and does not exhibit neutrino emission asymmetry. 
	In this case, the sensitivity primarily arises from the difference in events due to the enhanced emission for Kerr BHs.
	A similar behavior occurs for $\asi \gtrsim 0.8$, where the increased emission of neutrinos and antineutrinos results in a significantly larger number of events than expected for the chosen benchmark. 
	For instance, at $\asi = 0.9$ and $\theta=45^\circ$, the $N_{\nu_\mu} = 840.97$, and $N_{\overline{\nu}\mu} = 975.45$, making this value disfavored by approximately $\Delta\chi^2\approx 4\sigma$.
	Thus, by combining photon and neutrino measurements, the degeneracy present in the photon measurement is effectively resolved, leading to a more precise determination of the EPBH's initial characteristics.
	\begin{figure}
		\centering
		\includegraphics[width=\linewidth]{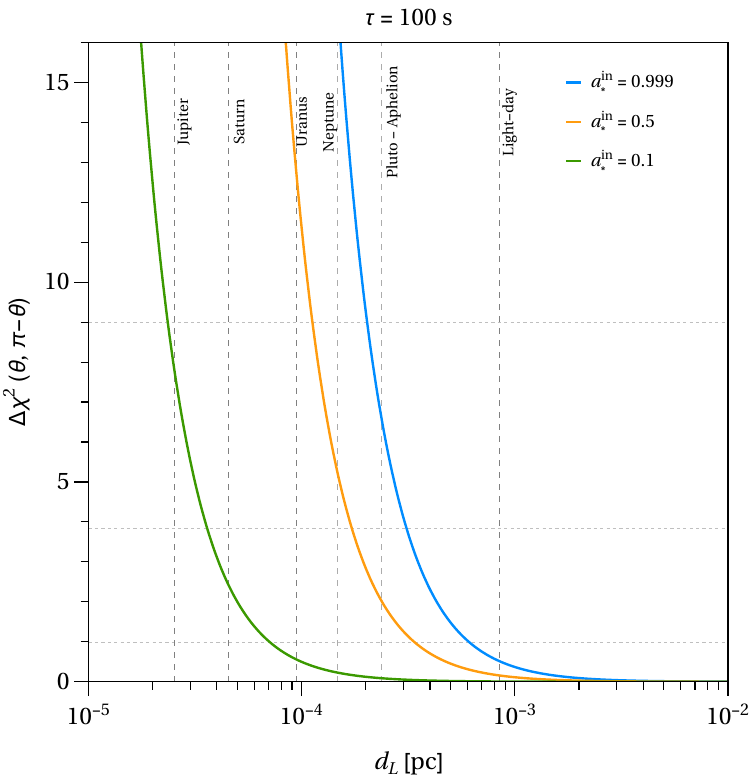}
		\caption{\label{fig:chisqdL} Chi-squared difference between the two degenerate angles $\theta$ and $\pi - \theta$ solutions in the photon measurement, after combination with the neutrino measurement, as function of the distance to the EPBH for various initial spin parameter values, $\asi=0.999$ (blue), $\asi=0.5$ (orange), and $\asi=0.1$ (green). The dashed vertical lines indicate the distances of the outer planets to the Sun and a light-day distance for reference.}
	\end{figure}
	
	It is then natural to question how close the EPBH must be for neutrinos to effectively break the degeneracy in determining the angle $\theta$. 
	To gain insights into this matter, we present in Fig.~\ref{fig:chisqdL} the profiled $\Delta\chi^2$ between the two degenerate solutions, $\theta$ and $\pi-\theta$, as a function of the EPBH-Earth distance for fixed values of the initial spin parameter and a burst duration of $\tau = 100$ s, assuming $\theta = 45^\circ$.
	We observe that, for an initially close-to-maximally rotating black hole, the neutrino emission asymmetry is capable of breaking the degeneracy at a $3\sigma$ confidence level when the EPBH-Earth distance is approximately $2\times 10^{-4}$ pc. 
	As expected, for lower spinning EPBHs, the required distances are smaller due to the decreased asymmetry in emission. 
	For $\asi = 0.5$, the necessary distance is roughly $1.1\times 10^{-4}$ pc, while for $\asi = 0.1$, the distance is approximately $2.3\times 10^{-5}$ pc.
	To provide a sense of scale, we have indicated in Fig.~\ref{fig:chisqdL} the distance to the Sun of the outer planets as vertical dashed lines. 
	From this, we observe that to determine the orientation of the EPBH with respect to Earth, a close-to-maximally rotating EPBH would need to be closer than Pluto's aphelion. 
	In contrast, an EPBH with an initial $\asi=0.1$ would need to be much closer, approximately as close as Jupiter is to the Sun, in order to measure the neutrino-antineutrino emission asymmetry and enable the determination of the EPBH hemisphere facing Earth.
	
	\emph{Future neutrino sensitivity. ---} Our previous estimates for determining the EPBH's initial characteristics using neutrino asymmetric emission were based on IceCube's current capabilities, particularly its effective area. 
	However, it is important to note that IceCube is set to undergo an upgrade, which is anticipated to increase its effective area by approximately five times, leading to a substantial enhancement in the detection of high-energy neutrinos~\cite{Clark:2021fkg}.
	Furthermore, the development of new techniques, such as deep learning, holds promise for refining the measurement of cascades. 
	This advancement has already demonstrated improved angular resolution for neutrino observations, as evidenced by recent findings from the Galactic Center~\cite{IceCube:2023ame}. 
	As a result, we can reasonably expect that these innovations will significantly bolster the statistics for nearby EPBH events, especially if they aid in measuring additional neutrino flavors.
	
	In addition to IceCube's upgrade, a plethora of new experiments, including KM3Net~\cite{KM3Net:2016zxf}, P-ONE~\cite{P-ONE:2020ljt}, Trident~\cite{Ye:2022vbk}, and Baikal-GVD~\cite{Baikal-GVD:2022fis}, are either currently being built or in the planning stages. 
	These experiments are also expected to measure neutrinos in the TeV scale, and in the event of an EPBH burst occurring near Earth, they will undoubtedly contribute to a substantial increase in the data.
	Furthermore, the distribution of these new facilities across different locations on Earth introduces the possibility of a independent neutrino measurement of the Earth-EPBH distance if a nearby event is observed. 
	Although we do not quantitatively assess the extent to which all these neutrino telescopes would improve the measurement of the initial EPBH characteristics at the onset of its burst, we can anticipate a substantial improvement in the measurement of the initial characteristics of an EPBH, should one be observed in the vicinity of Earth.

	\section{Conclusions}\label{sec:conc}

	The detection of an evaporating black hole near Earth would represent an extraordinary triumph for theoretical physics, validating our understanding of quantum fields in curved spacetimes and providing insights into the existing dofs in nature.
	Moreover, the detection of an evaporating black hole with an initial non-zero spin would hint the existence of physics beyond the Standard Model.
	
	Neutrino emission from Kerr black holes exhibits unique behavior, primarily due to the fact that only left-handed neutrinos and right-handed antineutrinos interact weakly. 
	Since particles with positive helicity are preferentially emitted along the rotation axis of a Kerr black hole, while those with negative helicity are predominantly emitted in the opposite direction, neutrinos present an asymmetric emission: neutrinos are preferentially emitted in the southern hemisphere, whereas antineutrinos are mostly emitted in the northern hemisphere.
	In this work, we have proposed to exploit this neutrino emission asymmetry as a powerful tool to determine the initial properties of an EPBH.
	After considering the distribution of neutrinos and photons, we have also derived, for the first time, the angular distribution of secondary neutrinos and photons.
	We have analysed the time evolution of EPBHs and calculated the full net neutrino flux integrated over the observed burst duration. 
	Our analysis assumed that the EPBH mass and spin follow the standard time evolution, considering the dofs in the SM.
	An interesting aspect to consider is that this time evolution may be influenced by additional BSM dofs, particularly if they manifest as scalars. 
	This could potentially result in a distinct integrated neutrino-antineutrino emission asymmetry. 
	Consequently, we might expect to observe specific correlations between the detected neutrino and antineutrino events, contingent on the specific BSM scenario at play.
	This will be investigated in detail in future work.
	
	We have computed the expected events in the IceCube observatory, finding that, for an initially close-to-maximally rotating EPBH, the number of antineutrino events would be larger than the neutrino events by a factor of $\sim 1.2$ ($\sim 0.9$), if Earth is located at an angle of $\theta = 15^\circ$ ($120^\circ$) with respect to the rotation axis for a distance of $10^{-4}$ pc.
	Furthermore, through simultaneous measurements of photons and neutrinos emitted from an EPBH, we have explored the possibility of not only determining the initial black hole angular momentum but also the orientation of its axis of rotation with respect to Earth. 
	While gamma-ray events are expected to significantly outnumber neutrino events, the symmetry in the angular eigenfunctions of photon emission under the transformation $\theta\to\pi-\theta$ makes it challenging to determine the black hole hemisphere facing Earth. 
	However, the net neutrino-antineutrino flux displays a definite dependence on the polar angle, aiding in breaking the degeneracy present in the photon measurement.
	
	Taking optimistic assumptions regarding backgrounds and discrimination between neutrinos and antineutrinos in IceCube, we found that, depending on the burst duration, a neutrino measurement could lift the degeneracy in determining the polar angle quadrant. 
	To exclude one of the degenerate angle solutions in the photon measurement at more than 95$\%$ CL, we deduced that the EPBH should be within a distance of approximately $1.78\times 10^{-4}$ pc for initial spin parameters of $\asi=0.5$ or $3.6\times 10^{-5}$ pc for $\asi=0.1$. These distances are smaller than the Neptune-Sun or Saturn-Sun distances, respectively.
	Consequently, perhaps an EPBH exists within our solar system, approaching its final stages of life, presenting us with a unique opportunity to directly observe Hawking radiation through a multimessenger approach and investigate its properties before the final burst.
	
	
	\section*{Acknowledgments}
	
	The author is immensely thankful to Jessica Turner for her encouragement to pursue this work and for engaging in several insightful discussions on various topics related to this work. 
	Moreover, the author would like to thank Jessica Turner and Pedro Machado for their meticulous review of the manuscript and their valuable comments, which significantly improved the quality of this paper.
	In addition, special thanks are owed to Ivan Mart\'inez-Soler for generously providing assistance with queries related to IceCube, and to Sam Dolan for valuable correspondence on the determination of the angular eigenvalues for fermions.
	Lastly, the author is deeply grateful for the warm hospitality extended by the Particle and Astroparticle Division of the Max-Planck-Institute f\"ur Kernphysik, where a portion of this research was finalized.
	This work has been funded by the UK Science and Technology Facilities Council (STFC) under grant ST/T001011/1. 
	This project has received funding/support from the European Union’s Horizon 2020 research and innovation programme under the Marie Sk\l{}odowska-Curie grant agreement No 860881-HIDDeN.
	This work has made use of the Hamilton HPC Service of Durham University.
	
	\appendix
	

	\section{Fermions in a Kerr Spacetime}\label{ap:FerKBs}
	
	Let us briefly describe the separation of variables in the Dirac equation for massive fermions on a Kerr background~\cite{Chandrasekhar:1976ap,Page:1976jj,Dolan:2009kj,Dolan:2015eua}.
	The Dirac equation for a fermion with mass $\mu$ is 
	\begin{align}
		(i\gamma^\alpha \hat{D}_\alpha - \mu)\Psi = 0,
	\end{align}
	where $\hat{D}_\alpha = \partial_\alpha - \Gamma_\alpha$, $\Gamma_\alpha$ being the spin connection matrices and $\gamma^\mu$ are the Dirac matrices in the curved spacetime satisfying the algebra
	\begin{align}
		\{\gamma^\mu,\gamma^\nu\}=2g^{\mu\nu}I_{4\times 4},
	\end{align}
	with $g^{\mu\nu}$ the inverse metric tensor associated to the Kerr metric.
	An appropriate choice for these matrices for the Kerr spacetime in the Boyer-Lindquist coordinates is
	\begin{align}
		\gamma^t &= \frac{r^2+a^2}{\rho\sqrt{\Delta}}\Tilde{\gamma}^0 + \frac{a\sin\theta}{\rho}\Tilde{\gamma}^2,& \gamma^r&=\frac{\sqrt{\Delta}}{\rho}\Tilde{\gamma}^3\notag\\
		\gamma^\phi &= \frac{a}{\rho\sqrt{\Delta}}\Tilde{\gamma}^0+\frac{1}{\rho\sin\theta}\Tilde{\gamma}^2, &\gamma^\theta&=\frac{1}{\sqrt{\rho}} \Tilde{\gamma}^1,
	\end{align}
	where $\rho^2=r^2+a^2\cos\theta$, and $\Tilde{\gamma}^a$ are the usual Dirac matrices in a flat spacetime. 
	We choose here the standard chiral representation for $\Tilde{\gamma}^a$,
	\begin{align}
		\Tilde{\gamma}^0=\begin{pmatrix}
			0 & I\\
			I & 0 
		\end{pmatrix}, \quad 
		\Tilde{\gamma}^i=\begin{pmatrix}
			0 & \sigma_i \\
			-\sigma_i & 0
		\end{pmatrix}.
	\end{align}
	Here $\sigma_i$ are the Pauli matrices, and $I$ is the $2 \times 2$ identity matrix.
	The chirality matrix $\gamma^5$ is defined in a standard manner,
	\begin{align}
		\gamma^5 = i\Tilde{\gamma}^0\Tilde{\gamma}^1\Tilde{\gamma}^2\Tilde{\gamma}^3=\begin{pmatrix}
			-1 & 0\\
			0 &  1  
		\end{pmatrix}.
	\end{align}
	The spin-connection matrices are 
	\begin{align}
		\Gamma_\alpha = \frac{1}{4}\omega_{\alpha bc}\Tilde{\gamma}^b\Tilde{\gamma}^c,
	\end{align}
	being $\omega_{\alpha bc}$ the spin connection. Explicitly, the spin connection matrices are
	\begin{widetext}
		\begin{subequations}
			\begin{align}
				\Gamma_t &= \frac{GM}{2}\begin{pmatrix}
					\varrho^{-2}\sigma_3 & 0 \\
					0 & -(\varrho^*)^{-2}\sigma_3
				\end{pmatrix},\\
				\Gamma_r &= -\frac{1}{2}\frac{a\sin\theta}{\sqrt{\Delta}}\begin{pmatrix}
					\varrho^{-1}\sigma_2 & 0 \\
					0 & -(\varrho^*)^{-1}\sigma_2
				\end{pmatrix},\\
				\Gamma_\theta &= -\frac{1}{2}\sqrt{\Delta}\begin{pmatrix}
					(i\varrho)^{-1}\sigma_2 & 0 \\
					0 & -(i\varrho^*)^{-1}\sigma_2
				\end{pmatrix},\\
				\Gamma_\phi &= \frac{1}{2}\sqrt{\Delta}\sin\theta\begin{pmatrix}
					(i\varrho)^{-1}\sigma_1 & 0 \\
					0 & -(i\varrho^*)^{-1}\sigma_1
				\end{pmatrix} + \frac{1}{2}\vartheta \begin{pmatrix}
					\vartheta \sigma_3 & 0 \\
					0 & -\vartheta^* \sigma_3
				\end{pmatrix},
			\end{align}
		\end{subequations}
	\end{widetext}
	where $\varrho = r + i a \cos\theta$, and $\vartheta = i\cos\theta - a \varrho^{-2}(\varrho+GM)\sin^2\theta$. 
	The contraction of the spin-connection with the Dirac matrices $\gamma^\alpha\Gamma_\alpha$, corresponding to the term which appears in the Dirac equation, is explicitly given by
	\begin{align}
		\gamma^\alpha\Gamma_\alpha=\frac{1}{2}\begin{pmatrix}
			0  & s_\theta^* \sigma_1 + s_r^*\sigma_3\\
			-s_\theta \sigma_1 - s_r^*\sigma_3 & 0
		\end{pmatrix},
	\end{align}
	where
	\begin{align*}
		s_r=\frac{\rho}{\sqrt{\Delta}}\frac{1}{\varrho \sqrt{\Delta}}\frac{\partial}{\partial r}(\varrho \sqrt{\Delta}),\quad 
		s_\theta=\rho\frac{1}{\varrho \sin\theta}\frac{\partial}{\partial \theta}(\varrho \sin\theta).
	\end{align*}
	Following Ref.~\cite{Dolan:2015eua}, we propose an ansatz to separate the Dirac equation,
	\begin{align}
		\Psi = \Delta^{-\frac{1}{4}} e^{-i \omega t + im\phi}
		\begin{pmatrix}
			\varrho^{-1/2} \eta_-(r,\theta) \\
			(\varrho^*)^{-1/2}\eta_+(r,\theta) \\
		\end{pmatrix},
	\end{align}
	with the spinors $\eta_\pm$
	\begin{subequations}
		\begin{align}
			\eta_-(r,\theta) &= -\begin{pmatrix}
				R_2(r)\, \!\,_{-\frac{1}{2}}S_{lm}(\theta)\\
				R_1(r)\, \!\, _{+\frac{1}{2}}S_{lm}(\theta)
			\end{pmatrix},\\
			\eta_+(r,\theta) &= ~~\begin{pmatrix}
				R_1(r)\, \!\,_{-\frac{1}{2}}S_{lm}(\theta)\\
				R_2(r)\, \!\, _{+\frac{1}{2}}S_{lm}(\theta)
			\end{pmatrix}.
		\end{align}
	\end{subequations}
	Here, $R_{1,2}(r)$, $\!\,_{\pm\frac{1}{2}}S_{lm}(\theta)$ are the radial and angular functions associated to massive fermions. 
	Substituting in the Dirac equation, one finds the Eqs.~\eqref{eq:DREqs} and \eqref{eq:DSEqs}.
	
	\section{Chandrashekar -- Detweiler Potentials}
	\label{ap:TeuKBs}
	
	In a series of works, Chandrashekar and Detweiler proposed a method to transform the general Teukolsky equations into Schr\"odinger-like equations~\cite{Chandrasekhar:1975zz,Chandrasekhar:1976zz,Chandrasekhar:1977kf}.
	They found explicit forms for the potentials depending on the spin of the particle.
	We have the potentials for scalars and vectors 
	\begin{subequations}
		\begin{align}
			V_0(r) &= \frac{\Delta}{\rho^4}\left(\lambda_0 + \frac{\Delta+2r(r-GM)}{\rho^2}-\frac{3r^2\Delta}{\rho^4}\right),\\
			V_1(r) &= \frac{\Delta}{\rho^4}\left[\lambda_1 + 2 -\alpha^2 \frac{\Delta}{\rho^4} \pm i\alpha\rho^2\frac{\dd}{\dd r}\left(\frac{\Delta}{\rho^4}\right)\right],
		\end{align}
	\end{subequations}
	where $\lambda_s$ are the angular eigenvalues, as defined in the text, and $\alpha^2 = a^2 + am/\omega$. For vectors, there are two different values of the potential, depending on the sign of the last term. We have chosen in our numerical code the minus sign.

	\bibliographystyle{apsrev4-1}
	\bibliography{main.bib}

\begin{thebibliography}{125}%
\makeatletter
\providecommand \@ifxundefined [1]{%
 \@ifx{#1\undefined}
}%
\providecommand \@ifnum [1]{%
 \ifnum #1\expandafter \@firstoftwo
 \else \expandafter \@secondoftwo
 \fi
}%
\providecommand \@ifx [1]{%
 \ifx #1\expandafter \@firstoftwo
 \else \expandafter \@secondoftwo
 \fi
}%
\providecommand \natexlab [1]{#1}%
\providecommand \enquote  [1]{``#1''}%
\providecommand \bibnamefont  [1]{#1}%
\providecommand \bibfnamefont [1]{#1}%
\providecommand \citenamefont [1]{#1}%
\providecommand \href@noop [0]{\@secondoftwo}%
\providecommand \href [0]{\begingroup \@sanitize@url \@href}%
\providecommand \@href[1]{\@@startlink{#1}\@@href}%
\providecommand \@@href[1]{\endgroup#1\@@endlink}%
\providecommand \@sanitize@url [0]{\catcode `\\12\catcode `\$12\catcode
  `\&12\catcode `\#12\catcode `\^12\catcode `\_12\catcode `\%12\relax}%
\providecommand \@@startlink[1]{}%
\providecommand \@@endlink[0]{}%
\providecommand \url  [0]{\begingroup\@sanitize@url \@url }%
\providecommand \@url [1]{\endgroup\@href {#1}{\urlprefix }}%
\providecommand \urlprefix  [0]{URL }%
\providecommand \Eprint [0]{\href }%
\providecommand \doibase [0]{http://dx.doi.org/}%
\providecommand \selectlanguage [0]{\@gobble}%
\providecommand \bibinfo  [0]{\@secondoftwo}%
\providecommand \bibfield  [0]{\@secondoftwo}%
\providecommand \translation [1]{[#1]}%
\providecommand \BibitemOpen [0]{}%
\providecommand \bibitemStop [0]{}%
\providecommand \bibitemNoStop [0]{.\EOS\space}%
\providecommand \EOS [0]{\spacefactor3000\relax}%
\providecommand \BibitemShut  [1]{\csname bibitem#1\endcsname}%
\let\auto@bib@innerbib\@empty
\bibitem [{\citenamefont {Abbott}\ \emph {et~al.}(2016)\citenamefont {Abbott}
  \emph {et~al.}}]{LIGOScientific:2016aoc}%
  \BibitemOpen
  \bibfield  {author} {\bibinfo {author} {\bibfnamefont {B.~P.}\ \bibnamefont
  {Abbott}} \emph {et~al.} (\bibinfo {collaboration} {LIGO Scientific,
  Virgo}),\ }\href {\doibase 10.1103/PhysRevLett.116.061102} {\bibfield
  {journal} {\bibinfo  {journal} {Phys. Rev. Lett.}\ }\textbf {\bibinfo
  {volume} {116}},\ \bibinfo {pages} {061102} (\bibinfo {year} {2016})},\
  \Eprint {http://arxiv.org/abs/1602.03837} {arXiv:1602.03837 [gr-qc]}
  \BibitemShut {NoStop}%
\bibitem [{\citenamefont {Abbott}\ \emph {et~al.}(2017)\citenamefont {Abbott}
  \emph {et~al.}}]{LIGOScientific:2017vwq}%
  \BibitemOpen
  \bibfield  {author} {\bibinfo {author} {\bibfnamefont {B.~P.}\ \bibnamefont
  {Abbott}} \emph {et~al.} (\bibinfo {collaboration} {LIGO Scientific,
  Virgo}),\ }\href {\doibase 10.1103/PhysRevLett.119.161101} {\bibfield
  {journal} {\bibinfo  {journal} {Phys. Rev. Lett.}\ }\textbf {\bibinfo
  {volume} {119}},\ \bibinfo {pages} {161101} (\bibinfo {year} {2017})},\
  \Eprint {http://arxiv.org/abs/1710.05832} {arXiv:1710.05832 [gr-qc]}
  \BibitemShut {NoStop}%
\bibitem [{\citenamefont {Akiyama}\ \emph {et~al.}(2019)\citenamefont {Akiyama}
  \emph {et~al.}}]{EventHorizonTelescope:2019dse}%
  \BibitemOpen
  \bibfield  {author} {\bibinfo {author} {\bibfnamefont {K.}~\bibnamefont
  {Akiyama}} \emph {et~al.} (\bibinfo {collaboration} {Event Horizon
  Telescope}),\ }\href {\doibase 10.3847/2041-8213/ab0ec7} {\bibfield
  {journal} {\bibinfo  {journal} {Astrophys. J. Lett.}\ }\textbf {\bibinfo
  {volume} {875}},\ \bibinfo {pages} {L1} (\bibinfo {year} {2019})},\ \Eprint
  {http://arxiv.org/abs/1906.11238} {arXiv:1906.11238 [astro-ph.GA]}
  \BibitemShut {NoStop}%
\bibitem [{\citenamefont {Zel'dovich}\ and\ \citenamefont
  {Novikov}(1967)}]{Zeldovich:1967lct}%
  \BibitemOpen
  \bibfield  {author} {\bibinfo {author} {\bibfnamefont {Y.~B.}\ \bibnamefont
  {Zel'dovich}}\ and\ \bibinfo {author} {\bibfnamefont {I.~D.}\ \bibnamefont
  {Novikov}},\ }\href@noop {} {\bibfield  {journal} {\bibinfo  {journal}
  {Soviet Astron. AJ (Engl. Transl. ),}\ }\textbf {\bibinfo {volume} {10}},\
  \bibinfo {pages} {602} (\bibinfo {year} {1967})}\BibitemShut {NoStop}%
\bibitem [{\citenamefont {Hawking}(1971)}]{Hawking:1971ei}%
  \BibitemOpen
  \bibfield  {author} {\bibinfo {author} {\bibfnamefont {S.}~\bibnamefont
  {Hawking}},\ }\href@noop {} {\bibfield  {journal} {\bibinfo  {journal} {Mon.
  Not. Roy. Astron. Soc.}\ }\textbf {\bibinfo {volume} {152}},\ \bibinfo
  {pages} {75} (\bibinfo {year} {1971})}\BibitemShut {NoStop}%
\bibitem [{\citenamefont {Carr}\ and\ \citenamefont
  {Hawking}(1974)}]{Carr:1974nx}%
  \BibitemOpen
  \bibfield  {author} {\bibinfo {author} {\bibfnamefont {B.~J.}\ \bibnamefont
  {Carr}}\ and\ \bibinfo {author} {\bibfnamefont {S.~W.}\ \bibnamefont
  {Hawking}},\ }\href@noop {} {\bibfield  {journal} {\bibinfo  {journal} {Mon.
  Not. Roy. Astron. Soc.}\ }\textbf {\bibinfo {volume} {168}},\ \bibinfo
  {pages} {399} (\bibinfo {year} {1974})}\BibitemShut {NoStop}%
\bibitem [{\citenamefont {Hawking}(1974)}]{Hawking:1974rv}%
  \BibitemOpen
  \bibfield  {author} {\bibinfo {author} {\bibfnamefont {S.~W.}\ \bibnamefont
  {Hawking}},\ }\href {\doibase 10.1038/248030a0} {\bibfield  {journal}
  {\bibinfo  {journal} {Nature}\ }\textbf {\bibinfo {volume} {248}},\ \bibinfo
  {pages} {30} (\bibinfo {year} {1974})}\BibitemShut {NoStop}%
\bibitem [{\citenamefont {Hawking}(1975)}]{Hawking:1974sw}%
  \BibitemOpen
  \bibfield  {author} {\bibinfo {author} {\bibfnamefont {S.~W.}\ \bibnamefont
  {Hawking}},\ }\bibfield  {booktitle} {\emph {\bibinfo {booktitle} {{Euclidean
  quantum gravity}}},\ }\href {\doibase 10.1007/BF02345020, 10.1007/BF01608497}
  {\bibfield  {journal} {\bibinfo  {journal} {Commun. Math. Phys.}\ }\textbf
  {\bibinfo {volume} {43}},\ \bibinfo {pages} {199} (\bibinfo {year} {1975})},\
  \bibinfo {note} {[,167(1975)]}\BibitemShut {NoStop}%
\bibitem [{\citenamefont {Carr}\ \emph {et~al.}(2010)\citenamefont {Carr},
  \citenamefont {Kohri}, \citenamefont {Sendouda},\ and\ \citenamefont
  {Yokoyama}}]{Carr:2009jm}%
  \BibitemOpen
  \bibfield  {author} {\bibinfo {author} {\bibfnamefont {B.~J.}\ \bibnamefont
  {Carr}}, \bibinfo {author} {\bibfnamefont {K.}~\bibnamefont {Kohri}},
  \bibinfo {author} {\bibfnamefont {Y.}~\bibnamefont {Sendouda}}, \ and\
  \bibinfo {author} {\bibfnamefont {J.}~\bibnamefont {Yokoyama}},\ }\href
  {\doibase 10.1103/PhysRevD.81.104019} {\bibfield  {journal} {\bibinfo
  {journal} {Phys. Rev. D}\ }\textbf {\bibinfo {volume} {81}},\ \bibinfo
  {pages} {104019} (\bibinfo {year} {2010})},\ \Eprint
  {http://arxiv.org/abs/0912.5297} {arXiv:0912.5297 [astro-ph.CO]} \BibitemShut
  {NoStop}%
\bibitem [{\citenamefont {Carr}\ \emph {et~al.}(2020)\citenamefont {Carr},
  \citenamefont {Kohri}, \citenamefont {Sendouda},\ and\ \citenamefont
  {Yokoyama}}]{Carr:2020gox}%
  \BibitemOpen
  \bibfield  {author} {\bibinfo {author} {\bibfnamefont {B.}~\bibnamefont
  {Carr}}, \bibinfo {author} {\bibfnamefont {K.}~\bibnamefont {Kohri}},
  \bibinfo {author} {\bibfnamefont {Y.}~\bibnamefont {Sendouda}}, \ and\
  \bibinfo {author} {\bibfnamefont {J.}~\bibnamefont {Yokoyama}},\ }\href@noop
  {} {\  (\bibinfo {year} {2020})},\ \Eprint {http://arxiv.org/abs/2002.12778}
  {arXiv:2002.12778 [astro-ph.CO]} \BibitemShut {NoStop}%
\bibitem [{\citenamefont {Keith}\ \emph {et~al.}(2020)\citenamefont {Keith},
  \citenamefont {Hooper}, \citenamefont {Blinov},\ and\ \citenamefont
  {McDermott}}]{Keith:2020jww}%
  \BibitemOpen
  \bibfield  {author} {\bibinfo {author} {\bibfnamefont {C.}~\bibnamefont
  {Keith}}, \bibinfo {author} {\bibfnamefont {D.}~\bibnamefont {Hooper}},
  \bibinfo {author} {\bibfnamefont {N.}~\bibnamefont {Blinov}}, \ and\ \bibinfo
  {author} {\bibfnamefont {S.~D.}\ \bibnamefont {McDermott}},\ }\href@noop {}
  {\  (\bibinfo {year} {2020})},\ \Eprint {http://arxiv.org/abs/2006.03608}
  {arXiv:2006.03608 [astro-ph.CO]} \BibitemShut {NoStop}%
\bibitem [{\citenamefont {He}\ and\ \citenamefont {Fang}(2002)}]{He:2002vz}%
  \BibitemOpen
  \bibfield  {author} {\bibinfo {author} {\bibfnamefont {P.}~\bibnamefont
  {He}}\ and\ \bibinfo {author} {\bibfnamefont {L.-Z.}\ \bibnamefont {Fang}},\
  }\href {\doibase 10.1086/340144} {\bibfield  {journal} {\bibinfo  {journal}
  {Astrophys. J. Lett.}\ }\textbf {\bibinfo {volume} {568}},\ \bibinfo {pages}
  {L1} (\bibinfo {year} {2002})},\ \Eprint
  {http://arxiv.org/abs/astro-ph/0202218} {arXiv:astro-ph/0202218} \BibitemShut
  {NoStop}%
\bibitem [{\citenamefont {Mack}\ and\ \citenamefont
  {Wesley}(2008)}]{Mack:2008nv}%
  \BibitemOpen
  \bibfield  {author} {\bibinfo {author} {\bibfnamefont {K.~J.}\ \bibnamefont
  {Mack}}\ and\ \bibinfo {author} {\bibfnamefont {D.~H.}\ \bibnamefont
  {Wesley}},\ }\href@noop {} {\  (\bibinfo {year} {2008})},\ \Eprint
  {http://arxiv.org/abs/0805.1531} {arXiv:0805.1531 [astro-ph]} \BibitemShut
  {NoStop}%
\bibitem [{\citenamefont {Cheek}\ \emph
  {et~al.}(2022{\natexlab{a}})\citenamefont {Cheek}, \citenamefont {Heurtier},
  \citenamefont {Perez-Gonzalez},\ and\ \citenamefont
  {Turner}}]{Cheek:2021odj}%
  \BibitemOpen
  \bibfield  {author} {\bibinfo {author} {\bibfnamefont {A.}~\bibnamefont
  {Cheek}}, \bibinfo {author} {\bibfnamefont {L.}~\bibnamefont {Heurtier}},
  \bibinfo {author} {\bibfnamefont {Y.~F.}\ \bibnamefont {Perez-Gonzalez}}, \
  and\ \bibinfo {author} {\bibfnamefont {J.}~\bibnamefont {Turner}},\ }\href
  {\doibase 10.1103/PhysRevD.105.015022} {\bibfield  {journal} {\bibinfo
  {journal} {Phys. Rev. D}\ }\textbf {\bibinfo {volume} {105}},\ \bibinfo
  {pages} {015022} (\bibinfo {year} {2022}{\natexlab{a}})},\ \Eprint
  {http://arxiv.org/abs/2107.00013} {arXiv:2107.00013 [hep-ph]} \BibitemShut
  {NoStop}%
\bibitem [{\citenamefont {Cheek}\ \emph
  {et~al.}(2022{\natexlab{b}})\citenamefont {Cheek}, \citenamefont {Heurtier},
  \citenamefont {Perez-Gonzalez},\ and\ \citenamefont
  {Turner}}]{Cheek:2021cfe}%
  \BibitemOpen
  \bibfield  {author} {\bibinfo {author} {\bibfnamefont {A.}~\bibnamefont
  {Cheek}}, \bibinfo {author} {\bibfnamefont {L.}~\bibnamefont {Heurtier}},
  \bibinfo {author} {\bibfnamefont {Y.~F.}\ \bibnamefont {Perez-Gonzalez}}, \
  and\ \bibinfo {author} {\bibfnamefont {J.}~\bibnamefont {Turner}},\ }\href
  {\doibase 10.1103/PhysRevD.105.015023} {\bibfield  {journal} {\bibinfo
  {journal} {Phys. Rev. D}\ }\textbf {\bibinfo {volume} {105}},\ \bibinfo
  {pages} {015023} (\bibinfo {year} {2022}{\natexlab{b}})},\ \Eprint
  {http://arxiv.org/abs/2107.00016} {arXiv:2107.00016 [hep-ph]} \BibitemShut
  {NoStop}%
\bibitem [{\citenamefont {Hooper}\ \emph {et~al.}(2019)\citenamefont {Hooper},
  \citenamefont {Krnjaic},\ and\ \citenamefont {McDermott}}]{Hooper:2019gtx}%
  \BibitemOpen
  \bibfield  {author} {\bibinfo {author} {\bibfnamefont {D.}~\bibnamefont
  {Hooper}}, \bibinfo {author} {\bibfnamefont {G.}~\bibnamefont {Krnjaic}}, \
  and\ \bibinfo {author} {\bibfnamefont {S.~D.}\ \bibnamefont {McDermott}},\
  }\href {\doibase 10.1007/JHEP08(2019)001} {\bibfield  {journal} {\bibinfo
  {journal} {JHEP}\ }\textbf {\bibinfo {volume} {08}},\ \bibinfo {pages} {001}
  (\bibinfo {year} {2019})},\ \Eprint {http://arxiv.org/abs/1905.01301}
  {arXiv:1905.01301 [hep-ph]} \BibitemShut {NoStop}%
\bibitem [{\citenamefont {Masina}(2020)}]{Masina:2020xhk}%
  \BibitemOpen
  \bibfield  {author} {\bibinfo {author} {\bibfnamefont {I.}~\bibnamefont
  {Masina}},\ }\href {\doibase 10.1140/epjp/s13360-020-00564-9} {\bibfield
  {journal} {\bibinfo  {journal} {Eur. Phys. J. Plus}\ }\textbf {\bibinfo
  {volume} {135}},\ \bibinfo {pages} {552} (\bibinfo {year} {2020})},\ \Eprint
  {http://arxiv.org/abs/2004.04740} {arXiv:2004.04740 [hep-ph]} \BibitemShut
  {NoStop}%
\bibitem [{\citenamefont {Morrison}\ \emph {et~al.}(2019)\citenamefont
  {Morrison}, \citenamefont {Profumo},\ and\ \citenamefont
  {Yu}}]{Morrison:2018xla}%
  \BibitemOpen
  \bibfield  {author} {\bibinfo {author} {\bibfnamefont {L.}~\bibnamefont
  {Morrison}}, \bibinfo {author} {\bibfnamefont {S.}~\bibnamefont {Profumo}}, \
  and\ \bibinfo {author} {\bibfnamefont {Y.}~\bibnamefont {Yu}},\ }\href
  {\doibase 10.1088/1475-7516/2019/05/005} {\bibfield  {journal} {\bibinfo
  {journal} {JCAP}\ }\textbf {\bibinfo {volume} {1905}},\ \bibinfo {pages}
  {005} (\bibinfo {year} {2019})},\ \Eprint {http://arxiv.org/abs/1812.10606}
  {arXiv:1812.10606 [astro-ph.CO]} \BibitemShut {NoStop}%
\bibitem [{\citenamefont {Auffinger}\ \emph {et~al.}(2021)\citenamefont
  {Auffinger}, \citenamefont {Masina},\ and\ \citenamefont
  {Orlando}}]{Auffinger:2020afu}%
  \BibitemOpen
  \bibfield  {author} {\bibinfo {author} {\bibfnamefont {J.}~\bibnamefont
  {Auffinger}}, \bibinfo {author} {\bibfnamefont {I.}~\bibnamefont {Masina}}, \
  and\ \bibinfo {author} {\bibfnamefont {G.}~\bibnamefont {Orlando}},\ }\href
  {\doibase 10.1140/epjp/s13360-021-01247-9} {\bibfield  {journal} {\bibinfo
  {journal} {Eur. Phys. J. Plus}\ }\textbf {\bibinfo {volume} {136}},\ \bibinfo
  {pages} {261} (\bibinfo {year} {2021})},\ \Eprint
  {http://arxiv.org/abs/2012.09867} {arXiv:2012.09867 [hep-ph]} \BibitemShut
  {NoStop}%
\bibitem [{\citenamefont {Khlopov}\ \emph {et~al.}(2006)\citenamefont
  {Khlopov}, \citenamefont {Barrau},\ and\ \citenamefont
  {Grain}}]{Khlopov:2004tn}%
  \BibitemOpen
  \bibfield  {author} {\bibinfo {author} {\bibfnamefont {M.~{\relax Yu}.}\
  \bibnamefont {Khlopov}}, \bibinfo {author} {\bibfnamefont {A.}~\bibnamefont
  {Barrau}}, \ and\ \bibinfo {author} {\bibfnamefont {J.}~\bibnamefont
  {Grain}},\ }\href {\doibase 10.1088/0264-9381/23/6/004} {\bibfield  {journal}
  {\bibinfo  {journal} {Class. Quant. Grav.}\ }\textbf {\bibinfo {volume}
  {23}},\ \bibinfo {pages} {1875} (\bibinfo {year} {2006})},\ \Eprint
  {http://arxiv.org/abs/astro-ph/0406621} {arXiv:astro-ph/0406621 [astro-ph]}
  \BibitemShut {NoStop}%
\bibitem [{\citenamefont {Allahverdi}\ \emph {et~al.}(2018)\citenamefont
  {Allahverdi}, \citenamefont {Dent},\ and\ \citenamefont
  {Osinski}}]{Allahverdi:2017sks}%
  \BibitemOpen
  \bibfield  {author} {\bibinfo {author} {\bibfnamefont {R.}~\bibnamefont
  {Allahverdi}}, \bibinfo {author} {\bibfnamefont {J.}~\bibnamefont {Dent}}, \
  and\ \bibinfo {author} {\bibfnamefont {J.}~\bibnamefont {Osinski}},\ }\href
  {\doibase 10.1103/PhysRevD.97.055013} {\bibfield  {journal} {\bibinfo
  {journal} {Phys. Rev.}\ }\textbf {\bibinfo {volume} {D97}},\ \bibinfo {pages}
  {055013} (\bibinfo {year} {2018})},\ \Eprint
  {http://arxiv.org/abs/1711.10511} {arXiv:1711.10511 [astro-ph.CO]}
  \BibitemShut {NoStop}%
\bibitem [{\citenamefont {Lennon}\ \emph {et~al.}(2018)\citenamefont {Lennon},
  \citenamefont {March-Russell}, \citenamefont {Petrossian-Byrne},\ and\
  \citenamefont {Tillim}}]{Lennon:2017tqq}%
  \BibitemOpen
  \bibfield  {author} {\bibinfo {author} {\bibfnamefont {O.}~\bibnamefont
  {Lennon}}, \bibinfo {author} {\bibfnamefont {J.}~\bibnamefont
  {March-Russell}}, \bibinfo {author} {\bibfnamefont {R.}~\bibnamefont
  {Petrossian-Byrne}}, \ and\ \bibinfo {author} {\bibfnamefont
  {H.}~\bibnamefont {Tillim}},\ }\href {\doibase 10.1088/1475-7516/2018/04/009}
  {\bibfield  {journal} {\bibinfo  {journal} {JCAP}\ }\textbf {\bibinfo
  {volume} {1804}},\ \bibinfo {pages} {009} (\bibinfo {year} {2018})},\ \Eprint
  {http://arxiv.org/abs/1712.07664} {arXiv:1712.07664 [hep-ph]} \BibitemShut
  {NoStop}%
\bibitem [{\citenamefont {Gondolo}\ \emph {et~al.}(2020)\citenamefont
  {Gondolo}, \citenamefont {Sandick},\ and\ \citenamefont {Shams
  Es~Haghi}}]{Gondolo:2020uqv}%
  \BibitemOpen
  \bibfield  {author} {\bibinfo {author} {\bibfnamefont {P.}~\bibnamefont
  {Gondolo}}, \bibinfo {author} {\bibfnamefont {P.}~\bibnamefont {Sandick}}, \
  and\ \bibinfo {author} {\bibfnamefont {B.}~\bibnamefont {Shams Es~Haghi}},\
  }\href {\doibase 10.1103/PhysRevD.102.095018} {\bibfield  {journal} {\bibinfo
   {journal} {Phys. Rev. D}\ }\textbf {\bibinfo {volume} {102}},\ \bibinfo
  {pages} {095018} (\bibinfo {year} {2020})},\ \Eprint
  {http://arxiv.org/abs/2009.02424} {arXiv:2009.02424 [hep-ph]} \BibitemShut
  {NoStop}%
\bibitem [{\citenamefont {Baldes}\ \emph {et~al.}(2020)\citenamefont {Baldes},
  \citenamefont {Decant}, \citenamefont {Hooper},\ and\ \citenamefont
  {Lopez-Honorez}}]{Baldes:2020nuv}%
  \BibitemOpen
  \bibfield  {author} {\bibinfo {author} {\bibfnamefont {I.}~\bibnamefont
  {Baldes}}, \bibinfo {author} {\bibfnamefont {Q.}~\bibnamefont {Decant}},
  \bibinfo {author} {\bibfnamefont {D.~C.}\ \bibnamefont {Hooper}}, \ and\
  \bibinfo {author} {\bibfnamefont {L.}~\bibnamefont {Lopez-Honorez}},\ }\href
  {\doibase 10.1088/1475-7516/2020/08/045} {\bibfield  {journal} {\bibinfo
  {journal} {JCAP}\ }\textbf {\bibinfo {volume} {08}},\ \bibinfo {pages} {045}
  (\bibinfo {year} {2020})},\ \Eprint {http://arxiv.org/abs/2004.14773}
  {arXiv:2004.14773 [astro-ph.CO]} \BibitemShut {NoStop}%
\bibitem [{\citenamefont {Bernal}\ and\ \citenamefont
  {Zapata}(2021{\natexlab{a}})}]{Bernal:2020bjf}%
  \BibitemOpen
  \bibfield  {author} {\bibinfo {author} {\bibfnamefont {N.}~\bibnamefont
  {Bernal}}\ and\ \bibinfo {author} {\bibfnamefont {O.}~\bibnamefont
  {Zapata}},\ }\href {\doibase 10.1088/1475-7516/2021/03/015} {\bibfield
  {journal} {\bibinfo  {journal} {JCAP}\ }\textbf {\bibinfo {volume} {03}},\
  \bibinfo {pages} {015} (\bibinfo {year} {2021}{\natexlab{a}})},\ \Eprint
  {http://arxiv.org/abs/2011.12306} {arXiv:2011.12306 [astro-ph.CO]}
  \BibitemShut {NoStop}%
\bibitem [{\citenamefont {Bernal}\ and\ \citenamefont
  {Zapata}(2021{\natexlab{b}})}]{Bernal:2020ili}%
  \BibitemOpen
  \bibfield  {author} {\bibinfo {author} {\bibfnamefont {N.}~\bibnamefont
  {Bernal}}\ and\ \bibinfo {author} {\bibfnamefont {O.}~\bibnamefont
  {Zapata}},\ }\href {\doibase 10.1016/j.physletb.2021.136129} {\bibfield
  {journal} {\bibinfo  {journal} {Phys. Lett. B}\ }\textbf {\bibinfo {volume}
  {815}},\ \bibinfo {pages} {136129} (\bibinfo {year} {2021}{\natexlab{b}})},\
  \Eprint {http://arxiv.org/abs/2011.02510} {arXiv:2011.02510 [hep-ph]}
  \BibitemShut {NoStop}%
\bibitem [{\citenamefont {Masina}(2021)}]{Masina:2021zpu}%
  \BibitemOpen
  \bibfield  {author} {\bibinfo {author} {\bibfnamefont {I.}~\bibnamefont
  {Masina}},\ }\href {\doibase 10.1134/S0202289321040101} {\bibfield  {journal}
  {\bibinfo  {journal} {Grav. Cosmol.}\ }\textbf {\bibinfo {volume} {27}},\
  \bibinfo {pages} {315} (\bibinfo {year} {2021})},\ \Eprint
  {http://arxiv.org/abs/2103.13825} {arXiv:2103.13825 [gr-qc]} \BibitemShut
  {NoStop}%
\bibitem [{\citenamefont {Kitabayashi}(2021)}]{Kitabayashi:2021hox}%
  \BibitemOpen
  \bibfield  {author} {\bibinfo {author} {\bibfnamefont {T.}~\bibnamefont
  {Kitabayashi}},\ }\href@noop {} {\  (\bibinfo {year} {2021})},\ \Eprint
  {http://arxiv.org/abs/2101.01921} {arXiv:2101.01921 [hep-ph]} \BibitemShut
  {NoStop}%
\bibitem [{\citenamefont {Bernal}\ \emph {et~al.}(2021)\citenamefont {Bernal},
  \citenamefont {Perez-Gonzalez}, \citenamefont {Xu},\ and\ \citenamefont
  {Zapata}}]{Bernal:2021bbv}%
  \BibitemOpen
  \bibfield  {author} {\bibinfo {author} {\bibfnamefont {N.}~\bibnamefont
  {Bernal}}, \bibinfo {author} {\bibfnamefont {Y.~F.}\ \bibnamefont
  {Perez-Gonzalez}}, \bibinfo {author} {\bibfnamefont {Y.}~\bibnamefont {Xu}},
  \ and\ \bibinfo {author} {\bibfnamefont {O.}~\bibnamefont {Zapata}},\ }\href
  {\doibase 10.1103/PhysRevD.104.123536} {\bibfield  {journal} {\bibinfo
  {journal} {Phys. Rev. D}\ }\textbf {\bibinfo {volume} {104}},\ \bibinfo
  {pages} {123536} (\bibinfo {year} {2021})},\ \Eprint
  {http://arxiv.org/abs/2110.04312} {arXiv:2110.04312 [hep-ph]} \BibitemShut
  {NoStop}%
\bibitem [{\citenamefont {Bernal}\ and\ \citenamefont
  {Zapata}(2021{\natexlab{c}})}]{Bernal:2020kse}%
  \BibitemOpen
  \bibfield  {author} {\bibinfo {author} {\bibfnamefont {N.}~\bibnamefont
  {Bernal}}\ and\ \bibinfo {author} {\bibfnamefont {O.}~\bibnamefont
  {Zapata}},\ }\href {\doibase 10.1088/1475-7516/2021/03/007} {\bibfield
  {journal} {\bibinfo  {journal} {JCAP}\ }\textbf {\bibinfo {volume} {03}},\
  \bibinfo {pages} {007} (\bibinfo {year} {2021}{\natexlab{c}})},\ \Eprint
  {http://arxiv.org/abs/2010.09725} {arXiv:2010.09725 [hep-ph]} \BibitemShut
  {NoStop}%
\bibitem [{\citenamefont {Bernal}\ \emph
  {et~al.}(2022{\natexlab{a}})\citenamefont {Bernal}, \citenamefont
  {Perez-Gonzalez},\ and\ \citenamefont {Xu}}]{Bernal:2022oha}%
  \BibitemOpen
  \bibfield  {author} {\bibinfo {author} {\bibfnamefont {N.}~\bibnamefont
  {Bernal}}, \bibinfo {author} {\bibfnamefont {Y.~F.}\ \bibnamefont
  {Perez-Gonzalez}}, \ and\ \bibinfo {author} {\bibfnamefont {Y.}~\bibnamefont
  {Xu}},\ }\href@noop {} {\  (\bibinfo {year} {2022}{\natexlab{a}})},\ \Eprint
  {http://arxiv.org/abs/2205.11522} {arXiv:2205.11522 [hep-ph]} \BibitemShut
  {NoStop}%
\bibitem [{\citenamefont {Cheek}\ \emph {et~al.}(2023)\citenamefont {Cheek},
  \citenamefont {Heurtier}, \citenamefont {Perez-Gonzalez},\ and\ \citenamefont
  {Turner}}]{Cheek:2022mmy}%
  \BibitemOpen
  \bibfield  {author} {\bibinfo {author} {\bibfnamefont {A.}~\bibnamefont
  {Cheek}}, \bibinfo {author} {\bibfnamefont {L.}~\bibnamefont {Heurtier}},
  \bibinfo {author} {\bibfnamefont {Y.~F.}\ \bibnamefont {Perez-Gonzalez}}, \
  and\ \bibinfo {author} {\bibfnamefont {J.}~\bibnamefont {Turner}},\ }\href
  {\doibase 10.1103/PhysRevD.108.015005} {\bibfield  {journal} {\bibinfo
  {journal} {Phys. Rev. D}\ }\textbf {\bibinfo {volume} {108}},\ \bibinfo
  {pages} {015005} (\bibinfo {year} {2023})},\ \Eprint
  {http://arxiv.org/abs/2212.03878} {arXiv:2212.03878 [hep-ph]} \BibitemShut
  {NoStop}%
\bibitem [{\citenamefont {Cheek}\ \emph
  {et~al.}(2022{\natexlab{c}})\citenamefont {Cheek}, \citenamefont {Heurtier},
  \citenamefont {Perez-Gonzalez},\ and\ \citenamefont
  {Turner}}]{Cheek:2022dbx}%
  \BibitemOpen
  \bibfield  {author} {\bibinfo {author} {\bibfnamefont {A.}~\bibnamefont
  {Cheek}}, \bibinfo {author} {\bibfnamefont {L.}~\bibnamefont {Heurtier}},
  \bibinfo {author} {\bibfnamefont {Y.~F.}\ \bibnamefont {Perez-Gonzalez}}, \
  and\ \bibinfo {author} {\bibfnamefont {J.}~\bibnamefont {Turner}},\ }\href
  {\doibase 10.1103/PhysRevD.106.103012} {\bibfield  {journal} {\bibinfo
  {journal} {Phys. Rev. D}\ }\textbf {\bibinfo {volume} {106}},\ \bibinfo
  {pages} {103012} (\bibinfo {year} {2022}{\natexlab{c}})},\ \Eprint
  {http://arxiv.org/abs/2207.09462} {arXiv:2207.09462 [astro-ph.CO]}
  \BibitemShut {NoStop}%
\bibitem [{\citenamefont {Hooper}\ \emph {et~al.}(2020)\citenamefont {Hooper},
  \citenamefont {Krnjaic}, \citenamefont {March-Russell}, \citenamefont
  {McDermott},\ and\ \citenamefont {Petrossian-Byrne}}]{Hooper:2020evu}%
  \BibitemOpen
  \bibfield  {author} {\bibinfo {author} {\bibfnamefont {D.}~\bibnamefont
  {Hooper}}, \bibinfo {author} {\bibfnamefont {G.}~\bibnamefont {Krnjaic}},
  \bibinfo {author} {\bibfnamefont {J.}~\bibnamefont {March-Russell}}, \bibinfo
  {author} {\bibfnamefont {S.~D.}\ \bibnamefont {McDermott}}, \ and\ \bibinfo
  {author} {\bibfnamefont {R.}~\bibnamefont {Petrossian-Byrne}},\ }\href@noop
  {} {\  (\bibinfo {year} {2020})},\ \Eprint {http://arxiv.org/abs/2004.00618}
  {arXiv:2004.00618 [astro-ph.CO]} \BibitemShut {NoStop}%
\bibitem [{\citenamefont {Arbey}\ \emph {et~al.}(2021)\citenamefont {Arbey},
  \citenamefont {Auffinger}, \citenamefont {Sandick}, \citenamefont {Shams
  Es~Haghi},\ and\ \citenamefont {Sinha}}]{Arbey:2021ysg}%
  \BibitemOpen
  \bibfield  {author} {\bibinfo {author} {\bibfnamefont {A.}~\bibnamefont
  {Arbey}}, \bibinfo {author} {\bibfnamefont {J.}~\bibnamefont {Auffinger}},
  \bibinfo {author} {\bibfnamefont {P.}~\bibnamefont {Sandick}}, \bibinfo
  {author} {\bibfnamefont {B.}~\bibnamefont {Shams Es~Haghi}}, \ and\ \bibinfo
  {author} {\bibfnamefont {K.}~\bibnamefont {Sinha}},\ }\href@noop {} {\
  (\bibinfo {year} {2021})},\ \Eprint {http://arxiv.org/abs/2104.04051}
  {arXiv:2104.04051 [astro-ph.CO]} \BibitemShut {NoStop}%
\bibitem [{\citenamefont {Perez-Gonzalez}\ and\ \citenamefont
  {Turner}(2020)}]{Perez-Gonzalez:2020vnz}%
  \BibitemOpen
  \bibfield  {author} {\bibinfo {author} {\bibfnamefont {Y.~F.}\ \bibnamefont
  {Perez-Gonzalez}}\ and\ \bibinfo {author} {\bibfnamefont {J.}~\bibnamefont
  {Turner}},\ }\href@noop {} {\  (\bibinfo {year} {2020})},\ \Eprint
  {http://arxiv.org/abs/2010.03565} {arXiv:2010.03565 [hep-ph]} \BibitemShut
  {NoStop}%
\bibitem [{\citenamefont {Jyoti~Das}\ \emph {et~al.}(2021)\citenamefont
  {Jyoti~Das}, \citenamefont {Mahanta},\ and\ \citenamefont
  {Borah}}]{JyotiDas:2021shi}%
  \BibitemOpen
  \bibfield  {author} {\bibinfo {author} {\bibfnamefont {S.}~\bibnamefont
  {Jyoti~Das}}, \bibinfo {author} {\bibfnamefont {D.}~\bibnamefont {Mahanta}},
  \ and\ \bibinfo {author} {\bibfnamefont {D.}~\bibnamefont {Borah}},\
  }\href@noop {} {\  (\bibinfo {year} {2021})},\ \Eprint
  {http://arxiv.org/abs/2104.14496} {arXiv:2104.14496 [hep-ph]} \BibitemShut
  {NoStop}%
\bibitem [{\citenamefont {Datta}\ \emph {et~al.}(2020)\citenamefont {Datta},
  \citenamefont {Ghosal},\ and\ \citenamefont {Samanta}}]{Datta:2020bht}%
  \BibitemOpen
  \bibfield  {author} {\bibinfo {author} {\bibfnamefont {S.}~\bibnamefont
  {Datta}}, \bibinfo {author} {\bibfnamefont {A.}~\bibnamefont {Ghosal}}, \
  and\ \bibinfo {author} {\bibfnamefont {R.}~\bibnamefont {Samanta}},\
  }\href@noop {} {\  (\bibinfo {year} {2020})},\ \Eprint
  {http://arxiv.org/abs/2012.14981} {arXiv:2012.14981 [hep-ph]} \BibitemShut
  {NoStop}%
\bibitem [{\citenamefont {Fujita}\ \emph {et~al.}(2014)\citenamefont {Fujita},
  \citenamefont {Kawasaki}, \citenamefont {Harigaya},\ and\ \citenamefont
  {Matsuda}}]{Fujita:2014hha}%
  \BibitemOpen
  \bibfield  {author} {\bibinfo {author} {\bibfnamefont {T.}~\bibnamefont
  {Fujita}}, \bibinfo {author} {\bibfnamefont {M.}~\bibnamefont {Kawasaki}},
  \bibinfo {author} {\bibfnamefont {K.}~\bibnamefont {Harigaya}}, \ and\
  \bibinfo {author} {\bibfnamefont {R.}~\bibnamefont {Matsuda}},\ }\href
  {\doibase 10.1103/PhysRevD.89.103501} {\bibfield  {journal} {\bibinfo
  {journal} {Phys. Rev.}\ }\textbf {\bibinfo {volume} {D89}},\ \bibinfo {pages}
  {103501} (\bibinfo {year} {2014})},\ \Eprint {http://arxiv.org/abs/1401.1909}
  {arXiv:1401.1909 [astro-ph.CO]} \BibitemShut {NoStop}%
\bibitem [{\citenamefont {Granelli}\ \emph {et~al.}(2021)\citenamefont
  {Granelli}, \citenamefont {Moffat}, \citenamefont {Perez-Gonzalez},
  \citenamefont {Schulz},\ and\ \citenamefont {Turner}}]{Granelli:2020pim}%
  \BibitemOpen
  \bibfield  {author} {\bibinfo {author} {\bibfnamefont {A.}~\bibnamefont
  {Granelli}}, \bibinfo {author} {\bibfnamefont {K.}~\bibnamefont {Moffat}},
  \bibinfo {author} {\bibfnamefont {Y.~F.}\ \bibnamefont {Perez-Gonzalez}},
  \bibinfo {author} {\bibfnamefont {H.}~\bibnamefont {Schulz}}, \ and\ \bibinfo
  {author} {\bibfnamefont {J.}~\bibnamefont {Turner}},\ }\href {\doibase
  10.1016/j.cpc.2020.107813} {\bibfield  {journal} {\bibinfo  {journal}
  {Comput. Phys. Commun.}\ }\textbf {\bibinfo {volume} {262}},\ \bibinfo
  {pages} {107813} (\bibinfo {year} {2021})},\ \Eprint
  {http://arxiv.org/abs/2007.09150} {arXiv:2007.09150 [hep-ph]} \BibitemShut
  {NoStop}%
\bibitem [{\citenamefont {Hook}(2014)}]{Hook:2014mla}%
  \BibitemOpen
  \bibfield  {author} {\bibinfo {author} {\bibfnamefont {A.}~\bibnamefont
  {Hook}},\ }\href {\doibase 10.1103/PhysRevD.90.083535} {\bibfield  {journal}
  {\bibinfo  {journal} {Phys. Rev. D}\ }\textbf {\bibinfo {volume} {90}},\
  \bibinfo {pages} {083535} (\bibinfo {year} {2014})},\ \Eprint
  {http://arxiv.org/abs/1404.0113} {arXiv:1404.0113 [hep-ph]} \BibitemShut
  {NoStop}%
\bibitem [{\citenamefont {Hamada}\ and\ \citenamefont
  {Iso}(2017)}]{Hamada:2016jnq}%
  \BibitemOpen
  \bibfield  {author} {\bibinfo {author} {\bibfnamefont {Y.}~\bibnamefont
  {Hamada}}\ and\ \bibinfo {author} {\bibfnamefont {S.}~\bibnamefont {Iso}},\
  }\href {\doibase 10.1093/ptep/ptx011} {\bibfield  {journal} {\bibinfo
  {journal} {PTEP}\ }\textbf {\bibinfo {volume} {2017}},\ \bibinfo {pages}
  {033B02} (\bibinfo {year} {2017})},\ \Eprint
  {http://arxiv.org/abs/1610.02586} {arXiv:1610.02586 [hep-ph]} \BibitemShut
  {NoStop}%
\bibitem [{\citenamefont {Hooper}\ and\ \citenamefont
  {Krnjaic}(2021)}]{Hooper:2020otu}%
  \BibitemOpen
  \bibfield  {author} {\bibinfo {author} {\bibfnamefont {D.}~\bibnamefont
  {Hooper}}\ and\ \bibinfo {author} {\bibfnamefont {G.}~\bibnamefont
  {Krnjaic}},\ }\href {\doibase 10.1103/PhysRevD.103.043504} {\bibfield
  {journal} {\bibinfo  {journal} {Phys. Rev. D}\ }\textbf {\bibinfo {volume}
  {103}},\ \bibinfo {pages} {043504} (\bibinfo {year} {2021})},\ \Eprint
  {http://arxiv.org/abs/2010.01134} {arXiv:2010.01134 [hep-ph]} \BibitemShut
  {NoStop}%
\bibitem [{\citenamefont {Chaudhuri}\ and\ \citenamefont
  {Dolgov}(2020)}]{Chaudhuri:2020wjo}%
  \BibitemOpen
  \bibfield  {author} {\bibinfo {author} {\bibfnamefont {A.}~\bibnamefont
  {Chaudhuri}}\ and\ \bibinfo {author} {\bibfnamefont {A.}~\bibnamefont
  {Dolgov}},\ }\href@noop {} {\  (\bibinfo {year} {2020})},\ \Eprint
  {http://arxiv.org/abs/2001.11219} {arXiv:2001.11219 [astro-ph.CO]}
  \BibitemShut {NoStop}%
\bibitem [{\citenamefont {Bernal}\ \emph
  {et~al.}(2022{\natexlab{b}})\citenamefont {Bernal}, \citenamefont {Fong},
  \citenamefont {Perez-Gonzalez},\ and\ \citenamefont
  {Turner}}]{Bernal:2022pue}%
  \BibitemOpen
  \bibfield  {author} {\bibinfo {author} {\bibfnamefont {N.}~\bibnamefont
  {Bernal}}, \bibinfo {author} {\bibfnamefont {C.~S.}\ \bibnamefont {Fong}},
  \bibinfo {author} {\bibfnamefont {Y.~F.}\ \bibnamefont {Perez-Gonzalez}}, \
  and\ \bibinfo {author} {\bibfnamefont {J.}~\bibnamefont {Turner}},\ }\href
  {\doibase 10.1103/PhysRevD.106.035019} {\bibfield  {journal} {\bibinfo
  {journal} {Phys. Rev. D}\ }\textbf {\bibinfo {volume} {106}},\ \bibinfo
  {pages} {035019} (\bibinfo {year} {2022}{\natexlab{b}})},\ \Eprint
  {http://arxiv.org/abs/2203.08823} {arXiv:2203.08823 [hep-ph]} \BibitemShut
  {NoStop}%
\bibitem [{\citenamefont {Barman}\ \emph {et~al.}(2022)\citenamefont {Barman},
  \citenamefont {Borah}, \citenamefont {Jyoti~Das},\ and\ \citenamefont
  {Roshan}}]{Barman:2022pdo}%
  \BibitemOpen
  \bibfield  {author} {\bibinfo {author} {\bibfnamefont {B.}~\bibnamefont
  {Barman}}, \bibinfo {author} {\bibfnamefont {D.}~\bibnamefont {Borah}},
  \bibinfo {author} {\bibfnamefont {S.}~\bibnamefont {Jyoti~Das}}, \ and\
  \bibinfo {author} {\bibfnamefont {R.}~\bibnamefont {Roshan}},\ }\href@noop {}
  {\  (\bibinfo {year} {2022})},\ \Eprint {http://arxiv.org/abs/2212.00052}
  {arXiv:2212.00052 [hep-ph]} \BibitemShut {NoStop}%
\bibitem [{\citenamefont {Gehrman}\ \emph {et~al.}(2022)\citenamefont
  {Gehrman}, \citenamefont {Shams Es~Haghi}, \citenamefont {Sinha},\ and\
  \citenamefont {Xu}}]{Gehrman:2022imk}%
  \BibitemOpen
  \bibfield  {author} {\bibinfo {author} {\bibfnamefont {T.~C.}\ \bibnamefont
  {Gehrman}}, \bibinfo {author} {\bibfnamefont {B.}~\bibnamefont {Shams
  Es~Haghi}}, \bibinfo {author} {\bibfnamefont {K.}~\bibnamefont {Sinha}}, \
  and\ \bibinfo {author} {\bibfnamefont {T.}~\bibnamefont {Xu}},\ }\href@noop
  {} {\  (\bibinfo {year} {2022})},\ \Eprint {http://arxiv.org/abs/2211.08431}
  {arXiv:2211.08431 [hep-ph]} \BibitemShut {NoStop}%
\bibitem [{\citenamefont {Agashe}\ \emph {et~al.}(2023)\citenamefont {Agashe},
  \citenamefont {Chang}, \citenamefont {Clark}, \citenamefont {Dutta},
  \citenamefont {Tsai},\ and\ \citenamefont {Xu}}]{Agashe:2022phd}%
  \BibitemOpen
  \bibfield  {author} {\bibinfo {author} {\bibfnamefont {K.}~\bibnamefont
  {Agashe}}, \bibinfo {author} {\bibfnamefont {J.~H.}\ \bibnamefont {Chang}},
  \bibinfo {author} {\bibfnamefont {S.~J.}\ \bibnamefont {Clark}}, \bibinfo
  {author} {\bibfnamefont {B.}~\bibnamefont {Dutta}}, \bibinfo {author}
  {\bibfnamefont {Y.}~\bibnamefont {Tsai}}, \ and\ \bibinfo {author}
  {\bibfnamefont {T.}~\bibnamefont {Xu}},\ }\href {\doibase
  10.1103/PhysRevD.108.023014} {\bibfield  {journal} {\bibinfo  {journal}
  {Phys. Rev. D}\ }\textbf {\bibinfo {volume} {108}},\ \bibinfo {pages}
  {023014} (\bibinfo {year} {2023})},\ \Eprint
  {http://arxiv.org/abs/2212.11980} {arXiv:2212.11980 [hep-ph]} \BibitemShut
  {NoStop}%
\bibitem [{\citenamefont {Gehrman}\ \emph {et~al.}(2023)\citenamefont
  {Gehrman}, \citenamefont {Shams Es~Haghi}, \citenamefont {Sinha},\ and\
  \citenamefont {Xu}}]{Gehrman:2023esa}%
  \BibitemOpen
  \bibfield  {author} {\bibinfo {author} {\bibfnamefont {T.~C.}\ \bibnamefont
  {Gehrman}}, \bibinfo {author} {\bibfnamefont {B.}~\bibnamefont {Shams
  Es~Haghi}}, \bibinfo {author} {\bibfnamefont {K.}~\bibnamefont {Sinha}}, \
  and\ \bibinfo {author} {\bibfnamefont {T.}~\bibnamefont {Xu}},\ }\href@noop
  {} {\  (\bibinfo {year} {2023})},\ \Eprint {http://arxiv.org/abs/2304.09194}
  {arXiv:2304.09194 [hep-ph]} \BibitemShut {NoStop}%
\bibitem [{\citenamefont {Friedlander}\ \emph {et~al.}(2023)\citenamefont
  {Friedlander}, \citenamefont {Song},\ and\ \citenamefont
  {Vincent}}]{Friedlander:2023qmc}%
  \BibitemOpen
  \bibfield  {author} {\bibinfo {author} {\bibfnamefont {A.}~\bibnamefont
  {Friedlander}}, \bibinfo {author} {\bibfnamefont {N.}~\bibnamefont {Song}}, \
  and\ \bibinfo {author} {\bibfnamefont {A.~C.}\ \bibnamefont {Vincent}},\
  }\href@noop {} {\  (\bibinfo {year} {2023})},\ \Eprint
  {http://arxiv.org/abs/2306.01520} {arXiv:2306.01520 [hep-ph]} \BibitemShut
  {NoStop}%
\bibitem [{\citenamefont {Carr}\ \emph {et~al.}(2016)\citenamefont {Carr},
  \citenamefont {Kuhnel},\ and\ \citenamefont {Sandstad}}]{Carr:2016drx}%
  \BibitemOpen
  \bibfield  {author} {\bibinfo {author} {\bibfnamefont {B.}~\bibnamefont
  {Carr}}, \bibinfo {author} {\bibfnamefont {F.}~\bibnamefont {Kuhnel}}, \ and\
  \bibinfo {author} {\bibfnamefont {M.}~\bibnamefont {Sandstad}},\ }\href
  {\doibase 10.1103/PhysRevD.94.083504} {\bibfield  {journal} {\bibinfo
  {journal} {Phys. Rev. D}\ }\textbf {\bibinfo {volume} {94}},\ \bibinfo
  {pages} {083504} (\bibinfo {year} {2016})},\ \Eprint
  {http://arxiv.org/abs/1607.06077} {arXiv:1607.06077 [astro-ph.CO]}
  \BibitemShut {NoStop}%
\bibitem [{\citenamefont {Green}\ and\ \citenamefont
  {Kavanagh}(2021)}]{Green:2020jor}%
  \BibitemOpen
  \bibfield  {author} {\bibinfo {author} {\bibfnamefont {A.~M.}\ \bibnamefont
  {Green}}\ and\ \bibinfo {author} {\bibfnamefont {B.~J.}\ \bibnamefont
  {Kavanagh}},\ }\href {\doibase 10.1088/1361-6471/abc534} {\bibfield
  {journal} {\bibinfo  {journal} {J. Phys. G}\ }\textbf {\bibinfo {volume}
  {48}},\ \bibinfo {pages} {4} (\bibinfo {year} {2021})},\ \Eprint
  {http://arxiv.org/abs/2007.10722} {arXiv:2007.10722 [astro-ph.CO]}
  \BibitemShut {NoStop}%
\bibitem [{\citenamefont {Villanueva-Domingo}\ \emph
  {et~al.}(2021)\citenamefont {Villanueva-Domingo}, \citenamefont {Mena},\ and\
  \citenamefont {Palomares-Ruiz}}]{Villanueva-Domingo:2021spv}%
  \BibitemOpen
  \bibfield  {author} {\bibinfo {author} {\bibfnamefont {P.}~\bibnamefont
  {Villanueva-Domingo}}, \bibinfo {author} {\bibfnamefont {O.}~\bibnamefont
  {Mena}}, \ and\ \bibinfo {author} {\bibfnamefont {S.}~\bibnamefont
  {Palomares-Ruiz}},\ }\href {\doibase 10.3389/fspas.2021.681084} {\bibfield
  {journal} {\bibinfo  {journal} {Front. Astron. Space Sci.}\ }\textbf
  {\bibinfo {volume} {8}},\ \bibinfo {pages} {87} (\bibinfo {year} {2021})},\
  \Eprint {http://arxiv.org/abs/2103.12087} {arXiv:2103.12087 [astro-ph.CO]}
  \BibitemShut {NoStop}%
\bibitem [{\citenamefont {Friedlander}\ \emph {et~al.}(2022)\citenamefont
  {Friedlander}, \citenamefont {Mack}, \citenamefont {Schon}, \citenamefont
  {Song},\ and\ \citenamefont {Vincent}}]{Friedlander:2022ttk}%
  \BibitemOpen
  \bibfield  {author} {\bibinfo {author} {\bibfnamefont {A.}~\bibnamefont
  {Friedlander}}, \bibinfo {author} {\bibfnamefont {K.~J.}\ \bibnamefont
  {Mack}}, \bibinfo {author} {\bibfnamefont {S.}~\bibnamefont {Schon}},
  \bibinfo {author} {\bibfnamefont {N.}~\bibnamefont {Song}}, \ and\ \bibinfo
  {author} {\bibfnamefont {A.~C.}\ \bibnamefont {Vincent}},\ }\href {\doibase
  10.1103/PhysRevD.105.103508} {\bibfield  {journal} {\bibinfo  {journal}
  {Phys. Rev. D}\ }\textbf {\bibinfo {volume} {105}},\ \bibinfo {pages}
  {103508} (\bibinfo {year} {2022})},\ \Eprint
  {http://arxiv.org/abs/2201.11761} {arXiv:2201.11761 [hep-ph]} \BibitemShut
  {NoStop}%
\bibitem [{\citenamefont {Auffinger}(2022)}]{Auffinger:2022khh}%
  \BibitemOpen
  \bibfield  {author} {\bibinfo {author} {\bibfnamefont {J.}~\bibnamefont
  {Auffinger}},\ }\href@noop {} {\  (\bibinfo {year} {2022})},\ \Eprint
  {http://arxiv.org/abs/2206.02672} {arXiv:2206.02672 [astro-ph.CO]}
  \BibitemShut {NoStop}%
\bibitem [{\citenamefont {Glicenstein}\ \emph {et~al.}(2013)\citenamefont
  {Glicenstein}, \citenamefont {Barnacka}, \citenamefont {Vivier},\ and\
  \citenamefont {Herr}}]{Glicenstein:2013vha}%
  \BibitemOpen
  \bibfield  {author} {\bibinfo {author} {\bibfnamefont {J.-F.}\ \bibnamefont
  {Glicenstein}}, \bibinfo {author} {\bibfnamefont {A.}~\bibnamefont
  {Barnacka}}, \bibinfo {author} {\bibfnamefont {M.}~\bibnamefont {Vivier}}, \
  and\ \bibinfo {author} {\bibfnamefont {T.}~\bibnamefont {Herr}} (\bibinfo
  {collaboration} {H.E.S.S.}),\ }in\ \href@noop {} {\emph {\bibinfo {booktitle}
  {{33rd International Cosmic Ray Conference}}}}\ (\bibinfo {year} {2013})\ p.\
  \bibinfo {pages} {0930},\ \Eprint {http://arxiv.org/abs/1307.4898}
  {arXiv:1307.4898 [astro-ph.HE]} \BibitemShut {NoStop}%
\bibitem [{\citenamefont {Tavernier}\ \emph {et~al.}(2020)\citenamefont
  {Tavernier}, \citenamefont {Glicenstein},\ and\ \citenamefont
  {Brun}}]{Tavernier:2019exh}%
  \BibitemOpen
  \bibfield  {author} {\bibinfo {author} {\bibfnamefont {T.}~\bibnamefont
  {Tavernier}}, \bibinfo {author} {\bibfnamefont {J.-F.}\ \bibnamefont
  {Glicenstein}}, \ and\ \bibinfo {author} {\bibfnamefont {F.}~\bibnamefont
  {Brun}},\ }\href {\doibase 10.22323/1.358.0804} {\bibfield  {journal}
  {\bibinfo  {journal} {PoS}\ }\textbf {\bibinfo {volume} {ICRC2019}},\
  \bibinfo {pages} {804} (\bibinfo {year} {2020})},\ \Eprint
  {http://arxiv.org/abs/1909.01620} {arXiv:1909.01620 [astro-ph.HE]}
  \BibitemShut {NoStop}%
\bibitem [{\citenamefont {Abdo}\ \emph {et~al.}(2015)\citenamefont {Abdo} \emph
  {et~al.}}]{Abdo:2014apa}%
  \BibitemOpen
  \bibfield  {author} {\bibinfo {author} {\bibfnamefont {A.~A.}\ \bibnamefont
  {Abdo}} \emph {et~al.},\ }\href {\doibase
  10.1016/j.astropartphys.2014.10.007} {\bibfield  {journal} {\bibinfo
  {journal} {Astropart. Phys.}\ }\textbf {\bibinfo {volume} {64}},\ \bibinfo
  {pages} {4} (\bibinfo {year} {2015})},\ \Eprint
  {http://arxiv.org/abs/1407.1686} {arXiv:1407.1686 [astro-ph.HE]} \BibitemShut
  {NoStop}%
\bibitem [{\citenamefont {Archambault}(2018)}]{Archambault:2017asc}%
  \BibitemOpen
  \bibfield  {author} {\bibinfo {author} {\bibfnamefont {S.}~\bibnamefont
  {Archambault}} (\bibinfo {collaboration} {VERITAS}),\ }\href {\doibase
  10.22323/1.301.0691} {\bibfield  {journal} {\bibinfo  {journal} {PoS}\
  }\textbf {\bibinfo {volume} {ICRC2017}},\ \bibinfo {pages} {691} (\bibinfo
  {year} {2018})},\ \Eprint {http://arxiv.org/abs/1709.00307} {arXiv:1709.00307
  [astro-ph.HE]} \BibitemShut {NoStop}%
\bibitem [{\citenamefont {Abeysekara}\ \emph {et~al.}(2013)\citenamefont
  {Abeysekara} \emph {et~al.}}]{HAWC:2013kzm}%
  \BibitemOpen
  \bibfield  {author} {\bibinfo {author} {\bibfnamefont {A.~U.}\ \bibnamefont
  {Abeysekara}} \emph {et~al.} (\bibinfo {collaboration} {HAWC}),\ }\href@noop
  {} {\  (\bibinfo {year} {2013})},\ \Eprint {http://arxiv.org/abs/1310.0073}
  {arXiv:1310.0073 [astro-ph.HE]} \BibitemShut {NoStop}%
\bibitem [{\citenamefont {Albert}\ \emph {et~al.}(2020)\citenamefont {Albert}
  \emph {et~al.}}]{HAWC:2019wla}%
  \BibitemOpen
  \bibfield  {author} {\bibinfo {author} {\bibfnamefont {A.}~\bibnamefont
  {Albert}} \emph {et~al.} (\bibinfo {collaboration} {HAWC}),\ }\href {\doibase
  10.1088/1475-7516/2020/04/026} {\bibfield  {journal} {\bibinfo  {journal}
  {JCAP}\ }\textbf {\bibinfo {volume} {04}},\ \bibinfo {pages} {026} (\bibinfo
  {year} {2020})},\ \Eprint {http://arxiv.org/abs/1911.04356} {arXiv:1911.04356
  [astro-ph.HE]} \BibitemShut {NoStop}%
\bibitem [{\citenamefont {Ackermann}\ \emph {et~al.}(2018)\citenamefont
  {Ackermann} \emph {et~al.}}]{Fermi-LAT:2018pfs}%
  \BibitemOpen
  \bibfield  {author} {\bibinfo {author} {\bibfnamefont {M.}~\bibnamefont
  {Ackermann}} \emph {et~al.} (\bibinfo {collaboration} {Fermi-LAT}),\ }\href
  {\doibase 10.3847/1538-4357/aaac7b} {\bibfield  {journal} {\bibinfo
  {journal} {Astrophys. J.}\ }\textbf {\bibinfo {volume} {857}},\ \bibinfo
  {pages} {49} (\bibinfo {year} {2018})},\ \Eprint
  {http://arxiv.org/abs/1802.00100} {arXiv:1802.00100 [astro-ph.HE]}
  \BibitemShut {NoStop}%
\bibitem [{\citenamefont {Halzen}\ \emph {et~al.}(1995)\citenamefont {Halzen},
  \citenamefont {Keszthelyi},\ and\ \citenamefont {Zas}}]{Halzen:1995hu}%
  \BibitemOpen
  \bibfield  {author} {\bibinfo {author} {\bibfnamefont {F.}~\bibnamefont
  {Halzen}}, \bibinfo {author} {\bibfnamefont {B.}~\bibnamefont {Keszthelyi}},
  \ and\ \bibinfo {author} {\bibfnamefont {E.}~\bibnamefont {Zas}},\ }\href
  {\doibase 10.1103/PhysRevD.52.3239} {\bibfield  {journal} {\bibinfo
  {journal} {Phys. Rev. D}\ }\textbf {\bibinfo {volume} {52}},\ \bibinfo
  {pages} {3239} (\bibinfo {year} {1995})},\ \Eprint
  {http://arxiv.org/abs/hep-ph/9502268} {arXiv:hep-ph/9502268} \BibitemShut
  {NoStop}%
\bibitem [{\citenamefont {Dave}\ and\ \citenamefont
  {Taboada}(2021)}]{Dave:2019epr}%
  \BibitemOpen
  \bibfield  {author} {\bibinfo {author} {\bibfnamefont {P.}~\bibnamefont
  {Dave}}\ and\ \bibinfo {author} {\bibfnamefont {I.}~\bibnamefont {Taboada}}
  (\bibinfo {collaboration} {IceCube}),\ }\href {\doibase 10.22323/1.358.0863}
  {\bibfield  {journal} {\bibinfo  {journal} {PoS}\ }\textbf {\bibinfo {volume}
  {ICRC2019}},\ \bibinfo {pages} {863} (\bibinfo {year} {2021})},\ \Eprint
  {http://arxiv.org/abs/1908.05403} {arXiv:1908.05403 [astro-ph.HE]}
  \BibitemShut {NoStop}%
\bibitem [{\citenamefont {Capanema}\ \emph {et~al.}(2021)\citenamefont
  {Capanema}, \citenamefont {Esmaeili},\ and\ \citenamefont
  {Esmaili}}]{Capanema:2021hnm}%
  \BibitemOpen
  \bibfield  {author} {\bibinfo {author} {\bibfnamefont {A.}~\bibnamefont
  {Capanema}}, \bibinfo {author} {\bibfnamefont {A.}~\bibnamefont {Esmaeili}},
  \ and\ \bibinfo {author} {\bibfnamefont {A.}~\bibnamefont {Esmaili}},\ }\href
  {\doibase 10.1088/1475-7516/2021/12/051} {\bibfield  {journal} {\bibinfo
  {journal} {JCAP}\ }\textbf {\bibinfo {volume} {12}},\ \bibinfo {pages} {051}
  (\bibinfo {year} {2021})},\ \Eprint {http://arxiv.org/abs/2110.05637}
  {arXiv:2110.05637 [hep-ph]} \BibitemShut {NoStop}%
\bibitem [{\citenamefont {Ukwatta}\ \emph {et~al.}(2016)\citenamefont
  {Ukwatta}, \citenamefont {Stump}, \citenamefont {Linnemann}, \citenamefont
  {MacGibbon}, \citenamefont {Marinelli}, \citenamefont {Yapici},\ and\
  \citenamefont {Tollefson}}]{Ukwatta:2015iba}%
  \BibitemOpen
  \bibfield  {author} {\bibinfo {author} {\bibfnamefont {T.~N.}\ \bibnamefont
  {Ukwatta}}, \bibinfo {author} {\bibfnamefont {D.~R.}\ \bibnamefont {Stump}},
  \bibinfo {author} {\bibfnamefont {J.~T.}\ \bibnamefont {Linnemann}}, \bibinfo
  {author} {\bibfnamefont {J.~H.}\ \bibnamefont {MacGibbon}}, \bibinfo {author}
  {\bibfnamefont {S.~S.}\ \bibnamefont {Marinelli}}, \bibinfo {author}
  {\bibfnamefont {T.}~\bibnamefont {Yapici}}, \ and\ \bibinfo {author}
  {\bibfnamefont {K.}~\bibnamefont {Tollefson}},\ }\href {\doibase
  10.1016/j.astropartphys.2016.03.007} {\bibfield  {journal} {\bibinfo
  {journal} {Astropart. Phys.}\ }\textbf {\bibinfo {volume} {80}},\ \bibinfo
  {pages} {90} (\bibinfo {year} {2016})},\ \Eprint
  {http://arxiv.org/abs/1510.04372} {arXiv:1510.04372 [astro-ph.HE]}
  \BibitemShut {NoStop}%
\bibitem [{\citenamefont {Baker}\ and\ \citenamefont
  {Thamm}(2022)}]{Baker:2021btk}%
  \BibitemOpen
  \bibfield  {author} {\bibinfo {author} {\bibfnamefont {M.~J.}\ \bibnamefont
  {Baker}}\ and\ \bibinfo {author} {\bibfnamefont {A.}~\bibnamefont {Thamm}},\
  }\href {\doibase 10.21468/SciPostPhys.12.5.150} {\bibfield  {journal}
  {\bibinfo  {journal} {SciPost Phys.}\ }\textbf {\bibinfo {volume} {12}},\
  \bibinfo {pages} {150} (\bibinfo {year} {2022})},\ \Eprint
  {http://arxiv.org/abs/2105.10506} {arXiv:2105.10506 [hep-ph]} \BibitemShut
  {NoStop}%
\bibitem [{\citenamefont {Baker}\ and\ \citenamefont
  {Thamm}(2023)}]{Baker:2022rkn}%
  \BibitemOpen
  \bibfield  {author} {\bibinfo {author} {\bibfnamefont {M.~J.}\ \bibnamefont
  {Baker}}\ and\ \bibinfo {author} {\bibfnamefont {A.}~\bibnamefont {Thamm}},\
  }\href {\doibase 10.1007/JHEP01(2023)063} {\bibfield  {journal} {\bibinfo
  {journal} {JHEP}\ }\textbf {\bibinfo {volume} {01}},\ \bibinfo {pages} {063}
  (\bibinfo {year} {2023})},\ \Eprint {http://arxiv.org/abs/2210.02805}
  {arXiv:2210.02805 [hep-ph]} \BibitemShut {NoStop}%
\bibitem [{\citenamefont {Boluna}\ \emph {et~al.}(2023)\citenamefont {Boluna},
  \citenamefont {Profumo}, \citenamefont {Bl\'e},\ and\ \citenamefont
  {Hennings}}]{Boluna:2023jlo}%
  \BibitemOpen
  \bibfield  {author} {\bibinfo {author} {\bibfnamefont {X.}~\bibnamefont
  {Boluna}}, \bibinfo {author} {\bibfnamefont {S.}~\bibnamefont {Profumo}},
  \bibinfo {author} {\bibfnamefont {J.}~\bibnamefont {Bl\'e}}, \ and\ \bibinfo
  {author} {\bibfnamefont {D.}~\bibnamefont {Hennings}},\ }\href@noop {} {\
  (\bibinfo {year} {2023})},\ \Eprint {http://arxiv.org/abs/2307.06467}
  {arXiv:2307.06467 [astro-ph.HE]} \BibitemShut {NoStop}%
\bibitem [{\citenamefont {Page}(1976{\natexlab{a}})}]{Page:1976df}%
  \BibitemOpen
  \bibfield  {author} {\bibinfo {author} {\bibfnamefont {D.~N.}\ \bibnamefont
  {Page}},\ }\href {\doibase 10.1103/PhysRevD.13.198} {\bibfield  {journal}
  {\bibinfo  {journal} {Phys. Rev. D}\ }\textbf {\bibinfo {volume} {13}},\
  \bibinfo {pages} {198} (\bibinfo {year} {1976}{\natexlab{a}})}\BibitemShut
  {NoStop}%
\bibitem [{\citenamefont {Page}(1976{\natexlab{b}})}]{Page:1976ki}%
  \BibitemOpen
  \bibfield  {author} {\bibinfo {author} {\bibfnamefont {D.~N.}\ \bibnamefont
  {Page}},\ }\href {\doibase 10.1103/PhysRevD.14.3260} {\bibfield  {journal}
  {\bibinfo  {journal} {Phys. Rev. D}\ }\textbf {\bibinfo {volume} {14}},\
  \bibinfo {pages} {3260} (\bibinfo {year} {1976}{\natexlab{b}})}\BibitemShut
  {NoStop}%
\bibitem [{\citenamefont {Arvanitaki}\ \emph {et~al.}(2010)\citenamefont
  {Arvanitaki}, \citenamefont {Dimopoulos}, \citenamefont {Dubovsky},
  \citenamefont {Kaloper},\ and\ \citenamefont
  {March-Russell}}]{Arvanitaki:2009fg}%
  \BibitemOpen
  \bibfield  {author} {\bibinfo {author} {\bibfnamefont {A.}~\bibnamefont
  {Arvanitaki}}, \bibinfo {author} {\bibfnamefont {S.}~\bibnamefont
  {Dimopoulos}}, \bibinfo {author} {\bibfnamefont {S.}~\bibnamefont
  {Dubovsky}}, \bibinfo {author} {\bibfnamefont {N.}~\bibnamefont {Kaloper}}, \
  and\ \bibinfo {author} {\bibfnamefont {J.}~\bibnamefont {March-Russell}},\
  }\href {\doibase 10.1103/PhysRevD.81.123530} {\bibfield  {journal} {\bibinfo
  {journal} {Phys. Rev. D}\ }\textbf {\bibinfo {volume} {81}},\ \bibinfo
  {pages} {123530} (\bibinfo {year} {2010})},\ \Eprint
  {http://arxiv.org/abs/0905.4720} {arXiv:0905.4720 [hep-th]} \BibitemShut
  {NoStop}%
\bibitem [{\citenamefont {Chambers}\ \emph {et~al.}(1997)\citenamefont
  {Chambers}, \citenamefont {Hiscock},\ and\ \citenamefont
  {Taylor}}]{Chambers:1997ai}%
  \BibitemOpen
  \bibfield  {author} {\bibinfo {author} {\bibfnamefont {C.~M.}\ \bibnamefont
  {Chambers}}, \bibinfo {author} {\bibfnamefont {W.~A.}\ \bibnamefont
  {Hiscock}}, \ and\ \bibinfo {author} {\bibfnamefont {B.}~\bibnamefont
  {Taylor}},\ }\href {\doibase 10.1103/PhysRevLett.78.3249} {\bibfield
  {journal} {\bibinfo  {journal} {Phys. Rev. Lett.}\ }\textbf {\bibinfo
  {volume} {78}},\ \bibinfo {pages} {3249} (\bibinfo {year} {1997})},\ \Eprint
  {http://arxiv.org/abs/gr-qc/9703018} {arXiv:gr-qc/9703018} \BibitemShut
  {NoStop}%
\bibitem [{\citenamefont {Taylor}\ \emph {et~al.}(1998)\citenamefont {Taylor},
  \citenamefont {Chambers},\ and\ \citenamefont {Hiscock}}]{Taylor:1998dk}%
  \BibitemOpen
  \bibfield  {author} {\bibinfo {author} {\bibfnamefont {B.~E.}\ \bibnamefont
  {Taylor}}, \bibinfo {author} {\bibfnamefont {C.~M.}\ \bibnamefont
  {Chambers}}, \ and\ \bibinfo {author} {\bibfnamefont {W.~A.}\ \bibnamefont
  {Hiscock}},\ }\href {\doibase 10.1103/PhysRevD.58.044012} {\bibfield
  {journal} {\bibinfo  {journal} {Phys. Rev. D}\ }\textbf {\bibinfo {volume}
  {58}},\ \bibinfo {pages} {044012} (\bibinfo {year} {1998})},\ \Eprint
  {http://arxiv.org/abs/gr-qc/9801044} {arXiv:gr-qc/9801044} \BibitemShut
  {NoStop}%
\bibitem [{\citenamefont {Calz\`a}\ \emph {et~al.}(2021)\citenamefont
  {Calz\`a}, \citenamefont {March-Russell},\ and\ \citenamefont
  {Rosa}}]{Calza:2021czr}%
  \BibitemOpen
  \bibfield  {author} {\bibinfo {author} {\bibfnamefont {M.}~\bibnamefont
  {Calz\`a}}, \bibinfo {author} {\bibfnamefont {J.}~\bibnamefont
  {March-Russell}}, \ and\ \bibinfo {author} {\bibfnamefont {J.~a.~G.}\
  \bibnamefont {Rosa}},\ }\href@noop {} {\  (\bibinfo {year} {2021})},\ \Eprint
  {http://arxiv.org/abs/2110.13602} {arXiv:2110.13602 [astro-ph.CO]}
  \BibitemShut {NoStop}%
\bibitem [{\citenamefont {Calz\`a}\ and\ \citenamefont
  {Rosa}(2022)}]{Calza:2022ljw}%
  \BibitemOpen
  \bibfield  {author} {\bibinfo {author} {\bibfnamefont {M.}~\bibnamefont
  {Calz\`a}}\ and\ \bibinfo {author} {\bibfnamefont {J.~a.~G.}\ \bibnamefont
  {Rosa}},\ }\href {\doibase 10.1007/JHEP12(2022)090} {\bibfield  {journal}
  {\bibinfo  {journal} {JHEP}\ }\textbf {\bibinfo {volume} {12}},\ \bibinfo
  {pages} {090} (\bibinfo {year} {2022})},\ \Eprint
  {http://arxiv.org/abs/2210.06500} {arXiv:2210.06500 [gr-qc]} \BibitemShut
  {NoStop}%
\bibitem [{\citenamefont {{NASA}}(1968)}]{nasa_earthrise}%
  \BibitemOpen
  \bibfield  {author} {\bibinfo {author} {\bibnamefont {{NASA}}},\ }\href
  {https://www.nasa.gov/image-feature/apollo-8-earthrise} {\enquote {\bibinfo
  {title} {Earthrise},}\ } (\bibinfo {year} {1968})\BibitemShut {NoStop}%
\bibitem [{\citenamefont {Arbey}\ and\ \citenamefont
  {Auffinger}(2019)}]{Arbey:2019mbc}%
  \BibitemOpen
  \bibfield  {author} {\bibinfo {author} {\bibfnamefont {A.}~\bibnamefont
  {Arbey}}\ and\ \bibinfo {author} {\bibfnamefont {J.}~\bibnamefont
  {Auffinger}},\ }\href {\doibase 10.1140/epjc/s10052-019-7161-1} {\bibfield
  {journal} {\bibinfo  {journal} {Eur. Phys. J. C}\ }\textbf {\bibinfo {volume}
  {79}},\ \bibinfo {pages} {693} (\bibinfo {year} {2019})},\ \Eprint
  {http://arxiv.org/abs/1905.04268} {arXiv:1905.04268 [gr-qc]} \BibitemShut
  {NoStop}%
\bibitem [{\citenamefont {Arbey}\ and\ \citenamefont
  {Auffinger}(2021)}]{Arbey:2021mbl}%
  \BibitemOpen
  \bibfield  {author} {\bibinfo {author} {\bibfnamefont {A.}~\bibnamefont
  {Arbey}}\ and\ \bibinfo {author} {\bibfnamefont {J.}~\bibnamefont
  {Auffinger}},\ }\href {\doibase 10.1140/epjc/s10052-021-09702-8} {\bibfield
  {journal} {\bibinfo  {journal} {Eur. Phys. J. C}\ }\textbf {\bibinfo {volume}
  {81}},\ \bibinfo {pages} {910} (\bibinfo {year} {2021})},\ \Eprint
  {http://arxiv.org/abs/2108.02737} {arXiv:2108.02737 [gr-qc]} \BibitemShut
  {NoStop}%
\bibitem [{\citenamefont {Unruh}(1973)}]{Unruh:1973bda}%
  \BibitemOpen
  \bibfield  {author} {\bibinfo {author} {\bibfnamefont {W.}~\bibnamefont
  {Unruh}},\ }\href {\doibase 10.1103/PhysRevLett.31.1265} {\bibfield
  {journal} {\bibinfo  {journal} {Phys. Rev. Lett.}\ }\textbf {\bibinfo
  {volume} {31}},\ \bibinfo {pages} {1265} (\bibinfo {year}
  {1973})}\BibitemShut {NoStop}%
\bibitem [{\citenamefont {Vilenkin}(1978)}]{Vilenkin:1978is}%
  \BibitemOpen
  \bibfield  {author} {\bibinfo {author} {\bibfnamefont {A.}~\bibnamefont
  {Vilenkin}},\ }\href {\doibase 10.1103/PhysRevLett.41.1575} {\bibfield
  {journal} {\bibinfo  {journal} {Phys. Rev. Lett.}\ }\textbf {\bibinfo
  {volume} {41}},\ \bibinfo {pages} {1575} (\bibinfo {year}
  {1978})}\BibitemShut {NoStop}%
\bibitem [{\citenamefont {Leahy}\ and\ \citenamefont
  {Unruh}(1979)}]{Leahy:1979xi}%
  \BibitemOpen
  \bibfield  {author} {\bibinfo {author} {\bibfnamefont {D.~A.}\ \bibnamefont
  {Leahy}}\ and\ \bibinfo {author} {\bibfnamefont {W.~G.}\ \bibnamefont
  {Unruh}},\ }\href {\doibase 10.1103/PhysRevD.19.3509} {\bibfield  {journal}
  {\bibinfo  {journal} {Phys. Rev. D}\ }\textbf {\bibinfo {volume} {19}},\
  \bibinfo {pages} {3509} (\bibinfo {year} {1979})}\BibitemShut {NoStop}%
\bibitem [{\citenamefont {Vilenkin}(1979)}]{Vilenkin:1979ui}%
  \BibitemOpen
  \bibfield  {author} {\bibinfo {author} {\bibfnamefont {A.}~\bibnamefont
  {Vilenkin}},\ }\href {\doibase 10.1103/PhysRevD.20.1807} {\bibfield
  {journal} {\bibinfo  {journal} {Phys. Rev. D}\ }\textbf {\bibinfo {volume}
  {20}},\ \bibinfo {pages} {1807} (\bibinfo {year} {1979})}\BibitemShut
  {NoStop}%
\bibitem [{\citenamefont {Vilenkin}(1980)}]{Vilenkin:1980ft}%
  \BibitemOpen
  \bibfield  {author} {\bibinfo {author} {\bibfnamefont {A.}~\bibnamefont
  {Vilenkin}},\ }\href {\doibase 10.1103/PhysRevD.22.3067} {\bibfield
  {journal} {\bibinfo  {journal} {Phys. Rev. D}\ }\textbf {\bibinfo {volume}
  {22}},\ \bibinfo {pages} {3067} (\bibinfo {year} {1980})}\BibitemShut
  {NoStop}%
\bibitem [{\citenamefont {Aartsen}\ \emph {et~al.}(2019)\citenamefont {Aartsen}
  \emph {et~al.}}]{IceCube:2018pgc}%
  \BibitemOpen
  \bibfield  {author} {\bibinfo {author} {\bibfnamefont {M.~G.}\ \bibnamefont
  {Aartsen}} \emph {et~al.} (\bibinfo {collaboration} {IceCube}),\ }\href
  {\doibase 10.1103/PhysRevD.99.032004} {\bibfield  {journal} {\bibinfo
  {journal} {Phys. Rev. D}\ }\textbf {\bibinfo {volume} {99}},\ \bibinfo
  {pages} {032004} (\bibinfo {year} {2019})},\ \Eprint
  {http://arxiv.org/abs/1808.07629} {arXiv:1808.07629 [hep-ex]} \BibitemShut
  {NoStop}%
\bibitem [{\citenamefont {Aartsen}\ \emph {et~al.}(2020)\citenamefont {Aartsen}
  \emph {et~al.}}]{IceCube:2019cia}%
  \BibitemOpen
  \bibfield  {author} {\bibinfo {author} {\bibfnamefont {M.~G.}\ \bibnamefont
  {Aartsen}} \emph {et~al.} (\bibinfo {collaboration} {IceCube}),\ }\href
  {\doibase 10.1103/PhysRevLett.124.051103} {\bibfield  {journal} {\bibinfo
  {journal} {Phys. Rev. Lett.}\ }\textbf {\bibinfo {volume} {124}},\ \bibinfo
  {pages} {051103} (\bibinfo {year} {2020})},\ \Eprint
  {http://arxiv.org/abs/1910.08488} {arXiv:1910.08488 [astro-ph.HE]}
  \BibitemShut {NoStop}%
\bibitem [{\citenamefont {Boyer}\ and\ \citenamefont
  {Lindquist}(1967)}]{Boyer:1966qh}%
  \BibitemOpen
  \bibfield  {author} {\bibinfo {author} {\bibfnamefont {R.~H.}\ \bibnamefont
  {Boyer}}\ and\ \bibinfo {author} {\bibfnamefont {R.~W.}\ \bibnamefont
  {Lindquist}},\ }\href {\doibase 10.1063/1.1705193} {\bibfield  {journal}
  {\bibinfo  {journal} {J. Math. Phys.}\ }\textbf {\bibinfo {volume} {8}},\
  \bibinfo {pages} {265} (\bibinfo {year} {1967})}\BibitemShut {NoStop}%
\bibitem [{\citenamefont {Teukolsky}(1973)}]{Teukolsky:1973ha}%
  \BibitemOpen
  \bibfield  {author} {\bibinfo {author} {\bibfnamefont {S.~A.}\ \bibnamefont
  {Teukolsky}},\ }\href {\doibase 10.1086/152444} {\bibfield  {journal}
  {\bibinfo  {journal} {Astrophys. J.}\ }\textbf {\bibinfo {volume} {185}},\
  \bibinfo {pages} {635} (\bibinfo {year} {1973})}\BibitemShut {NoStop}%
\bibitem [{\citenamefont {Press}\ and\ \citenamefont
  {Teukolsky}(1973)}]{Press:1973zz}%
  \BibitemOpen
  \bibfield  {author} {\bibinfo {author} {\bibfnamefont {W.~H.}\ \bibnamefont
  {Press}}\ and\ \bibinfo {author} {\bibfnamefont {S.~A.}\ \bibnamefont
  {Teukolsky}},\ }\href {\doibase 10.1086/152445} {\bibfield  {journal}
  {\bibinfo  {journal} {Astrophys. J.}\ }\textbf {\bibinfo {volume} {185}},\
  \bibinfo {pages} {649} (\bibinfo {year} {1973})}\BibitemShut {NoStop}%
\bibitem [{\citenamefont {Teukolsky}\ and\ \citenamefont
  {Press}(1974)}]{Teukolsky:1974yv}%
  \BibitemOpen
  \bibfield  {author} {\bibinfo {author} {\bibfnamefont {S.~A.}\ \bibnamefont
  {Teukolsky}}\ and\ \bibinfo {author} {\bibfnamefont {W.~H.}\ \bibnamefont
  {Press}},\ }\href {\doibase 10.1086/153180} {\bibfield  {journal} {\bibinfo
  {journal} {Astrophys. J.}\ }\textbf {\bibinfo {volume} {193}},\ \bibinfo
  {pages} {443} (\bibinfo {year} {1974})}\BibitemShut {NoStop}%
\bibitem [{\citenamefont {Lunardini}\ and\ \citenamefont
  {Perez-Gonzalez}(2020)}]{Lunardini:2019zob}%
  \BibitemOpen
  \bibfield  {author} {\bibinfo {author} {\bibfnamefont {C.}~\bibnamefont
  {Lunardini}}\ and\ \bibinfo {author} {\bibfnamefont {Y.~F.}\ \bibnamefont
  {Perez-Gonzalez}},\ }\href {\doibase 10.1088/1475-7516/2020/08/014}
  {\bibfield  {journal} {\bibinfo  {journal} {JCAP}\ }\textbf {\bibinfo
  {volume} {08}},\ \bibinfo {pages} {014} (\bibinfo {year} {2020})},\ \Eprint
  {http://arxiv.org/abs/1910.07864} {arXiv:1910.07864 [hep-ph]} \BibitemShut
  {NoStop}%
\bibitem [{\citenamefont {Chandrasekhar}(1976)}]{Chandrasekhar:1976ap}%
  \BibitemOpen
  \bibfield  {author} {\bibinfo {author} {\bibfnamefont {S.}~\bibnamefont
  {Chandrasekhar}},\ }\href {\doibase 10.1098/rspa.1976.0090} {\bibfield
  {journal} {\bibinfo  {journal} {Proc. Roy. Soc. Lond. A}\ }\textbf {\bibinfo
  {volume} {349}},\ \bibinfo {pages} {571} (\bibinfo {year}
  {1976})}\BibitemShut {NoStop}%
\bibitem [{\citenamefont {Page}(1976{\natexlab{c}})}]{Page:1976jj}%
  \BibitemOpen
  \bibfield  {author} {\bibinfo {author} {\bibfnamefont {D.~N.}\ \bibnamefont
  {Page}},\ }\href {\doibase 10.1103/PhysRevD.14.1509} {\bibfield  {journal}
  {\bibinfo  {journal} {Phys. Rev. D}\ }\textbf {\bibinfo {volume} {14}},\
  \bibinfo {pages} {1509} (\bibinfo {year} {1976}{\natexlab{c}})}\BibitemShut
  {NoStop}%
\bibitem [{\citenamefont {Dolan}\ and\ \citenamefont
  {Gair}(2009)}]{Dolan:2009kj}%
  \BibitemOpen
  \bibfield  {author} {\bibinfo {author} {\bibfnamefont {S.}~\bibnamefont
  {Dolan}}\ and\ \bibinfo {author} {\bibfnamefont {J.}~\bibnamefont {Gair}},\
  }\href {\doibase 10.1088/0264-9381/26/17/175020} {\bibfield  {journal}
  {\bibinfo  {journal} {Class. Quant. Grav.}\ }\textbf {\bibinfo {volume}
  {26}},\ \bibinfo {pages} {175020} (\bibinfo {year} {2009})},\ \Eprint
  {http://arxiv.org/abs/0905.2974} {arXiv:0905.2974 [gr-qc]} \BibitemShut
  {NoStop}%
\bibitem [{\citenamefont {Dolan}\ and\ \citenamefont
  {Dempsey}(2015)}]{Dolan:2015eua}%
  \BibitemOpen
  \bibfield  {author} {\bibinfo {author} {\bibfnamefont {S.~R.}\ \bibnamefont
  {Dolan}}\ and\ \bibinfo {author} {\bibfnamefont {D.}~\bibnamefont
  {Dempsey}},\ }\href {\doibase 10.1088/0264-9381/32/18/184001} {\bibfield
  {journal} {\bibinfo  {journal} {Class. Quant. Grav.}\ }\textbf {\bibinfo
  {volume} {32}},\ \bibinfo {pages} {184001} (\bibinfo {year} {2015})},\
  \Eprint {http://arxiv.org/abs/1504.03190} {arXiv:1504.03190 [gr-qc]}
  \BibitemShut {NoStop}%
\bibitem [{\citenamefont {Page}(1977)}]{Page:1977um}%
  \BibitemOpen
  \bibfield  {author} {\bibinfo {author} {\bibfnamefont {D.~N.}\ \bibnamefont
  {Page}},\ }\href {\doibase 10.1103/PhysRevD.16.2402} {\bibfield  {journal}
  {\bibinfo  {journal} {Phys. Rev. D}\ }\textbf {\bibinfo {volume} {16}},\
  \bibinfo {pages} {2402} (\bibinfo {year} {1977})}\BibitemShut {NoStop}%
\bibitem [{\citenamefont {Batic}\ \emph {et~al.}(2005)\citenamefont {Batic},
  \citenamefont {Schmid},\ and\ \citenamefont {Winklmeier}}]{Batic:2004sz}%
  \BibitemOpen
  \bibfield  {author} {\bibinfo {author} {\bibfnamefont {D.}~\bibnamefont
  {Batic}}, \bibinfo {author} {\bibfnamefont {H.}~\bibnamefont {Schmid}}, \
  and\ \bibinfo {author} {\bibfnamefont {M.}~\bibnamefont {Winklmeier}},\
  }\href {\doibase 10.1063/1.1818720} {\bibfield  {journal} {\bibinfo
  {journal} {J. Math. Phys.}\ }\textbf {\bibinfo {volume} {46}},\ \bibinfo
  {pages} {012504} (\bibinfo {year} {2005})},\ \Eprint
  {http://arxiv.org/abs/math-ph/0402047} {arXiv:math-ph/0402047} \BibitemShut
  {NoStop}%
\bibitem [{\citenamefont {Batic}\ and\ \citenamefont
  {Schmid}(2008)}]{Batic:2005va}%
  \BibitemOpen
  \bibfield  {author} {\bibinfo {author} {\bibfnamefont {D.}~\bibnamefont
  {Batic}}\ and\ \bibinfo {author} {\bibfnamefont {H.}~\bibnamefont {Schmid}},\
  }\href@noop {} {\bibfield  {journal} {\bibinfo  {journal} {Rev. Col. Mat.}\
  }\textbf {\bibinfo {volume} {42}},\ \bibinfo {pages} {183} (\bibinfo {year}
  {2008})},\ \Eprint {http://arxiv.org/abs/gr-qc/0512112} {arXiv:gr-qc/0512112}
  \BibitemShut {NoStop}%
\bibitem [{\citenamefont {Neznamov}\ and\ \citenamefont
  {Safronov}(2016)}]{Neznamov:2016qej}%
  \BibitemOpen
  \bibfield  {author} {\bibinfo {author} {\bibfnamefont {V.~P.}\ \bibnamefont
  {Neznamov}}\ and\ \bibinfo {author} {\bibfnamefont {I.~I.}\ \bibnamefont
  {Safronov}},\ }\href {\doibase 10.1142/S0218271816500917} {\bibfield
  {journal} {\bibinfo  {journal} {Int. J. Mod. Phys. D}\ }\textbf {\bibinfo
  {volume} {25}},\ \bibinfo {pages} {1650091} (\bibinfo {year} {2016})},\
  \Eprint {http://arxiv.org/abs/1605.07450} {arXiv:1605.07450 [gr-qc]}
  \BibitemShut {NoStop}%
\bibitem [{\citenamefont {Casals}\ \emph {et~al.}(2007)\citenamefont {Casals},
  \citenamefont {Dolan}, \citenamefont {Kanti},\ and\ \citenamefont
  {Winstanley}}]{Casals:2006xp}%
  \BibitemOpen
  \bibfield  {author} {\bibinfo {author} {\bibfnamefont {M.}~\bibnamefont
  {Casals}}, \bibinfo {author} {\bibfnamefont {S.~R.}\ \bibnamefont {Dolan}},
  \bibinfo {author} {\bibfnamefont {P.}~\bibnamefont {Kanti}}, \ and\ \bibinfo
  {author} {\bibfnamefont {E.}~\bibnamefont {Winstanley}},\ }\href {\doibase
  10.1088/1126-6708/2007/03/019} {\bibfield  {journal} {\bibinfo  {journal}
  {JHEP}\ }\textbf {\bibinfo {volume} {03}},\ \bibinfo {pages} {019} (\bibinfo
  {year} {2007})},\ \Eprint {http://arxiv.org/abs/hep-th/0608193}
  {arXiv:hep-th/0608193} \BibitemShut {NoStop}%
\bibitem [{\citenamefont {Casals}\ \emph {et~al.}(2013)\citenamefont {Casals},
  \citenamefont {Dolan}, \citenamefont {Nolan}, \citenamefont {Ottewill},\ and\
  \citenamefont {Winstanley}}]{Casals:2012es}%
  \BibitemOpen
  \bibfield  {author} {\bibinfo {author} {\bibfnamefont {M.}~\bibnamefont
  {Casals}}, \bibinfo {author} {\bibfnamefont {S.~R.}\ \bibnamefont {Dolan}},
  \bibinfo {author} {\bibfnamefont {B.~C.}\ \bibnamefont {Nolan}}, \bibinfo
  {author} {\bibfnamefont {A.~C.}\ \bibnamefont {Ottewill}}, \ and\ \bibinfo
  {author} {\bibfnamefont {E.}~\bibnamefont {Winstanley}},\ }\href {\doibase
  10.1103/PhysRevD.87.064027} {\bibfield  {journal} {\bibinfo  {journal} {Phys.
  Rev. D}\ }\textbf {\bibinfo {volume} {87}},\ \bibinfo {pages} {064027}
  (\bibinfo {year} {2013})},\ \Eprint {http://arxiv.org/abs/1207.7089}
  {arXiv:1207.7089 [gr-qc]} \BibitemShut {NoStop}%
\bibitem [{\citenamefont {Winstanley}(2016)}]{Winstanley:2013hua}%
  \BibitemOpen
  \bibfield  {author} {\bibinfo {author} {\bibfnamefont {E.}~\bibnamefont
  {Winstanley}},\ }\href {\doibase 10.1007/978-3-319-20046-0_35} {\bibfield
  {journal} {\bibinfo  {journal} {Springer Proc. Phys.}\ }\textbf {\bibinfo
  {volume} {170}},\ \bibinfo {pages} {291} (\bibinfo {year} {2016})},\ \Eprint
  {http://arxiv.org/abs/1310.5064} {arXiv:1310.5064 [gr-qc]} \BibitemShut
  {NoStop}%
\bibitem [{\citenamefont {Unruh}(1976)}]{Unruh:1976db}%
  \BibitemOpen
  \bibfield  {author} {\bibinfo {author} {\bibfnamefont {W.~G.}\ \bibnamefont
  {Unruh}},\ }\href {\doibase 10.1103/PhysRevD.14.870} {\bibfield  {journal}
  {\bibinfo  {journal} {Phys. Rev. D}\ }\textbf {\bibinfo {volume} {14}},\
  \bibinfo {pages} {870} (\bibinfo {year} {1976})}\BibitemShut {NoStop}%
\bibitem [{\citenamefont {Kayser}(1982)}]{Kayser:1982br}%
  \BibitemOpen
  \bibfield  {author} {\bibinfo {author} {\bibfnamefont {B.}~\bibnamefont
  {Kayser}},\ }\href {\doibase 10.1103/PhysRevD.26.1662} {\bibfield  {journal}
  {\bibinfo  {journal} {Phys. Rev. D}\ }\textbf {\bibinfo {volume} {26}},\
  \bibinfo {pages} {1662} (\bibinfo {year} {1982})}\BibitemShut {NoStop}%
\bibitem [{\citenamefont {Kayser}\ and\ \citenamefont
  {Shrock}(1982)}]{Kayser:1981nw}%
  \BibitemOpen
  \bibfield  {author} {\bibinfo {author} {\bibfnamefont {B.}~\bibnamefont
  {Kayser}}\ and\ \bibinfo {author} {\bibfnamefont {R.~E.}\ \bibnamefont
  {Shrock}},\ }\href {\doibase 10.1016/0370-2693(82)90314-8} {\bibfield
  {journal} {\bibinfo  {journal} {Phys. Lett. B}\ }\textbf {\bibinfo {volume}
  {112}},\ \bibinfo {pages} {137} (\bibinfo {year} {1982})}\BibitemShut
  {NoStop}%
\bibitem [{\citenamefont {Chandrasekhar}\ and\ \citenamefont
  {Detweiler}(1975)}]{Chandrasekhar:1975zz}%
  \BibitemOpen
  \bibfield  {author} {\bibinfo {author} {\bibfnamefont {S.}~\bibnamefont
  {Chandrasekhar}}\ and\ \bibinfo {author} {\bibfnamefont {S.~L.}\ \bibnamefont
  {Detweiler}},\ }\href {\doibase 10.1098/rspa.1975.0130} {\bibfield  {journal}
  {\bibinfo  {journal} {Proc. Roy. Soc. Lond. A}\ }\textbf {\bibinfo {volume}
  {345}},\ \bibinfo {pages} {145} (\bibinfo {year} {1975})}\BibitemShut
  {NoStop}%
\bibitem [{\citenamefont {Chandrasekhar}\ and\ \citenamefont
  {Detweiler}(1976)}]{Chandrasekhar:1976zz}%
  \BibitemOpen
  \bibfield  {author} {\bibinfo {author} {\bibfnamefont {S.}~\bibnamefont
  {Chandrasekhar}}\ and\ \bibinfo {author} {\bibfnamefont {S.~L.}\ \bibnamefont
  {Detweiler}},\ }\href {\doibase 10.1098/rspa.1976.0101} {\bibfield  {journal}
  {\bibinfo  {journal} {Proc. Roy. Soc. Lond. A}\ }\textbf {\bibinfo {volume}
  {350}},\ \bibinfo {pages} {165} (\bibinfo {year} {1976})}\BibitemShut
  {NoStop}%
\bibitem [{\citenamefont {Chandrasekhar}\ and\ \citenamefont
  {Detweiler}(1977)}]{Chandrasekhar:1977kf}%
  \BibitemOpen
  \bibfield  {author} {\bibinfo {author} {\bibfnamefont {S.}~\bibnamefont
  {Chandrasekhar}}\ and\ \bibinfo {author} {\bibfnamefont {S.~L.}\ \bibnamefont
  {Detweiler}},\ }\href {\doibase 10.1098/rspa.1977.0002} {\bibfield  {journal}
  {\bibinfo  {journal} {Proc. Roy. Soc. Lond. A}\ }\textbf {\bibinfo {volume}
  {352}},\ \bibinfo {pages} {325} (\bibinfo {year} {1977})}\BibitemShut
  {NoStop}%
\bibitem [{\citenamefont {Leaver}(1985)}]{Leaver:1985ax}%
  \BibitemOpen
  \bibfield  {author} {\bibinfo {author} {\bibfnamefont {E.~W.}\ \bibnamefont
  {Leaver}},\ }\href {\doibase 10.1098/rspa.1985.0119} {\bibfield  {journal}
  {\bibinfo  {journal} {Proc. Roy. Soc. Lond. A}\ }\textbf {\bibinfo {volume}
  {402}},\ \bibinfo {pages} {285} (\bibinfo {year} {1985})}\BibitemShut
  {NoStop}%
\bibitem [{\citenamefont {Berti}\ \emph {et~al.}(2006)\citenamefont {Berti},
  \citenamefont {Cardoso},\ and\ \citenamefont {Casals}}]{Berti:2005gp}%
  \BibitemOpen
  \bibfield  {author} {\bibinfo {author} {\bibfnamefont {E.}~\bibnamefont
  {Berti}}, \bibinfo {author} {\bibfnamefont {V.}~\bibnamefont {Cardoso}}, \
  and\ \bibinfo {author} {\bibfnamefont {M.}~\bibnamefont {Casals}},\ }\href
  {\doibase 10.1103/PhysRevD.73.109902} {\bibfield  {journal} {\bibinfo
  {journal} {Phys. Rev. D}\ }\textbf {\bibinfo {volume} {73}},\ \bibinfo
  {pages} {024013} (\bibinfo {year} {2006})},\ \bibinfo {note} {[Erratum:
  Phys.Rev.D 73, 109902 (2006)]},\ \Eprint {http://arxiv.org/abs/gr-qc/0511111}
  {arXiv:gr-qc/0511111} \BibitemShut {NoStop}%
\bibitem [{\citenamefont {Duffy}\ \emph {et~al.}(2005)\citenamefont {Duffy},
  \citenamefont {Harris}, \citenamefont {Kanti},\ and\ \citenamefont
  {Winstanley}}]{Duffy:2005ns}%
  \BibitemOpen
  \bibfield  {author} {\bibinfo {author} {\bibfnamefont {G.}~\bibnamefont
  {Duffy}}, \bibinfo {author} {\bibfnamefont {C.}~\bibnamefont {Harris}},
  \bibinfo {author} {\bibfnamefont {P.}~\bibnamefont {Kanti}}, \ and\ \bibinfo
  {author} {\bibfnamefont {E.}~\bibnamefont {Winstanley}},\ }\href {\doibase
  10.1088/1126-6708/2005/09/049} {\bibfield  {journal} {\bibinfo  {journal}
  {JHEP}\ }\textbf {\bibinfo {volume} {09}},\ \bibinfo {pages} {049} (\bibinfo
  {year} {2005})},\ \Eprint {http://arxiv.org/abs/hep-th/0507274}
  {arXiv:hep-th/0507274} \BibitemShut {NoStop}%
\bibitem [{\citenamefont {Casals}\ \emph {et~al.}(2006)\citenamefont {Casals},
  \citenamefont {Kanti},\ and\ \citenamefont {Winstanley}}]{Casals:2005sa}%
  \BibitemOpen
  \bibfield  {author} {\bibinfo {author} {\bibfnamefont {M.}~\bibnamefont
  {Casals}}, \bibinfo {author} {\bibfnamefont {P.}~\bibnamefont {Kanti}}, \
  and\ \bibinfo {author} {\bibfnamefont {E.}~\bibnamefont {Winstanley}},\
  }\href {\doibase 10.1088/1126-6708/2006/02/051} {\bibfield  {journal}
  {\bibinfo  {journal} {JHEP}\ }\textbf {\bibinfo {volume} {02}},\ \bibinfo
  {pages} {051} (\bibinfo {year} {2006})},\ \Eprint
  {http://arxiv.org/abs/hep-th/0511163} {arXiv:hep-th/0511163} \BibitemShut
  {NoStop}%
\bibitem [{\citenamefont {Eingorn}\ \emph {et~al.}(2018)\citenamefont
  {Eingorn}, \citenamefont {Fernando}, \citenamefont {Vlahovic}, \citenamefont
  {Ilie}, \citenamefont {Wojtsekhowski}, \citenamefont {Urciuoli},
  \citenamefont {Persio}, \citenamefont {Meddi},\ and\ \citenamefont
  {Nelyubin}}]{Eingorn:2015oga}%
  \BibitemOpen
  \bibfield  {author} {\bibinfo {author} {\bibfnamefont {M.}~\bibnamefont
  {Eingorn}}, \bibinfo {author} {\bibfnamefont {L.}~\bibnamefont {Fernando}},
  \bibinfo {author} {\bibfnamefont {B.}~\bibnamefont {Vlahovic}}, \bibinfo
  {author} {\bibfnamefont {C.}~\bibnamefont {Ilie}}, \bibinfo {author}
  {\bibfnamefont {B.}~\bibnamefont {Wojtsekhowski}}, \bibinfo {author}
  {\bibfnamefont {G.~M.}\ \bibnamefont {Urciuoli}}, \bibinfo {author}
  {\bibfnamefont {F.~D.}\ \bibnamefont {Persio}}, \bibinfo {author}
  {\bibfnamefont {F.}~\bibnamefont {Meddi}}, \ and\ \bibinfo {author}
  {\bibfnamefont {V.}~\bibnamefont {Nelyubin}},\ }\href {\doibase
  10.1117/1.JATIS.4.1.011006} {\bibfield  {journal} {\bibinfo  {journal} {J.
  Astron. Telesc. Instrum. Syst.}\ }\textbf {\bibinfo {volume} {4}},\ \bibinfo
  {pages} {011006} (\bibinfo {year} {2018})},\ \Eprint
  {http://arxiv.org/abs/1501.05592} {arXiv:1501.05592 [astro-ph.HE]}
  \BibitemShut {NoStop}%
\bibitem [{\citenamefont {Depaola}\ \emph {et~al.}(1999)\citenamefont
  {Depaola}, \citenamefont {Kozameh},\ and\ \citenamefont
  {Tiglio}}]{DEPAOLA1999175}%
  \BibitemOpen
  \bibfield  {author} {\bibinfo {author} {\bibfnamefont {G.~O.}\ \bibnamefont
  {Depaola}}, \bibinfo {author} {\bibfnamefont {C.~N.}\ \bibnamefont
  {Kozameh}}, \ and\ \bibinfo {author} {\bibfnamefont {M.~H.}\ \bibnamefont
  {Tiglio}},\ }\href {\doibase https://doi.org/10.1016/S0927-6505(98)00054-1}
  {\bibfield  {journal} {\bibinfo  {journal} {Astroparticle Physics}\ }\textbf
  {\bibinfo {volume} {10}},\ \bibinfo {pages} {175} (\bibinfo {year}
  {1999})}\BibitemShut {NoStop}%
\bibitem [{\citenamefont {Morselli}\ \emph {et~al.}(2013)\citenamefont
  {Morselli} \emph {et~al.}}]{Morselli:2013ntc}%
  \BibitemOpen
  \bibfield  {author} {\bibinfo {author} {\bibfnamefont {A.}~\bibnamefont
  {Morselli}} \emph {et~al.},\ }\href {\doibase
  10.1016/j.nuclphysbps.2013.05.030} {\bibfield  {journal} {\bibinfo  {journal}
  {Nucl. Phys. B Proc. Suppl.}\ }\textbf {\bibinfo {volume} {239-240}},\
  \bibinfo {pages} {193} (\bibinfo {year} {2013})},\ \Eprint
  {http://arxiv.org/abs/1406.1071} {arXiv:1406.1071 [astro-ph.IM]} \BibitemShut
  {NoStop}%
\bibitem [{\citenamefont {Bernard}(2013)}]{Bernard:2013jea}%
  \BibitemOpen
  \bibfield  {author} {\bibinfo {author} {\bibfnamefont {D.}~\bibnamefont
  {Bernard}} (\bibinfo {collaboration} {HARPO}),\ }\href {\doibase
  10.1016/j.nima.2013.07.047} {\bibfield  {journal} {\bibinfo  {journal} {Nucl.
  Instrum. Meth. A}\ }\textbf {\bibinfo {volume} {729}},\ \bibinfo {pages}
  {765} (\bibinfo {year} {2013})},\ \Eprint {http://arxiv.org/abs/1307.3892}
  {arXiv:1307.3892 [astro-ph.IM]} \BibitemShut {NoStop}%
\bibitem [{\citenamefont {Gros}\ and\ \citenamefont
  {Bernard}(2017)}]{GROS201730}%
  \BibitemOpen
  \bibfield  {author} {\bibinfo {author} {\bibfnamefont {P.}~\bibnamefont
  {Gros}}\ and\ \bibinfo {author} {\bibfnamefont {D.}~\bibnamefont {Bernard}},\
  }\href {\doibase https://doi.org/10.1016/j.astropartphys.2016.12.006}
  {\bibfield  {journal} {\bibinfo  {journal} {Astroparticle Physics}\ }\textbf
  {\bibinfo {volume} {88}},\ \bibinfo {pages} {30} (\bibinfo {year}
  {2017})}\BibitemShut {NoStop}%
\bibitem [{\citenamefont {Bierlich}\ \emph {et~al.}(2022)\citenamefont
  {Bierlich} \emph {et~al.}}]{Bierlich:2022pfr}%
  \BibitemOpen
  \bibfield  {author} {\bibinfo {author} {\bibfnamefont {C.}~\bibnamefont
  {Bierlich}} \emph {et~al.},\ }\href {\doibase 10.21468/SciPostPhysCodeb.8} {\
   (\bibinfo {year} {2022}),\ 10.21468/SciPostPhysCodeb.8},\ \Eprint
  {http://arxiv.org/abs/2203.11601} {arXiv:2203.11601 [hep-ph]} \BibitemShut
  {NoStop}%
\bibitem [{\citenamefont {Carr}\ and\ \citenamefont
  {Kuhnel}(2020)}]{Carr:2020xqk}%
  \BibitemOpen
  \bibfield  {author} {\bibinfo {author} {\bibfnamefont {B.}~\bibnamefont
  {Carr}}\ and\ \bibinfo {author} {\bibfnamefont {F.}~\bibnamefont {Kuhnel}},\
  }\href {\doibase 10.1146/annurev-nucl-050520-125911} {\bibfield  {journal}
  {\bibinfo  {journal} {Ann. Rev. Nucl. Part. Sci.}\ }\textbf {\bibinfo
  {volume} {70}},\ \bibinfo {pages} {355} (\bibinfo {year} {2020})},\ \Eprint
  {http://arxiv.org/abs/2006.02838} {arXiv:2006.02838 [astro-ph.CO]}
  \BibitemShut {NoStop}%
\bibitem [{\citenamefont {Clark}(2021)}]{Clark:2021fkg}%
  \BibitemOpen
  \bibfield  {author} {\bibinfo {author} {\bibfnamefont {B.}~\bibnamefont
  {Clark}} (\bibinfo {collaboration} {IceCube-Gen2}),\ }\href {\doibase
  10.1088/1748-0221/16/10/C10007} {\bibfield  {journal} {\bibinfo  {journal}
  {JINST}\ }\textbf {\bibinfo {volume} {16}},\ \bibinfo {pages} {C10007}
  (\bibinfo {year} {2021})},\ \Eprint {http://arxiv.org/abs/2108.05292}
  {arXiv:2108.05292 [astro-ph.HE]} \BibitemShut {NoStop}%
\bibitem [{\citenamefont {Abbasi}\ \emph {et~al.}(2023)\citenamefont {Abbasi}
  \emph {et~al.}}]{IceCube:2023ame}%
  \BibitemOpen
  \bibfield  {author} {\bibinfo {author} {\bibfnamefont {R.}~\bibnamefont
  {Abbasi}} \emph {et~al.} (\bibinfo {collaboration} {IceCube}),\ }\href
  {\doibase 10.1126/science.adc9818} {\bibfield  {journal} {\bibinfo  {journal}
  {Science}\ }\textbf {\bibinfo {volume} {380}},\ \bibinfo {pages} {6652}
  (\bibinfo {year} {2023})},\ \Eprint {http://arxiv.org/abs/2307.04427}
  {arXiv:2307.04427 [astro-ph.HE]} \BibitemShut {NoStop}%
\bibitem [{\citenamefont {Adrian-Martinez}\ \emph {et~al.}(2016)\citenamefont
  {Adrian-Martinez} \emph {et~al.}}]{KM3Net:2016zxf}%
  \BibitemOpen
  \bibfield  {author} {\bibinfo {author} {\bibfnamefont {S.}~\bibnamefont
  {Adrian-Martinez}} \emph {et~al.} (\bibinfo {collaboration} {KM3Net}),\
  }\href {\doibase 10.1088/0954-3899/43/8/084001} {\bibfield  {journal}
  {\bibinfo  {journal} {J. Phys. G}\ }\textbf {\bibinfo {volume} {43}},\
  \bibinfo {pages} {084001} (\bibinfo {year} {2016})},\ \Eprint
  {http://arxiv.org/abs/1601.07459} {arXiv:1601.07459 [astro-ph.IM]}
  \BibitemShut {NoStop}%
\bibitem [{\citenamefont {Agostini}\ \emph {et~al.}(2020)\citenamefont
  {Agostini} \emph {et~al.}}]{P-ONE:2020ljt}%
  \BibitemOpen
  \bibfield  {author} {\bibinfo {author} {\bibfnamefont {M.}~\bibnamefont
  {Agostini}} \emph {et~al.} (\bibinfo {collaboration} {P-ONE}),\ }\href
  {\doibase 10.1038/s41550-020-1182-4} {\bibfield  {journal} {\bibinfo
  {journal} {Nature Astron.}\ }\textbf {\bibinfo {volume} {4}},\ \bibinfo
  {pages} {913} (\bibinfo {year} {2020})},\ \Eprint
  {http://arxiv.org/abs/2005.09493} {arXiv:2005.09493 [astro-ph.HE]}
  \BibitemShut {NoStop}%
\bibitem [{\citenamefont {Ye}\ \emph {et~al.}(2022)\citenamefont {Ye} \emph
  {et~al.}}]{Ye:2022vbk}%
  \BibitemOpen
  \bibfield  {author} {\bibinfo {author} {\bibfnamefont {Z.~P.}\ \bibnamefont
  {Ye}} \emph {et~al.},\ }\href@noop {} {\  (\bibinfo {year} {2022})},\ \Eprint
  {http://arxiv.org/abs/2207.04519} {arXiv:2207.04519 [astro-ph.HE]}
  \BibitemShut {NoStop}%
\bibitem [{\citenamefont {Allakhverdyan}\ \emph {et~al.}(2023)\citenamefont
  {Allakhverdyan} \emph {et~al.}}]{Baikal-GVD:2022fis}%
  \BibitemOpen
  \bibfield  {author} {\bibinfo {author} {\bibfnamefont {V.~A.}\ \bibnamefont
  {Allakhverdyan}} \emph {et~al.} (\bibinfo {collaboration} {Baikal-GVD}),\
  }\href {\doibase 10.1103/PhysRevD.107.042005} {\bibfield  {journal} {\bibinfo
   {journal} {Phys. Rev. D}\ }\textbf {\bibinfo {volume} {107}},\ \bibinfo
  {pages} {042005} (\bibinfo {year} {2023})},\ \Eprint
  {http://arxiv.org/abs/2211.09447} {arXiv:2211.09447 [astro-ph.HE]}
  \BibitemShut {NoStop}%
\end{thebibliography}%
	
\end{document}